\documentclass[10pt,aps,prd,preprintnumbers,showpacs,twocolumn,nofootinbib,notitlepage,longbibliography,superscriptaddress]{revtex4-1}

\usepackage{amsmath, amssymb,braket}
\usepackage[dvipsnames]{xcolor}
\usepackage{color}
\usepackage{float}
\usepackage{graphicx}
\usepackage{subfigure}
\usepackage{natbib}
\usepackage{tikz}
\usetikzlibrary{shapes,arrows}
\usepackage[compat=1.1.0]{tikz-feynman}
\usepackage[colorlinks=true
,urlcolor=blue
,anchorcolor=blue
,citecolor=blue
,filecolor=blue
,linkcolor=red
,menucolor=blue
,hyperfootnotes=false,
,linktocpage=true
,pdfproducer=medialab
,pdfa=true
]{hyperref}
\usepackage{siunitx}
\usepackage{tabularx}
\usepackage{multirow}
\usepackage{dcolumn}
\usepackage{makecell}
\usepackage{fontawesome}
\usepackage{adjustbox}

\newcolumntype{C}[1]{>{\centering\let\newline\\\arraybackslash\hspace{0pt}}m{#1}}

\DeclareSIUnit \h {\ensuremath{\mathit{h}}}
\DeclareSIUnit\eV{e\kern-.05em V}
\DeclareSIUnit\parsec{pc}
\DeclareSIUnit\year{yr}
\DeclareSIUnit\erg{erg}
\DeclareSIUnit\sr{sr}

\def\aj{\rm{AJ}}                   
\def\apj{\rm{ApJ}}                 
\def\apjl{\rm{ApJ}}                
\def\aap{\rm{A\&A}}                

\def\mnras{\rm{MNRAS}}             
\def\prd{\rm{Phys.~Rev.~D}}        
\makeatletter 
    
\renewcommand\onecolumngrid{
\do@columngrid{one}{\@ne}%
\def\set@footnotewidth{\onecolumngrid}
\def\footnoterule{\kern-6pt\hrule width 1.5in\kern6pt}%
}

\renewcommand\twocolumngrid{
        \def\footnoterule{
        \dimen@\skip\footins\divide\dimen@\thr@@
        \kern-\dimen@\hrule width.5in\kern\dimen@}
        \do@columngrid{mlt}{\tw@}
}%

\makeatother    

\newcommand{\githubDH}{\href{https://github.com/hongwanliu/DarkHistory/tree/lowengelec_upgrade}{\faGithub}} 
\newcommand{\githubZeus}{\href{https://github.com/JulianBMunoz/Zeus21}{\faGithub}} 
\newcommand\orcid[1]{\href{http://orcid.org/#1}{\adjustbox{trim={-.15\width} {0\height} {-.15\width} {0\height},clip}{\includegraphics[height=10pt]{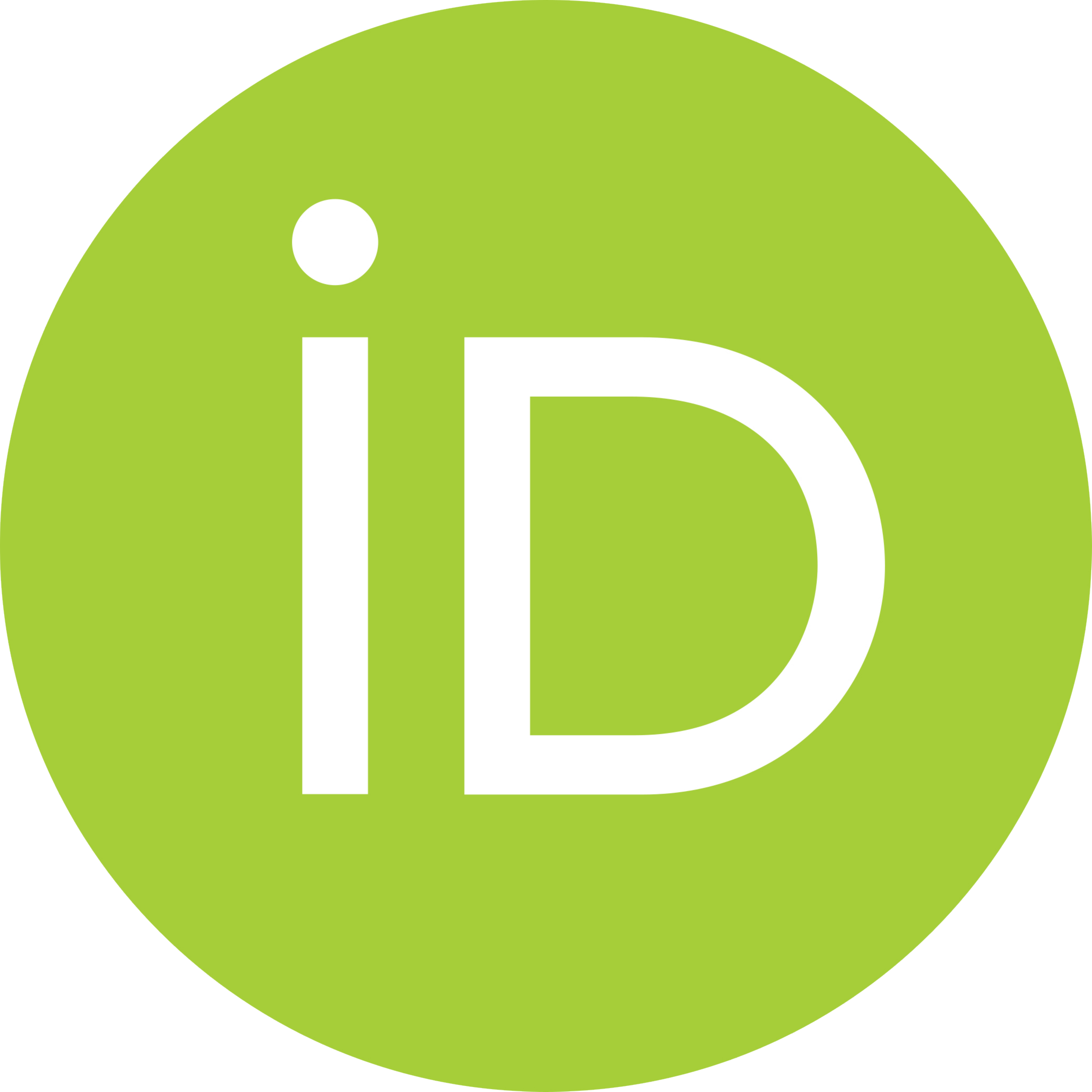}}}}

\def\n{\nonumber \\}

\begin{document}

\preprint{MIT-CTP/5596} 

\title{Birth of the first stars amidst decaying and annihilating dark matter}

\author{Wenzer Qin\orcid{0000-0001-7849-6585}}
\affiliation{Center for Theoretical Physics, Massachusetts Institute of Technology, Cambridge, Massachusetts 02139, USA}

\author{Julian~B. Mu\~{n}oz\orcid{0000-0002-8984-0465}}
\affiliation{Department of Astronomy, The University of Texas at Austin, 2515 Speedway, Stop C1400, Austin, TX 78712, USA}

\author{Hongwan Liu\orcid{0000-0003-2486-0681}}
\affiliation{Center for Cosmology and Particle Physics, Department of Physics, New York University, New York, NY 10003, U.S.A.}
\affiliation{Department of Physics, Princeton University, Princeton, New Jersey, 08544, U.S.A.}

\author{Tracy~R. Slatyer\orcid{0000-0001-9699-9047}}
\affiliation{Center for Theoretical Physics, Massachusetts Institute of Technology, Cambridge, Massachusetts 02139, USA}

\begin{abstract}
    The first stars are expected to form through molecular-hydrogen (H$_2$) cooling, a channel that is especially sensitive to the thermal and ionization state of gas, and can thus act as a probe of exotic energy injection from decaying or annihilating dark matter (DM).
    Here, we use a toy halo model to study the impact of DM-sourced energy injection on the H$_2$ content of the first galaxies, and thus estimate the threshold mass required for a halo to form stars at high redshifts.
    We find that currently allowed DM models can significantly change this threshold, producing both positive and negative feedback.
    In some scenarios, the extra heating of the gas raises the halo mass required for collapse, whereas in others, energy injection lowers the threshold by increasing the free-electron fraction and catalyzing H$_2$ formation.
    The direction of the effect can be redshift-dependent. 
    We also bracket the self-shielding uncertainties on the impact of the Lyman-Werner radiation from DM.
    Hence, exotic energy injection can both delay and accelerate the onset of star formation; we show how this can impact the timing of 21\,cm signals at cosmic dawn.
    We encourage detailed simulation follow-ups in the most promising regions of parameter space identified in this work.
\end{abstract}

\maketitle

\section{Introduction}
\label{sec:intro}

Exotic energy injection by e.g.\ decaying or annihilating dark matter (DM) is capable of heating and ionizing the baryonic gas in our universe, as well as altering the cosmic electromagnetic radiation background.
Such injection is a generic expectation of many proposed DM models, including weakly interacting massive particles (WIMPs).
Since the effects of such exotic energy injection can accumulate over cosmological timescales, interactions beyond the reach of terrestrial experiments can still lead to detectable changes in the global temperature and ionization level of the intergalactic medium (IGM), as well as in the cosmic electromagnetic background, detectable as e.g.\ distortions to the blackbody spectrum of the cosmic microwave background (CMB).

Many searches have been conducted to look for the effect of exotic energy injection on all three of these quantities.
For example, changes to the global ionization history can alter the cosmic microwave background anisotropies; measurements of the CMB power spectrum thus set strong limits on decaying or annihilating DM in the sub-GeV mass range~\cite{Adams:1998nr, Chen:2003gz, Padmanabhan:2005es, Slatyer:2009yq, Kanzaki:2009hf, Slatyer:2015jla, Slatyer:2016qyl, Poulin:2016anj, Cang:2020exa}.
Excess heating of the IGM can be constrained by measurements of the Lyman-$\alpha$ forest~\cite{Hiss:2017qyw, Walther:2018pnn, Gaikwad:2020art, Gaikwad:2020eip}, which has been used to set limits on DM velocity-dependent annihilation and decay~\cite{Cirelli:2009bb, Diamanti:2013bia, Liu:2016cnk, Liu:2020wqz}, as well as dark-photon DM~\cite{McDermott:2019lch, Caputo:2020bdy, Witte:2020rvb}; heating by dark photons has also been invoked to reconcile low- and high-redshift Lyman-$\alpha$ observations of the IGM~\cite{Bolton:2022hpt}. 
In addition, exotic heating has been shown to modify the 21\,cm brightness temperature, such that future observations could also set constraints on energy injection by decaying or annihilating DM, and primordial black holes~\cite{Evoli:2014pva, Lopez-Honorez:2016sur, DAmico:2018sxd, Liu:2018uzy, Cheung:2018vww, Mitridate:2018iag, Clark:2018ghm}.
Lastly, new contributions to background radiation could be observed as distortions to the CMB blackbody spectrum~\cite{Liu:2023fgu,Liu:2023nct}, which has been used to set limits on a range of DM models before recombination~\cite{Chluba:2013vsa,Ali-Haimoud:2015pwa,Chluba:2015hma,Acharya:2018iwh}.

Current constraints still allow exotic energy injection at a level as large as \SI{1}{\eV} of energy per baryon (corresponding to a temperature of $\sim 10^4$ K) to the IGM at $2 \lesssim z \lesssim 5$~\cite{Liu:2020wqz}; however, such high gas temperatures may also affect gas collapse and subsequent star formation during the earlier epoch of cosmic dawn, when the gas temperature is expected to be much colder~\cite{Munoz:2018pzp}. 
The prospect of new observational probes of this epoch, e.g.\ 21\,cm telescopes, motivates understanding how the formation of the first stars will be impacted by exotic energy injection.

The first luminous objects are expected to have formed in pristine environments of primordial gas that have not yet been reprocessed by stars and galaxies.
For the small halos that collapse first, the only available coolants are atomic hydrogen, which is only effective at temperatures above $\sim\SI{e4}{\kelvin}$, and molecular hydrogen (H$_2$), which is thus thought to be crucial for early star formation.
The formation of H$_2$ depends on the temperature, ionization, and light irradiating the host halo; since exotic energy injection can affect all of these quantities, it may be possible to observe signatures of decaying or annihilating DM as we sharpen our understanding of the first stellar formation.
However, while we have made much progress in understanding a number of effects that are important for early star formation (such as the various channels by which H$_2$ forms, its dissociation by Lyman-Werner (LW) radiation, self-shielding of H$_2$, and baryon streaming velocities) there are still a number of astrophysical uncertainties regarding this process~\cite{2016MNRAS.462..601R,2018MNRAS.474..443G,Mebane:2020jwl,Munoz:2021psm,Hartwig:2022lon,Nebrin:2023yzm}.
Observations at high redshift by the James Webb Space Telescope (JWST)~\cite{Gardner:2006ky,2013MNRAS.429.3658R,2011ApJ...740...13Z,2022ApJ...937L...6R,2023AJ....165....2L}, Giant Magellan Telescope (GMT)~\cite{2012SPIE.8444E..1HJ}, or radio interferometers such as the Hydrogen Epoch of Reionization Array (HERA)~\cite{DeBoer:2016tnn,HERA:2021noe} and Square Kilometre Array (SKA)~\cite{Weltman:2018zrl,SKA:2018ckk} will improve our understanding of these effects and thus also help us constrain the impact of exotic energy injection. 

In this work, we seek a qualitative understanding of the effect of exotic energy injection on the first star-forming halos using a simple calculation for the molecular hydrogen formation rate and a toy model for collapsing halos.
A precise study will eventually require hydrodynamic simulations including all of the aforementioned effects (see e.g.\ Refs~\cite{Machacek:2000us,Yoshida:2003rw,Latif:2019zdi,Schauer:2018iig,2020MNRAS.492.4386S,Kulkarni:2020ovu,Park:2021sbs,Schauer:2020gvx,Schauer:2022cgf}). 
However, such simulations are prohibitively expensive to run for each DM scenario. 
Our work aims to pave the way for simulations by analytically finding regions of parameter space that will likely have the most interesting effects.

This paper is structured as follows.
In Section~\ref{sec:H2}, we discuss the processes by which H$_2$ forms in the early universe.
Section~\ref{sec:injection} describes how we track the effects of exotic energy injection, and Section~\ref{sec:halo} outlines our toy halo model.
We then present our main results in Section~\ref{sec:results} before concluding with Section~\ref{sec:conclusion}.
In addition, Appendix~\ref{app:LW} discusses the effect of self-shielding from LW radiation for certain energy injection channels, Appendix~\ref{app:IGM_vs_halo} estimates the relative contribution of decays and annihilations from within and beyond the halo, and Appendix~\ref{app:fs_halo} examines our assumptions on energy deposition within the halo.
Throughout this work, we will use natural units and set $\hbar = c = k_B = 1$.

\section{H$_2$ Formation}
\label{sec:H2}

The dominant pathway through which H$_2$ is formed in the first halos involves the formation of an intermediate H$^-$ ion. All other pathways are responsible for less than $\sim 2\%$ of the final abundance, and we neglect them in this analysis~\cite{Hirata:2006bt}. The H$^-$ mechanism for forming molecular hydrogen begins with radiative attachment of an electron to hydrogen:
\begin{equation}
    \mathrm{H} + e^- \rightarrow \mathrm{H}^- + \gamma \,.
    \label{eqn:H-_REA}
\end{equation}
H$_2$ is then formed through the detachment reaction
\begin{equation}
    \mathrm{H} + \mathrm{H}^- \rightarrow \mathrm{H}_2 + e^- \,.
    \label{eqn:H2_detach}
\end{equation}
The intermediate H$^-$ ion can be destroyed through mutual neutralization,
\begin{equation}
    \mathrm{H}^+ + \mathrm{H}^- \rightarrow 2\mathrm{H} \,,
\end{equation}
or through photodetachment, i.e.\ the reverse reaction of Eqn.~\eqref{eqn:H-_REA}.\footnote{We can neglect the reverse reaction of Eqn.~\eqref{eqn:H2_detach} since this is highly suppressed at temperatures below about $10^4$ K~\cite{1998A&A...335..403G}.}
H$_2$ can also be destroyed by LW photons, which have energies between 11.2 and \SI{13.6}{\eV}~\cite{Haiman:1996rc,Tegmark:1996yt,Machacek:2000us,Yoshida:2003rw,Mesinger:2006pa,Wise:2007cf,OShea:2007ita,Schauer:2020gvx,Kulkarni:2020ovu}.

Using the same notation as in Ref.~\cite{Hirata:2006bt}, we denote the rates for each of these reactions by $k_1$, $k_2$, $k_3$, respectively; $k_{-1}$ for photodetachment; and $k_\mathrm{LW}$ for photodissociation by LW photons.
Then the abundances of H$^-$ and H$_2$ are given by 
\begin{alignat}{2}
    \frac{dx_{\mathrm{H}^-}}{dt} &=&& \, k_1 x_e n_\mathrm{HI} - k_{-1} x_{\mathrm{H}^-} \nonumber \\
    & && \, - k_2 x_{\mathrm{H}^-} n_\mathrm{HI} - k_3 x_{\mathrm{H}^-} n_{\mathrm{HII}} \,, \label{eqn:dxHminusdt}\\
    \frac{dx_\mathrm{H_2}}{dt} &=&& \, k_2 x_\mathrm{H^-} n_\mathrm{HI} - k_\mathrm{LW} x_{\mathrm{H}_2} \,. \label{eqn:dxH2dt_general}
\end{alignat}
$x_i$ denotes the abundance of species $i$, given by its number density $n_i$ relative to the total number density of hydrogen nuclei (ionized, atomic, and in H$_2$). 
Note that we track the free-electron fraction $x_e$ and the ionized hydrogen fraction $x_\mathrm{HII}$ separately, since $x_e$ also receives contributions from ionized helium.
Ref.~\cite{1998ApJ...509....1S} provides fits for the rates $k_1$, $k_2$, and $k_3$ as a function of the gas temperature, $T$:
\begin{align}
    k_1 (T) &= \SI{3e-16}{\centi\meter\cubed\per\second} \left( \frac{T}{\SI{300}{\kelvin}} \right)^{0.95} \!\!\!\!\!\!\! \exp \left(- \frac{T}{\SI{9320}{\kelvin}} \right)  \,, \\
    k_2 (T) &= \SI{1.5e-9}{\centi\meter\cubed\per\second} \left( \frac{T}{\SI{300}{\kelvin}} \right)^{-0.1} \,, \\
    k_3 (T) &= \SI{4e-8}{\centi\meter\cubed\per\second} \left( \frac{T}{\SI{300}{\kelvin}} \right)^{-0.5} \,.
\end{align}

The photodetachment rate receives contributions from CMB photons, including a term from nonthermal radiation, i.e. radiation that is not part of the CMB blackbody~\cite{Hirata:2006bt,Coppola:2013qza},
\begin{alignat}{2}
    k_{-1} & = && \, 4 \left( \frac{m_e T_\mathrm{CMB}}{2\pi} \right)^{3/2} e^{-B(\mathrm{H}^-) / T_\mathrm{CMB}} k_1 (T_\mathrm{CMB}) \n
    & && \, + \int_{B(\mathrm{H}^-)}^{\infty} d\omega \frac{dn_\gamma}{d \omega} \sigma_{-1} (\omega) \,.
    \label{eqn:k-1}
\end{alignat}
In the above equations, $m_e$ is the electron mass, $T_\mathrm{CMB}$ is the CMB temperature, $B(\mathrm{H}^-) = 0.754$ eV is the threshold energy for photodetachment, $dn_\gamma/d \omega$ is the number density of distortion photons per unit energy, and $\sigma_{-1}$ is the cross-section for photodetachment, which is well fit by the expression~\cite{Tegmark:1996yt}
\begin{equation}
    \sigma_{-1} (\omega) \approx 
    \SI{3.486e-16}{\centi\meter\squared}
    \times \frac{(\varpi-1)^{3/2}}{\varpi^{3.11}},
\end{equation}
where $\varpi = \omega / \SI{0.74}{\eV}$.

The LW photodissociation rate $k_{\rm LW}$ is highly dependent on self-shielding, i.e. whether or not the outermost shell of H$_2$ in a halo shields the H$_2$ at the center of the halo from the LW flux.
Given the current uncertainties on this process, we will bracket this effect by showing results for either complete self-shielding or no self-shielding.
However, recent hydrodynamical simulations find evidence for strong self-shielding~\cite{Schauer:2020gvx,Kulkarni:2020ovu,2020MNRAS.492.4386S}, so throughout the text we will focus on that limit by setting $k_\mathrm{LW} = 0$.
In the opposite limit of no self-shielding the photodissociation rate can be approximated by~\cite{Wolcott-Green:2016grm,Friedlander:2022ovf}
\begin{align}
    k_\mathrm{LW} \approx\, &\SI{1.39e-12}{\per\second} \n 
    &\times \left( \frac{J_\mathrm{LW}}{\SI{e-21}{\erg\per\second\per\hertz\per\centi\meter\squared\per\sr}} \right),
    \label{eqn:kLW}
\end{align}
where $J_\mathrm{LW}$ is the average intensity in the LW band.

Finally, since the H$^-$ destruction rate is much faster than the Hubble rate, we can treat $x_{\mathrm{H}^-}$ assuming a steady-state approximation, i.e.\ $dx_{\mathrm{H}^-} / dt = 0$.
Setting the right-hand side of Eqn.~\eqref{eqn:dxHminusdt} to zero and substituting the resulting expression for $x_{\mathrm{H}^-}$ into Eqn.~\eqref{eqn:dxH2dt_general} gives~\cite{Hirata:2006bt} 
\begin{equation}
    \frac{d x_{\mathrm{H}_2}}{dt} = \frac{k_1 k_2 x_e n_\mathrm{HI}^2}{k_2 n_\mathrm{HI} + k_{-1} + k_3 n_\mathrm{HII}} - k_\mathrm{LW} x_{\mathrm{H}_2} \,.
    \label{eqn:dxh2dt}
\end{equation}
From the equations listed above, we see that the formation rate for H$_2$ depends on the matter temperature $T$, the spectrum of nonthermal radiation $dn_\gamma / d\omega$, the ionized hydrogen fraction $x_\mathrm{HII}$, and the free-electron fraction $x_e$.
The values for all of these quantities in the homogeneous IGM can be calculated in the presence of exotic energy injection using the \texttt{DarkHistory} code~\cite{DH,Liu:2023fgu,Liu:2023nct}, which we will describe in the next section.

There are a number of additional processes that affect H$_2$ formation~\cite{Nebrin:2023yzm}, such as the other pathways mentioned above, H$_2$ formation on dust grains~\cite{2020ApJ...905..151N}, and baryon streaming velocities~\cite{Tseliakhovich:2010bj,Stacy:2010gg,Greif:2011iv,Fialkov:2011iw,Naoz:2012fr,Fialkov:2014rba,Hirano:2017wbu,Schauer:2022cgf,Schauer:2020gvx,Schauer:2018iig,Kulkarni:2020ovu}.
These effects should be included in careful simulation studies; however, our focus here is the effect of exotic energy injection.
Moreover, given that the transfer functions used in \texttt{DarkHistory} rely on approximations that affect the changes to the temperature and ionization histories from energy injection at the level of about 10\%~\cite{Galli:2013dna,DH}, ignoring these effects is sufficient for an initial study of the effect of exotic energy injection on early star formation.

\section{Exotic energy injection in the IGM}
\label{sec:injection}

We now review our treatment of exotic energy injection, and the impact of these processes on the formation of H$_2$ in the IGM, which we treat as homogeneous and at mean cosmological energy density.
This will serve as a warmup to the calculation inside the first galaxies, which will follow.
Exotic energy injection, such as annihilating or decaying DM, can heat and ionize the universe more than would be expected in a universe without such injections~\cite{Adams:1998nr, Chen:2003gz, Padmanabhan:2005es, Zhang:2007zzh, Slatyer:2009yq, Cirelli:2009bb, Kanzaki:2009hf, Galli:2009zc, Hisano:2011dc, Hutsi:2011vx, Galli:2011rz, Finkbeiner:2011dx, Slatyer:2012yq, Galli:2013dna, Diamanti:2013bia, Madhavacheril:2013cna, Evoli:2014pva, Slatyer:2015jla, Slatyer:2015kla, Lopez-Honorez:2016sur, Liu:2016cnk, Slatyer:2016qyl, Poulin:2016anj, DAmico:2018sxd, Liu:2018uzy, Cheung:2018vww, Mitridate:2018iag, Clark:2018ghm, McDermott:2019lch, Caputo:2020bdy, Cang:2020exa, Witte:2020rvb, Liu:2020wqz, Bolton:2022hpt}, and also modify the background radiation spectrum~\cite{Ali-Haimoud:2015pwa,Acharya:2018iwh,McDermott:2019lch,Bolliet:2020ofj,Bernal:2022wsu}.
\texttt{DarkHistory}\, \githubDH\,~\cite{DH,Liu:2023fgu,Liu:2023nct} calculates the global temperature and ionization, and evolves background radiation while self-consistently including the effect of homogeneous energy injection; hence, we use \texttt{DarkHistory} to evolve the properties of the IGM and background radiation in which the first halos and stars will form.

\texttt{DarkHistory} tracks how injected particles cool and deposit their energy; the rate of energy deposited into a channel $c$ can be parametrized relative to the rate of energy injected as 
\begin{equation}
    \left( \frac{dE}{dV dt} \right)^\mathrm{dep}_c = f_c (z) \left( \frac{dE}{dV dt} \right)^\mathrm{inj} .
\end{equation}
These $f_c$'s can then be used in the equations determining the IGM temperature, ionization, and radiation spectrum to include the effects from injected particles. 
Since the $f_c$'s are calculated at each redshift step after having evolved the background equations, the $f_c$'s take backreaction into account; in other words, as exotic injections change the background ionization, and radiation spectrum, these changes modify $f_c$'s at later redshifts and subsequent energy deposition.

The evolution of the IGM gas temperature, $T_\mathrm{IGM}$, is given by
\begin{equation}
    \dot{T}_\mathrm{IGM} = 
    \dot{T}_\mathrm{IGM}^\mathrm{adia} + \dot{T}_\mathrm{IGM}^\mathrm{comp} + \dot{T}_\mathrm{IGM}^\mathrm{inj},
    \label{eqn:evol_Tm}
\end{equation}
where the adiabatic cooling term is given by
\begin{equation}
    \dot{T}^\mathrm{adia} = \frac{2}{3} \frac{\dot{n}_\mathrm{H}}{n_\mathrm{H}} T,
    \label{eqn:T_adia}
\end{equation}
and $n_\mathrm{H}$ is the hydrogen number density.\footnote{In principle, $n_\mathrm{H}$ should be replaced with the total number of particles in the system, but since $x_e$ remains small and we assume that helium evolves the same way as hydrogen, we treat $n_\mathrm{H}$ as a proxy for the total particle number evolution}
Here, we drop the subscript on temperature since the expression is general and also applies for halos.
In the IGM, this expression simplifies to $\dot{T}_\mathrm{IGM}^\mathrm{adia} = - 2 H T_\mathrm{IGM}$, where $H$ is the Hubble parameter, since the background density evolves as $n_\mathrm{H} \propto (1+z)^3$.
The Compton scattering term is
\begin{equation}
    \dot{T}^\mathrm{comp} = \Gamma_C(T_\mathrm{CMB} - T) .
    \label{eqn:T_comp}
\end{equation}
In the above equation, $\Gamma_C$ is the Compton scattering rate (see e.g.\ Ref.~\cite{DH} for its definition).
$\dot{T}^\mathrm{inj}$ represents sources of exotic heating, and its homogeneous contribution can be written in terms of $f_\mathrm{heat}$ as
\begin{equation}
    \dot{T}_\mathrm{IGM}^\mathrm{inj} = \frac{2 f_\mathrm{heat}(z)}{3 \bar{n}_\mathrm{H} (1 + x_e + \mathcal{F}_\mathrm{He})} \left( \frac{dE}{dV dt} \right)^\mathrm{inj},
    \label{eqn:T_inj}
\end{equation}
where $\bar{n}_\mathrm{H}$ is the total number density of hydrogen nuclei(ionized, neutral, and molecular) in the IGM and $\mathcal{F}_\mathrm{He} = \overline{n}_\mathrm{He}/\bar{n}_\mathrm{H}$ is the relative abundance of helium nuclei by number.

To obtain the history of the global free-electron fraction $x_e$ and the spectrum of background radiation, we treat hydrogen as a modified Multi-Level Atom (MLA)~\cite{Seager:1999km,Chluba:2006bc,2010PhRvD..81h3005G,Rubino-Martin:2006hng} and track the energy levels of hydrogen up to the principal quantum number $n=200$.
The photon spectrum then takes into account the absorption and emission of photons from bound-bound transitions, recombination, photoionization, and exotic injection.
Details of how the MLA is implemented can be found in Ref.~\cite{Liu:2023fgu}.
Since we are concerned with the formation of the very first stars, we neglect astrophysical sources of radiation when calculating the spectrum of photons, e.g.\ for the LW background, we only consider the contribution from exotic energy injection.

\begin{figure}
    \includegraphics[scale=0.5]{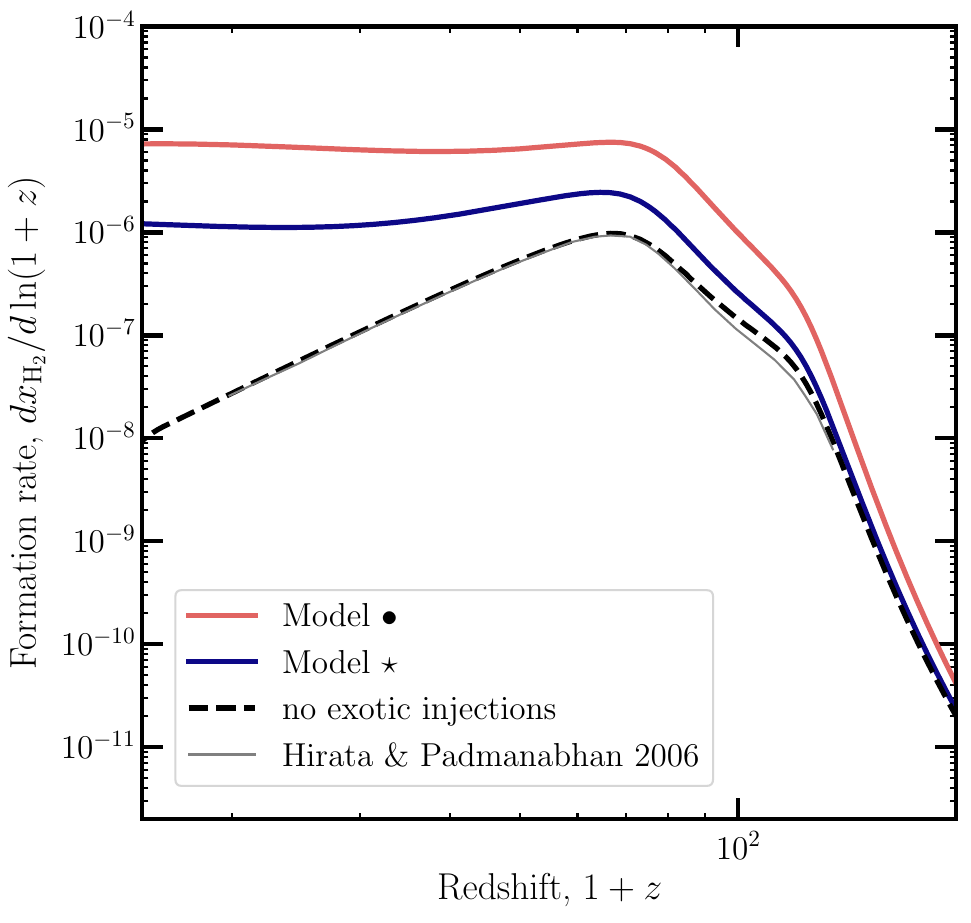}
    \hspace{5mm}
    \includegraphics[scale=0.5]{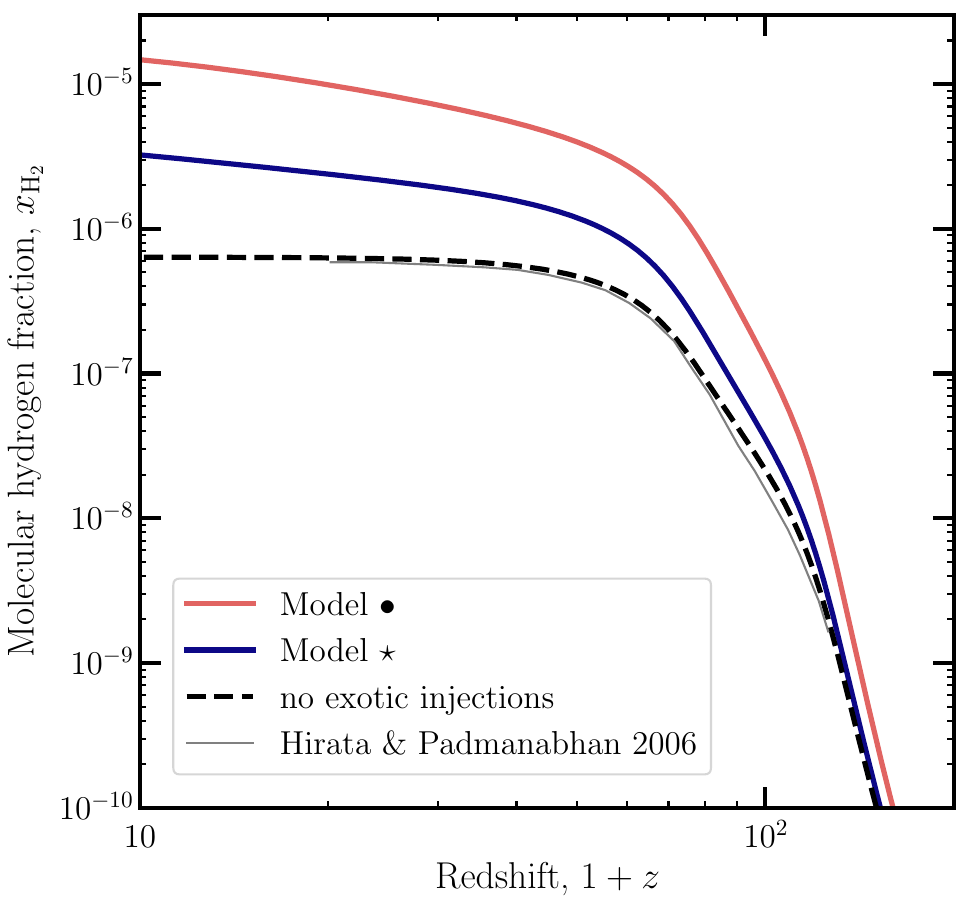}
    \caption{
    H$_2$ formation rate (top) and fraction (bottom) in the average IGM as a function of redshift.
    The black dashed curves show the results assuming standard temperature and ionization histories, including spectral distortions from standard cosmology and no exotic energy injection.
    For comparison, we show the results of Ref.~\cite{Hirata:2006bt} in the thin gray line.
    The blue and pink curves show the results if we include decays by DM to $e^+ e^-$ pairs; both models have $m_\chi = 185$ MeV, but different lifetimes.
    In both cases, the dominant effect is to \textit{enhance} the formation of H$_2$ due to the higher free-electron fraction caused by exotic ionizations.
    }
    \label{fig:global_H2}
\end{figure}

Together, Eqn.~\eqref{eqn:evol_Tm} and the MLA form a closed system of equations that \texttt{DarkHistory} solves to self-consistently determine the evolution of these quantities in the presence of sources of exotic energy injection, in particular decaying or annihilating DM.
The outputs of \texttt{DarkHistory} can then be used to calculate the abundance of H$_2$ in light of these effects.

\begin{table}[t!]
  \renewcommand{\arraystretch}{1.5}
  \begin{tabular}{|c|c|c|c|}
    \hline
    & Channel & Mass [MeV] & $\log_{10} (\tau / [\mathrm{s}])$ \\ 
    \hline
    Model $\bullet$ & $\chi \rightarrow e^+ e^-$ & 185 & 25.6 \\
    Model $\star$ & $\chi \rightarrow e^+ e^-$ & 185 & 26.4 \\
    \hline
  \end{tabular} 
  \caption{
  Parameters for two fiducial DM models. 
  }
  \label{tab:fiducial}
\end{table}

As an initial validation of our molecular hydrogen treatment, Fig.~\ref{fig:global_H2} shows the formation rate and total abundance of H$_2$ in the IGM in terms of the molecular hydrogen fraction, $x_{\mathrm{H}_2}$; ultimately, for star formation, we will replace $T_\mathrm{IGM}$, $x_e$, and $x_{\mathrm{H}_2}$ with the analogous quantities in a halo in Section~\ref{sec:halo}, but for comparison to other works we only consider the IGM contribution here.
We also set $k_\mathrm{LW}$ to zero here for the purpose of comparison with results from the literature.
The black dashed curved shows our calculation in the absence of exotic energy injection; i.e. using the global temperature history, ionization history, and spectral distortions from standard cosmology.
We compare our results against Ref.~\cite{Hirata:2006bt}, finding excellent agreement.

We also show the results including two fiducial models of exotic energy injection, whose parameters are listed in Table~\ref{tab:fiducial}.
In both models, DM has a mass $m_\chi = 185$ MeV and decays to $e^+ e^-$ pairs (throughout this text, $\chi$ denotes the hypothetical DM particle).
We focus on sub-GeV decaying DM in this paper, since this mass range is less constrained by existing probes, and the impact of the halo on energy deposition within the halo is expected to be more straightforward.
Two different lifetimes are chosen to reflect two distinct, opposite effects that energy injection can have on star formation.
Further discussion on our choice of models can be found in Sec.~\ref{sec:collapse}.
Model $\bullet$\, assumes a lifetime of $\log_{10} (\tau / [\mathrm{s}]) = 25.6$, which is excluded by Voyager constraints~\cite{Boudaud:2016mos, Boudaud:2018oya} but unconstrained by CMB anisotropies~\cite{Slatyer:2016qyl};\footnote{While Model $\bullet$\, is nominally ruled out, we include it for illustrative purposes because we show in Section~\ref{sec:halo} that the effect on star formation is in the opposite direction of Model $\star$.}
Model $\star$\, assumes an even longer lifetime $\log_{10} (\tau / [\mathrm{s}]) = 26.4$ is beyond current limits from Voyager and the CMB.
For the IGM, in both models, the H$_2$ abundance and formation rate is larger than for a standard cosmology; the effect is greater for Model $\bullet$\, because the lifetime is shorter.
This effect is due to the enhanced free-electron fraction from exotic ionization; the additional electrons serve to catalyze the formation of molecular hydrogen through the reaction in Eqn.~\eqref{eqn:H-_REA}.

\section{Halo evolution}
\label{sec:halo}

To study the effect of exotic energy injection on the first star-forming halos, we modify the simple procedure in Ref.~\cite{Tegmark:1996yt} to evolve the free-electron fraction $x_e$, the H$_2$ fraction $x_{\mathrm{H}_2}$, and the gas temperature $T_\mathrm{halo}$ inside a halo in the presence of energy injection from DM.
To summarize the method, we assume the halo has a mass $M_\mathrm{halo}$, and can be approximated as a spherical ``top-hat'' as it collapses, i.e. the halo has a uniform density $\rho$ within a spherical volume.
When the halo reaches either the virial density $\rho_\mathrm{vir}$ or virial temperature $T_\mathrm{vir}$, then the halo is considered virialized; we label the redshift at which this occurs as $z_\mathrm{vir}$.
Upon virialization, we assume the halo becomes an isothermal sphere; then the virial temperature is approximately related to the halo mass by~\cite{Barkana:2000fd}
\begin{equation}
    T_\mathrm{vir} \approx 224 \,\mathrm{K} \, \left( \frac{\Omega_m h^2}{0.14} \right)^{1/3}  \left( \frac{M_\mathrm{halo}}{10^4 M_\odot} \right)^{2/3} \left( \frac{1+z_\mathrm{vir}}{100} \right) .
    \label{eqn:Tvir}
\end{equation}
In the above equation, $h$ is defined by $H_0 = \SI[parse-numbers=false]{\num[parse-numbers=true]{100}h}{\kilo\meter\per\second\per\mega\parsec}$, where we use $H_0 = \SI{67.36}{\kilo\meter\per\second\per\mega\parsec}$ for the Hubble parameter today, $\Omega_m$ is the matter density parameter, and $M_\odot$ is a solar mass.
We take $\rho_\mathrm{vir} = 18 \pi^2 \rho_0 (1+z)^3$, where $\rho_0$ is the mean matter energy density today.
After the halo virializes, we hold its density fixed and continue to evolve the other quantities.
If the temperature of the gas falls quickly enough, then we infer that it is capable of collapsing to form stars; the precise conditions we employ are described in Sec.~\ref{sec:collapse}.
In the following sections, we describe the ingredients for this calculation in more detail.

In principle, decays and annihilations from within the star-forming halos themselves should be considered separately from the IGM contribution.
Moreover, one should account for the fact that energy deposition can be different in the overdense halo (even if illuminated with the background emitted from the average IGM).
Exotic energy injection from halos has been studied using Monte Carlo methods~\cite{Schon:2014xoa,Schon:2017bvu}, but is beyond the scope of this work.
Instead, we expect the energy deposited per particle inside halos to be comparable to that in the IGM for DM decay; we justify this assumption in two ways.

First, in Appendix~\ref{app:IGM_vs_halo}, we estimate the geometrical factors for decay and annihilation from early halos vs. from the IGM and find that the contributions from the IGM for decays dominate over the halo contribution for free-streaming decay/annihilation products.
Second, in Appendix~\ref{app:fs_halo}, we also follow simplified particle cascades to show that for most injected particles, the presence of the halo has a small impact on energy injection and deposition.
Intuitively, this happens because the path lengths of many of the particles produced in the cascade from the initial DM process tend to be much longer than the halo itself, leading to the intensity being dominated by the IGM contribution. 
If these long-path-length particles deposit their energy through interactions with the target gas particles, equal intensity of particles received in the halo and in the IGM translates into equal energy deposited \emph{per gas particle} in the halo, which is the relevant parameter for the effect of exotic energy injection on the free-electron fraction and temperature.
In the opposite limit, where products of decay/annihilation lose their energy promptly, for decays the enhanced injection in the halo is canceled by the higher density of targets, when computing the power deposited per target.
A more in-depth exploration of this question, while unnecessary for this paper, would be an interesting subject for future work.

\subsection{Density}
\label{sec:halo_density}

Following Ref.~\cite{Tegmark:1996yt}, we take a simple approximation for the temporal evolution of the density of a collapsing spherical top-hat:\footnote{This expression corrects a sign error in Ref.~\cite{Tegmark:1996yt}, providing a good fit to the spherical collapse model.}
\begin{gather}
    \rho(z) \approx \rho_0 (1 + z)^3 \exp \left( \frac{1.9 A}{1 - 0.75A^2} \right), \quad A = \frac{1 + z_\mathrm{vir}}{1+z} \,,
    \label{eqn:rho_TH}
\end{gather}
At early times, the density of the halo is diluted by the expansion of the universe and closely follows the evolution of the IGM mean density; however, as $z$ approaches $z_\mathrm{vir}$, the halo begins to collapse and the density increases.
Once the halo passes the condition for virialization (see Sec.~\ref{sec:collapse}), the density is held constant at $\rho (z_\mathrm{vir})$.

The number density of hydrogen within the halo is similarly assumed to be given by
\begin{equation}
    n_\mathrm{H} (z) \approx \bar{n}_{\mathrm{H}, 0} (1 + z)^3 \exp \left( \frac{1.9 A}{1 - 0.75A^2} \right) ,
    \label{eqn:n_TH}
\end{equation}
where $\bar{n}_{\mathrm{H}, 0}$ is the mean number density of all hydrogen nuclei today.

\subsection{Temperature}
\label{sec:halo_temp}

The halo temperature is affected by adiabatic cooling/heating as the volume of the halo changes, as well as Compton scattering off the CMB, atomic hydrogen line cooling, molecular hydrogen cooling, and exotic energy injection.
Hence, we can write the temperature evolution of gas in a halo as
\begin{equation}
    \dot{T}_\mathrm{halo} = \dot{T}_\mathrm{halo}^\mathrm{adia} + \dot{T}_\mathrm{halo}^\mathrm{comp} + \dot{T}_\mathrm{halo}^\mathrm{line} + \dot{T}_\mathrm{halo}^{\mathrm{H}_2} + \dot{T}_\mathrm{halo}^\mathrm{inj},
    \label{eqn:dTmdt}
\end{equation}
which has two new terms compared to its IGM counterpart.
The adiabatic and Compton-scattering contributions are still given by Eqn.~\eqref{eqn:T_adia} and \eqref{eqn:T_comp}, respectively, where the number density is given in Eqn.~\eqref{eqn:n_TH}, thus accounting for the collapse of the halo.
All instances of the background density, temperature, or ionization fraction should be replaced with the corresponding quantities in the halo.
The terms for atomic-line cooling and molecular cooling are respectively given by
\begin{multline}
    \dot{T}_\mathrm{halo}^\mathrm{line} = -\SI{7.5e-19}{\erg\centi\meter\cubed\per\second} \\
    \times \frac{2 n_\mathrm{HI} n_e}{3 n_\mathrm{H} (1 + x_e + \mathcal{F}_\mathrm{He})} \exp \left(- \frac{\SI{118348}{\kelvin}}{T_\mathrm{halo}} \right) \,,
\end{multline}
where $n_\mathrm{H}$ denotes the total number density of hydrogen in the halo in this case, and
\begin{equation}
    \dot{T}_\mathrm{halo}^{\mathrm{H}_2} = - \Lambda_\mathrm{H_2} n_\mathrm{H_2},
\end{equation}
where the cooling rate $\Lambda_\mathrm{H_2}$ is given by Eqn.~(A.5) of Ref.~\cite{1998A&A...335..403G}. 

The exotic injection term $\dot{T}_\mathrm{halo}^\mathrm{inj}$ in Eqn.~\eqref{eqn:dTmdt} is taken to be identical to the expression in Eqn.~\eqref{eqn:T_inj}, with $\bar{n}_\mathrm{H}$ and the injection rate taking their homogeneous value, and $f_\mathrm{heat}(z)$ taken from \texttt{DarkHistory}, using the IGM temperature and ionization level. 
We justify this assumption in Appendix~\ref{app:fs_halo}, where we show that under reasonable simplifying assumptions, the intensity of particles received from exotic energy injection at the center of halos is roughly equal to the intensity in the homogeneous IGM for our fiducial models, and across much of the relevant parameter space for decaying DM.

\subsection{Ionization fraction}
\label{sec:halo_ion}

For the ionization fraction inside the halo, we use the following evolution equation. 
\begin{equation}
    \dot{x}_e = - \mathcal{C} \left[ x_e n_e \alpha_\mathrm{B} + 4 (1 - x_e) \beta_\mathrm{B} e^{-E_{21} / T_\mathrm{CMB}} \right] + \dot{x}_e^\mathrm{inj},
    \label{eqn:dxedt}
\end{equation}
In this equation, $\mathcal{C}$ is the Peebles-C factor, $\alpha_\mathrm{B}$ and $\beta_\mathrm{B}$ are the case-B recombination and photoionization coefficients, which are calculated using $T_\mathrm{halo}$, and $E_{21} = \SI{10.2}{\eV}$ is the energy of the Lyman-$\alpha$ transition.
This is the equation for a Three-Level Atom (TLA) modified to include exotic energy injection; it is appropriate to use the TLA here instead of the MLA since we are not tracking the contribution of the halo to spectral distortions (see the discussion in Ref.~\cite{Liu:2023fgu}), and the MLA takes much longer to solve than the TLA.

Under the assumptions of the TLA, the term from exotic energy injection term, $\dot{x}_e^\mathrm{inj}$, is written as
\begin{equation}
    \dot{x}_e^\mathrm{inj} = \left[ \frac{f_\mathrm{H ion}}{\mathcal{R} n_\mathrm{H}} + \frac{(1-\mathcal{C}) f_\mathrm{exc}}{E_{21} n_\mathrm{H}} \right] \left( \frac{dE}{dV dt} \right)^\mathrm{inj},
    \label{eqn:x_inj}
\end{equation}
where $\mathcal{R} = \SI{13.6}{\eV}$ is the energy required to ionize hydrogen.
The first term represents photoionization and collisional ionization caused by exotic energy injection.
The second term represents excitations followed by ionization; the $(1 - \mathcal{C})$ factor ensures that this term does not include excitations that are immediately followed by recombination to the ground state.
Once again, we use the IGM result for $f_c / \bar{n}_\mathrm{H}$, since the energy deposited per baryon in the halo is expected to be similar to the same quantity in the homogeneous IGM for our fiducial models and for most of the parameter space for DM decay (as discussed in Appendix~\ref{app:fs_halo}).

\subsection{Full evolution and collapse}
\label{sec:collapse}

We simultaneously solve Eqns.~\eqref{eqn:dxh2dt},~\eqref{eqn:dTmdt}, and~\eqref{eqn:dxedt} to obtain the H$_2$ abundance, temperature, and ionization fraction inside the halo, while using Eqn.~\eqref{eqn:rho_TH} for the halo density.
There are two free parameters that parametrize the halo properties: \emph{1)} the virialization redshift $z_\mathrm{vir}$, and \emph{2)} the halo mass $M_\mathrm{halo}$, or equivalently the virial temperature $T_\mathrm{vir}$, which is related to $M_\mathrm{halo}$ via Eqn.~\eqref{eqn:Tvir}.
We also require the $f_c$'s and spectral distortions output by \texttt{DarkHistory} for a given DM model, which are calculated using the temperature and ionization from the homogeneous IGM.

For the initial conditions, we set $x_\mathrm{H_2} = 0$ and the halo temperature and ionization fraction to the IGM values at $1+z = 3000$.
Prior to $1+z = 900$, the system of differential equations is very stiff; since the halo density and temperature are very similar to the mean IGM density and temperature at these times, we fix $x_\mathrm{HII}$ and $T_\mathrm{halo}$ to the IGM values for $1+z > 900$, solving only the evolution equation for $x_\mathrm{H_2}$, and use the value of $x_\mathrm{H_2}$ at $1+z = 900$ as an initial condition for solving the full set of equations thereafter.

As we evolve the halo forward in time, the halo begins to collapse and increase in density. 
Eventually, at some redshift $z_\mathrm{vir}$, the halo is considered to be virialized when one of the following conditions is met:
\begin{itemize}
    \item the density reaches the value expected of a virialized halo, $\rho_\mathrm{vir} = 18\pi^2 \rho_0 (1+z)^3$, or

    \item $T_\mathrm{halo} = T_\mathrm{vir}$, and the halo is prevented from collapsing further by gas pressure~\cite{Tegmark:1996yt}.
\end{itemize}
Once one of these two conditions is met, we hold the gas density fixed to its collapsed value and raise the temperature to $T_\mathrm{vir}$ if it is below this value.
We continue to evolve the halo past $z_\mathrm{vir}$ down to $1+z=4$ in order to evaluate if it is cooling sufficiently quickly to collapse; however, we only use the results for studying halo cooling shortly after $z_\mathrm{vir}$, since our assumption that the halo density remains constant after virialization breaks down if the halo can cool and collapse further. 
During this collapsing phase, we only want to evaluate the ability of molecular and line cooling to lead to halo collapse, since these are crucial processes for star formation; hence, we neglect the Compton cooling term $\dot{T}_\mathrm{halo}^\mathrm{comp}$ after virialization.
This also ensures that the threshold we find is the minimum mass above which we expect \textit{all} halos to undergo runaway collapse.

We adopt the criterion in Ref.~\cite{Tegmark:1996yt}, and consider a halo to be cooling sufficiently quickly to collapse and form stars if
\begin{equation}
    T_\mathrm{halo} (\eta z_\mathrm{vir}) \leq \eta T_\mathrm{halo} (z_\mathrm{vir}),
    \label{eqn:criteria}
\end{equation}
where we choose $\eta = 0.75$.
As explained in Ref.~\cite{Tegmark:1996yt}, this condition comes from requiring the halo to cool by a significant amount within a Hubble time; since the scale factor is proportional to $t^{2/3}$ during matter domination, this roughly corresponds to the redshift dropping by $2^{2/3} \approx 0.63$.
We have checked that this condition is not particularly sensitive to the choice of $\eta$ by varying it between $0.6$ to $0.9$.

\section{Results}
\label{sec:results}

%
\begin{figure*}
    \includegraphics[scale=0.5]{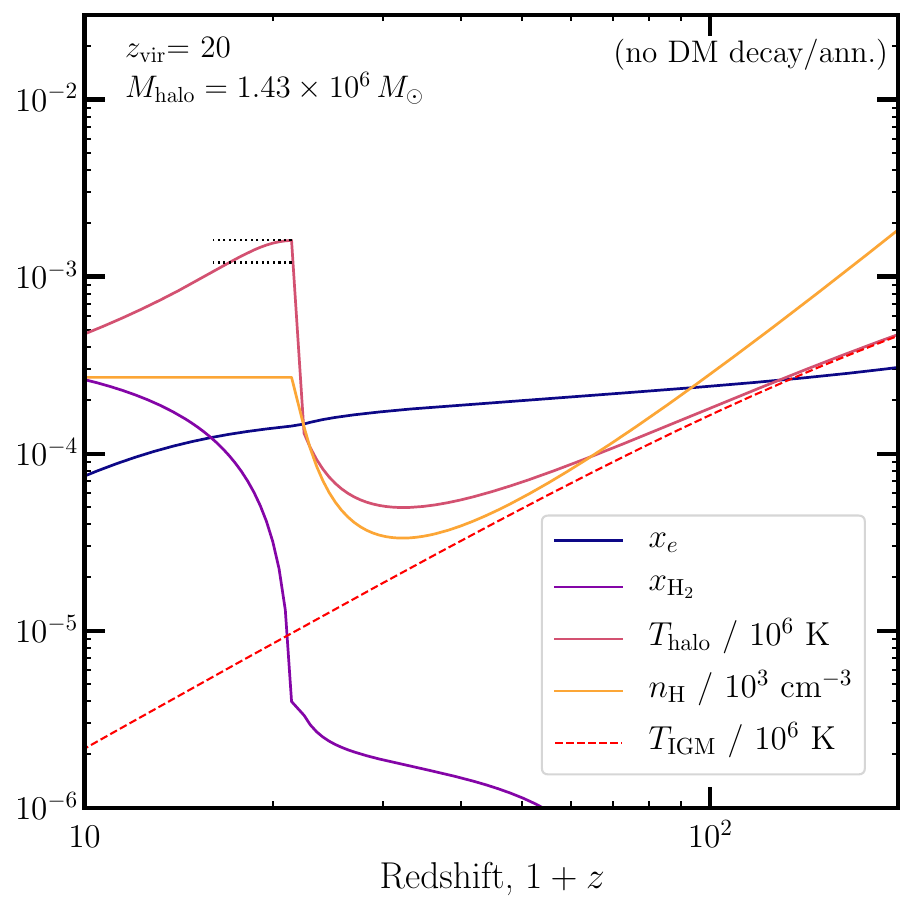}
    \\
    \includegraphics[scale=0.5]{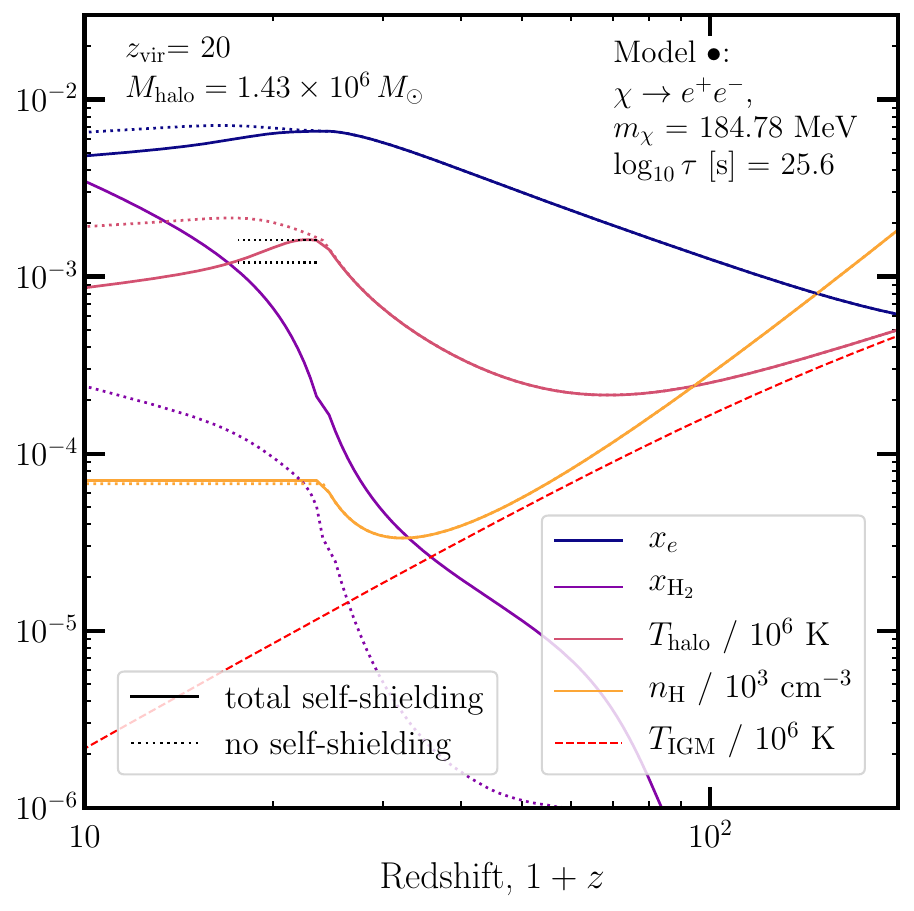}
    \hspace{5mm}
    \includegraphics[scale=0.5]{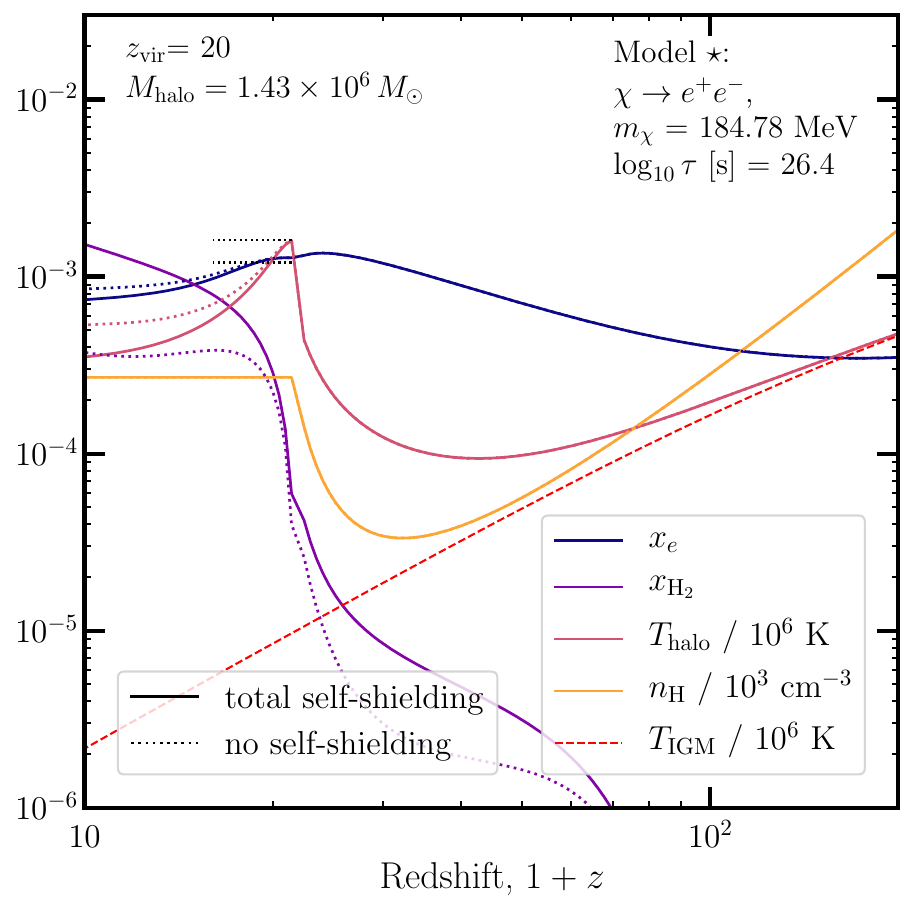}
    \caption{
    Three examples of halo evolution.
    In the top panel there is no DM decay or annihilation, and the halo virializes at $z_\mathrm{vir} = 20$ with a mass of $M_\mathrm{halo} = 1.43 \times 10^6 M_\odot$, corresponding to a virial temperature of $T_\mathrm{vir} = \SI{1600}{\kelvin}$.
    The bottom two panels include the effects of exotic energy injection, with Model $\bullet$\, on the left and Model $\star$\, on the right.
    In all plots, we show the free-electron fraction (blue), the molecular hydrogen fraction (purple), the halo temperature (magenta), and the number density of hydrogen nuclei (neutral or ionized, orange).
    Dotted lines include photodissociation by LW photons.
    The red dashed line indicates the average IGM temperature when we do not include exotic energy injection.
    The horizontal line segments on each panel indicate $T_\mathrm{vir}$ and $0.75 T_\mathrm{vir}$, and span the redshift range $(z_\mathrm{vir}, 0.75 z_\mathrm{vir})$; if the temperature curve crosses the lower line segment after virialization, the halo passes the criterion for collapsing and forming stars.
    In the top panel, the halo succeeds in collapsing.
    Model $\star$\, gives rise to even faster cooling, whereas Model $\bullet$\, cools more slowly, failing to meet our criterion for star formation.
    }
    \label{fig:halo_examples}
\end{figure*}

With the prescription described in Section~\ref{sec:collapse}, given $z_\mathrm{vir}$ and an exotic energy-injection model, we can determine the minimum $M_\mathrm{halo}$ or $T_\mathrm{vir}$ above which halos will collapse.
Fig.~\ref{fig:halo_examples} shows a few examples of evolving halos.
In each panel, we show $x_e$, $x_{\mathrm{H}_2}$, $T_\mathrm{halo}$, and $n_\mathrm{H}$ as a function of redshift, for both very efficient H$_2$ self-shielding and no self-shielding.
We also show two horizontal dotted line segments at $T_\mathrm{vir}$ and $0.75 T_\mathrm{vir}$.
The lines span the redshift range $z_\mathrm{vir}$ to $0.75 z_\mathrm{vir}$; hence, models where the temperature history crosses the lower line segment after virialization succeed in forming stars.
The top panel shows a halo that virializes at $z_\mathrm{vir} = 20$ with a mass of $M_\mathrm{halo} = 1.39 \times 10^6 M_\odot$, corresponding to a virial temperature of $T_\mathrm{vir} = 1600$ K.
The bottom two panels show halos with the same $z_\mathrm{vir}$ and $T_\mathrm{vir}$, but including the same energy injection models shown in Fig.~\ref{fig:global_H2}; Model $\bullet$\, is on the left and Model $\star$\, is shown on the right.

In the top panel at early times, the halo density decreases and closely follows the IGM density; at later times, as we approach $z_\mathrm{vir}$, the halo begins to collapse and the density increases until it reaches $\rho_\mathrm{vir}$.
At this point, the halo has virialized and the density is held fixed.
Similarly, at early times, the halo temperature follows the standard IGM temperature evolution, but starts to deviate around $z \sim 100$ as the overdensity begins to collapse.
Close to $z_\mathrm{vir}$, the temperature begins to increase and is raised to $T_\mathrm{vir}$ once the halo meets the conditions for virialization.
After this, the temperature decreases due to molecular cooling.
The curve crosses back through the lower line segment; hence, by the criteria given in Eqn.~\eqref{eqn:criteria}, this halo has succeeded in forming stars.

The bottom panels of Fig.~\ref{fig:halo_examples} demonstrate that exotic energy injection has multiple competing effects on halo collapse. 
On the one hand, the exotic heating impedes cooling, preventing halos from collapsing. 
If self-shielding of the halos is inefficient, then the effect of LW photons can further suppress cooling by reducing the H$_2$ abundance.
On the other hand, the increase in ionization levels catalyzes the formation of H$_2$, enhancing the molecular cooling rate. 
The effect of heating dominates for Model $\bullet$ \ in the lower-left plot, causing the halo temperature to reach $T_\mathrm{vir}$ before the density reaches $\rho_\mathrm{vir}$; after $z_\mathrm{vir}$, the exotic heating rate is also large enough that the halo cannot cool efficiently. 
The effect of ionization dominates for Model $\star$, leading to enhanced cooling relative to the standard cosmological expectation shown in the top plot.
This halo forms stars successfully by the criterion in Eqn.~\eqref{eqn:criteria}.
In both cases, inefficient self-shielding delays the cooling of the halo; the effect is larger for Model $\bullet$, where dark matter decays more quickly and therefore contributes a larger LW background.

\begin{figure*}
    \includegraphics[scale=0.5]{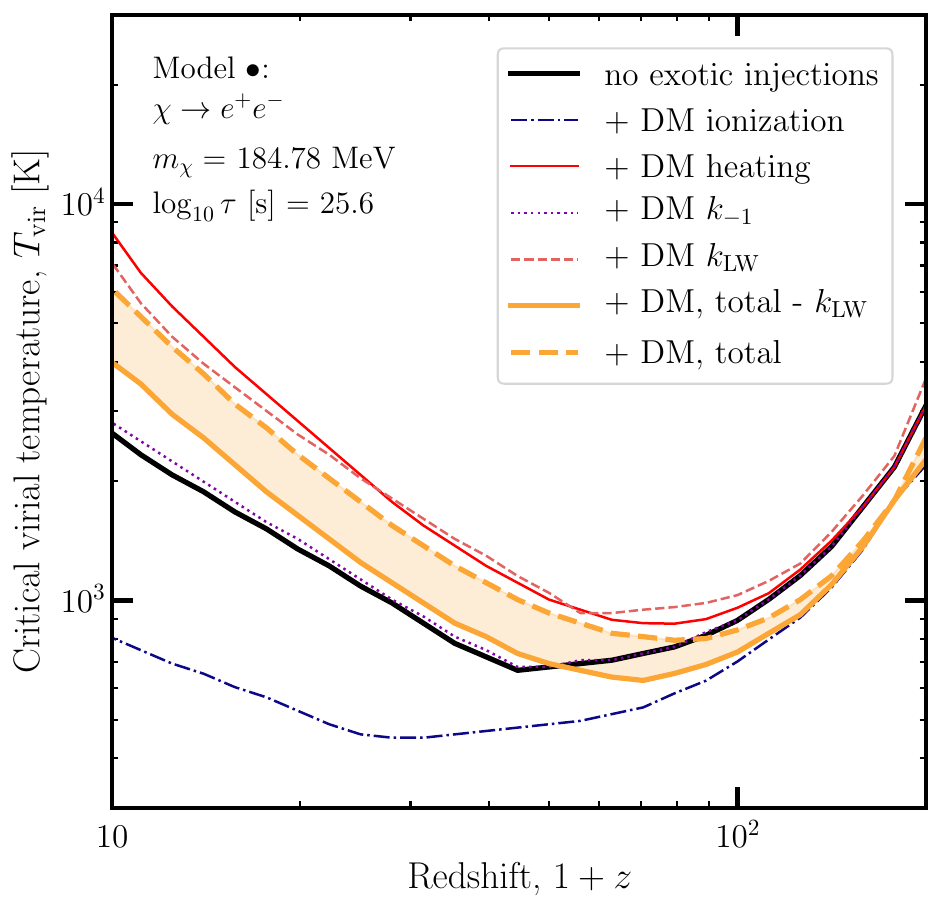}
    \hspace{5mm}
    \includegraphics[scale=0.5]{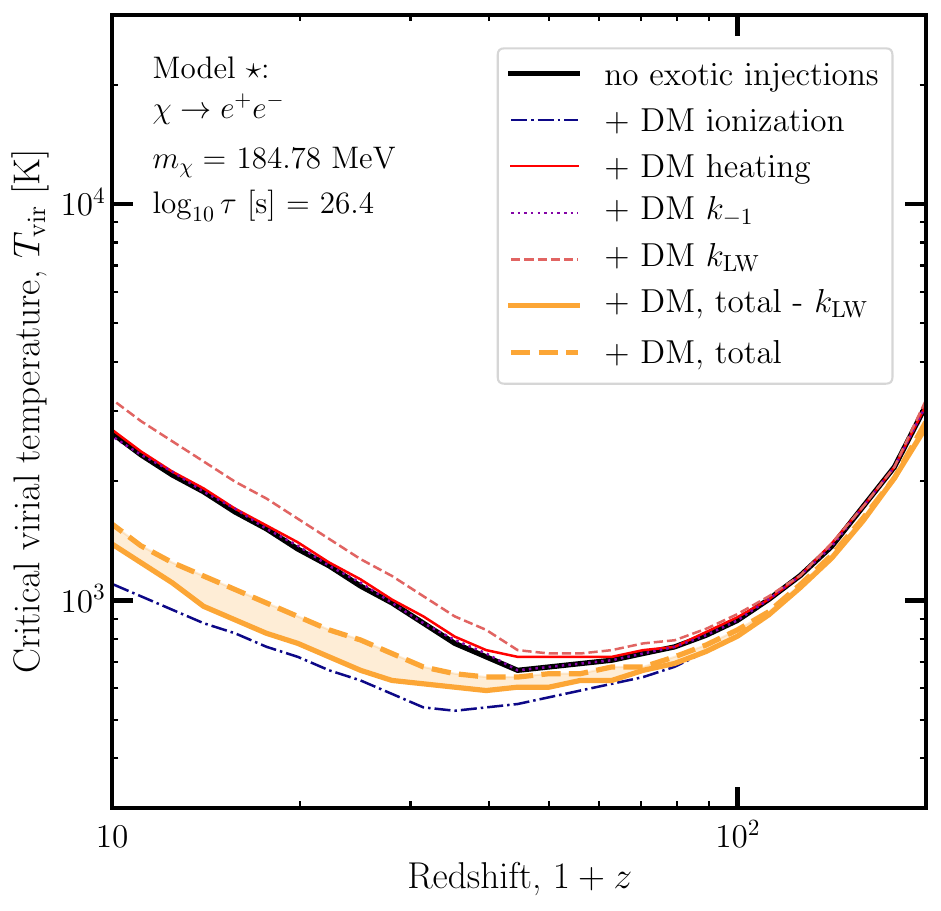}
    \caption{
    Critical virial temperature for collapse as a function of redshift, both in a standard cosmology without exotic injections (black) and including the fiducial DM energy injection models (gold); the contours bracket the effect of H$_2$ self-shielding.
    The left panel shows Model $\bullet$\, and the right shows Model $\star$.
    We also show the results from considering one effect at a time, including exotic ionization (blue dot dashed), exotic heating (red solid), H$^-$ photodetachment (purple dotted), and H$_2$ photodissociation (pink dashed).
    }
    \label{fig:critical_collapse_Tvir}
\end{figure*}
\begin{figure*}
    \includegraphics[scale=0.5]{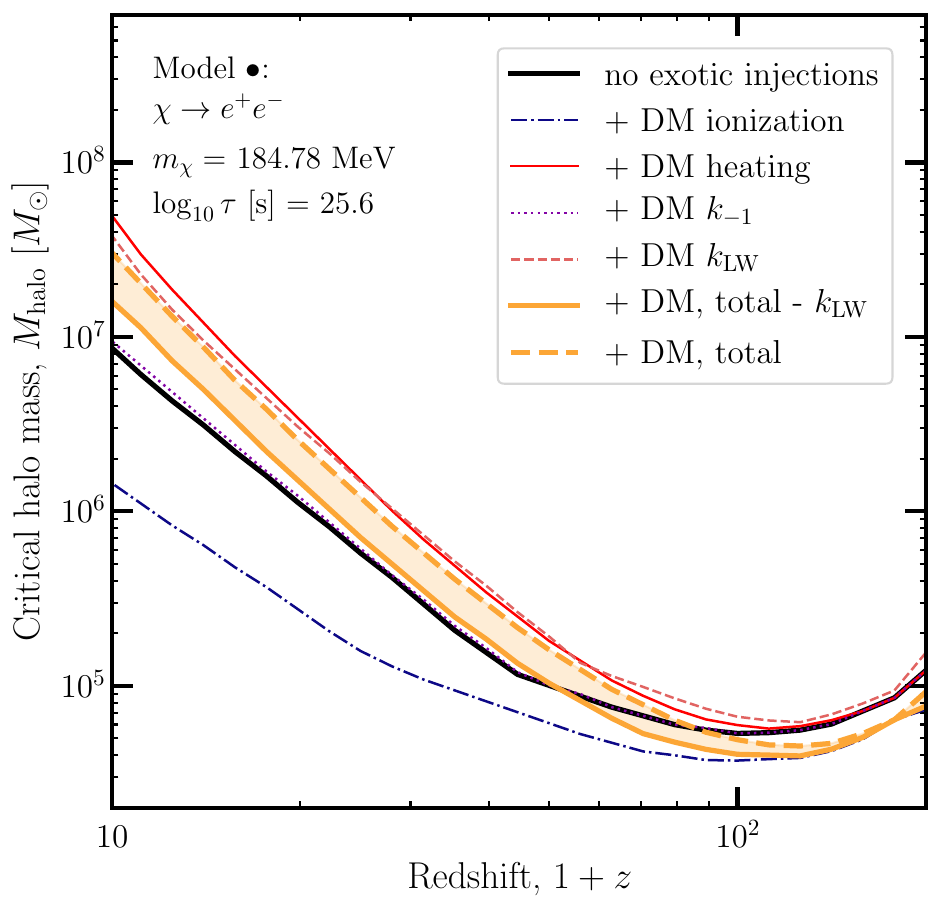}
    \hspace{5mm}
    \includegraphics[scale=0.5]{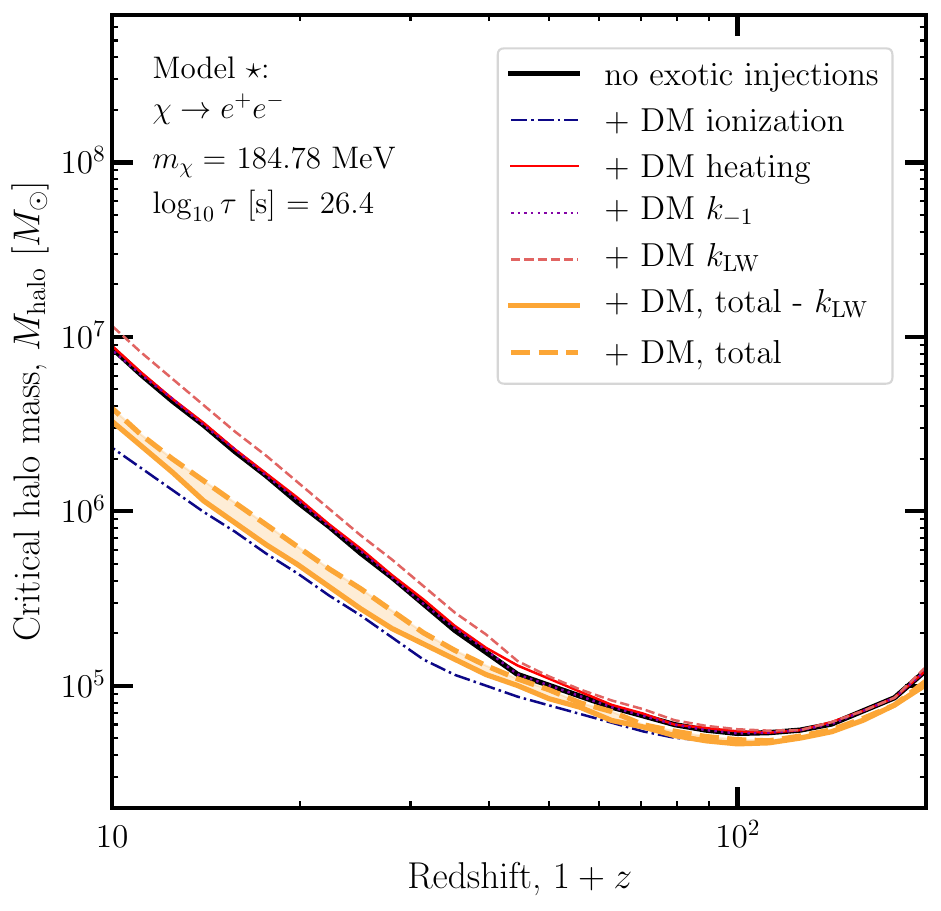}
    \caption{
    Same as Fig.~\ref{fig:critical_collapse_Tvir}, but showing results for the critical halo mass.
    }
    \label{fig:critical_collapse_Mhalo}
\end{figure*}

The threshold for halos to collapse is a redshift-dependent condition.
For example, Fig.~\ref{fig:critical_collapse_Tvir} shows the critical virial temperature as a function of the virialization redshift; Fig.~\ref{fig:critical_collapse_Mhalo} shows the same results, but in terms of the halo mass.
We show the critical values if we do not include exotic injections, and how the curves change if we include Model $\bullet$\, or Model $\star$, bracketing the effect of H$_2$ self-shielding.
We also show the resulting curves if in the evolution equations we only include one of the terms for exotic ionization (Eqn.~\eqref{eqn:x_inj}), exotic heating (Eqn.~\eqref{eqn:T_inj}), H$^-$ photodetachment (Eqn.~\eqref{eqn:k-1}), or H$_2$ photodissociation (Eqn.~\eqref{eqn:kLW}) from exotic injections.

From Figs.~\ref{fig:critical_collapse_Tvir} and~\ref{fig:critical_collapse_Mhalo}, we see that additional heating and LW photons generally raise the critical value of $T_\mathrm{vir}$ or equivalently $M_\mathrm{halo}$, making it more difficult for halos to collapse, whereas ionization caused by exotic energy injection lowers the critical value due to the enhanced production of H$_2$, allowing more halos to collapse.
Because of this interplay of effects, it is nontrivial to determine the direction of the effect on the critical virial temperature/halo mass for a given redshift and energy injection model.
For Model $\bullet$\, (left panels), it is easier for halos to collapse prior to $z \sim 50$ and harder after this, whereas for Model $\star$\, (right), the overall effect is to lower the critical masses or virial temperatures needed for collapse at all redshifts.
This complicated interplay of heating and ionization on collapsing halos has been noted in previous work studying energy injection by primordial black holes on the formation of supermassive black holes~\cite{Pandey:2018jun}; however, to our knowledge, our results are the first to show that the direction of the effect can change as one increases the energy injection rate.

\subsection{Scanning over DM parameter space}
\label{sec:scans}

Based on the results in the previous section, it is clear that different regions of DM parameter space can have opposite effects on halo collapse and star formation; moreover, the direction of the effect can also change with redshift.
We would like to now scan over the parameter space of various energy injection channels to understand where we get the strongest and most interesting signals.

In describing our methods up to this point, we have focused on the example of decaying dark matter; these largely apply to annihilations as well.
For $p$-wave annihilation, since the cross-section is velocity dependent, we define the cross-section using 
\begin{equation}
    \sigma v = (\sigma v)_\mathrm{ref} \left( \frac{v}{v_\mathrm{ref}} \right)^2
\end{equation}
and choose a reference velocity of $v_\mathrm{ref} = 100$ km/s, which is on the order of the dispersion velocity of DM in a Milky Way-like halo today.
We again assume that the $f_c$'s parametrizing energy deposition are the same in the halos as in the IGM.
This assumption is less likely to be valid for annihilations since the annihilation rate depends on density squared, and in particular may break down for $p$-wave annihilations since there is an additional velocity dependence.
Since the rate of energy injection from annihilations depends on density squared, this rate is boosted once structure formation begins; the average of the density squared exceeds the square of the average density.
We use the boost factor prescription included in \texttt{DarkHistory}~\cite{DH} under the assumption that halos have an NFW profile and no substructure; see Ref.~\cite{Liu:2016cnk} for more details about how this is calculated.

For both decays and annihilations, we will study photon and $e^+ e^-$ final states.
In principle, we could study decays or annihilations into other Standard Model particles; however, even when we inject other types of particles, the prompt decays of these particles primarily result in showers of secondary electrons, positrons, and photons.
Hence, one can usually understand the results for other final states by taking linear combinations of the results for injections of photons and $e^+ e^-$ pairs.

\subsection{Comparison to existing constraints}
\label{sec:constraints}

\begin{figure*}
    \includegraphics[scale=0.4]{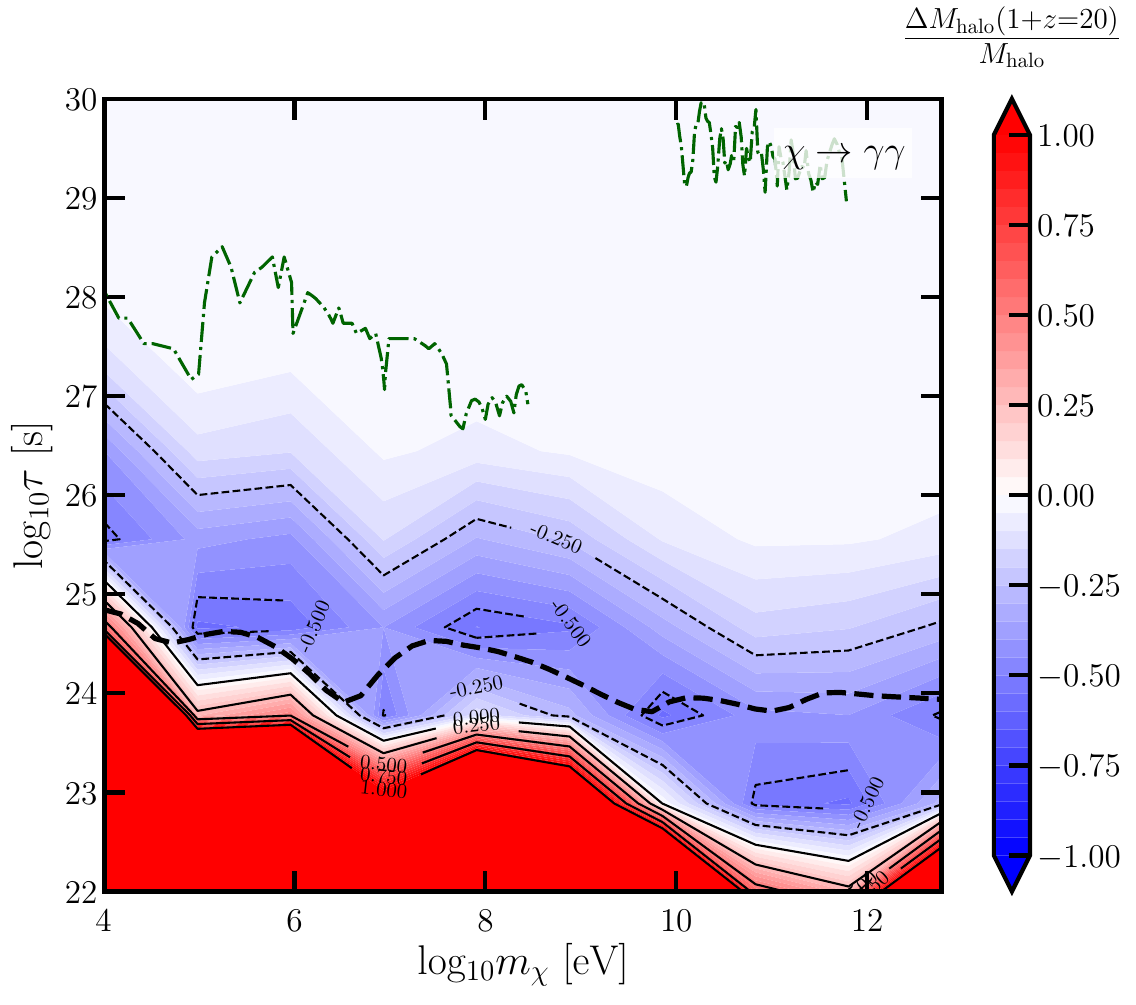}
    \includegraphics[scale=0.4]{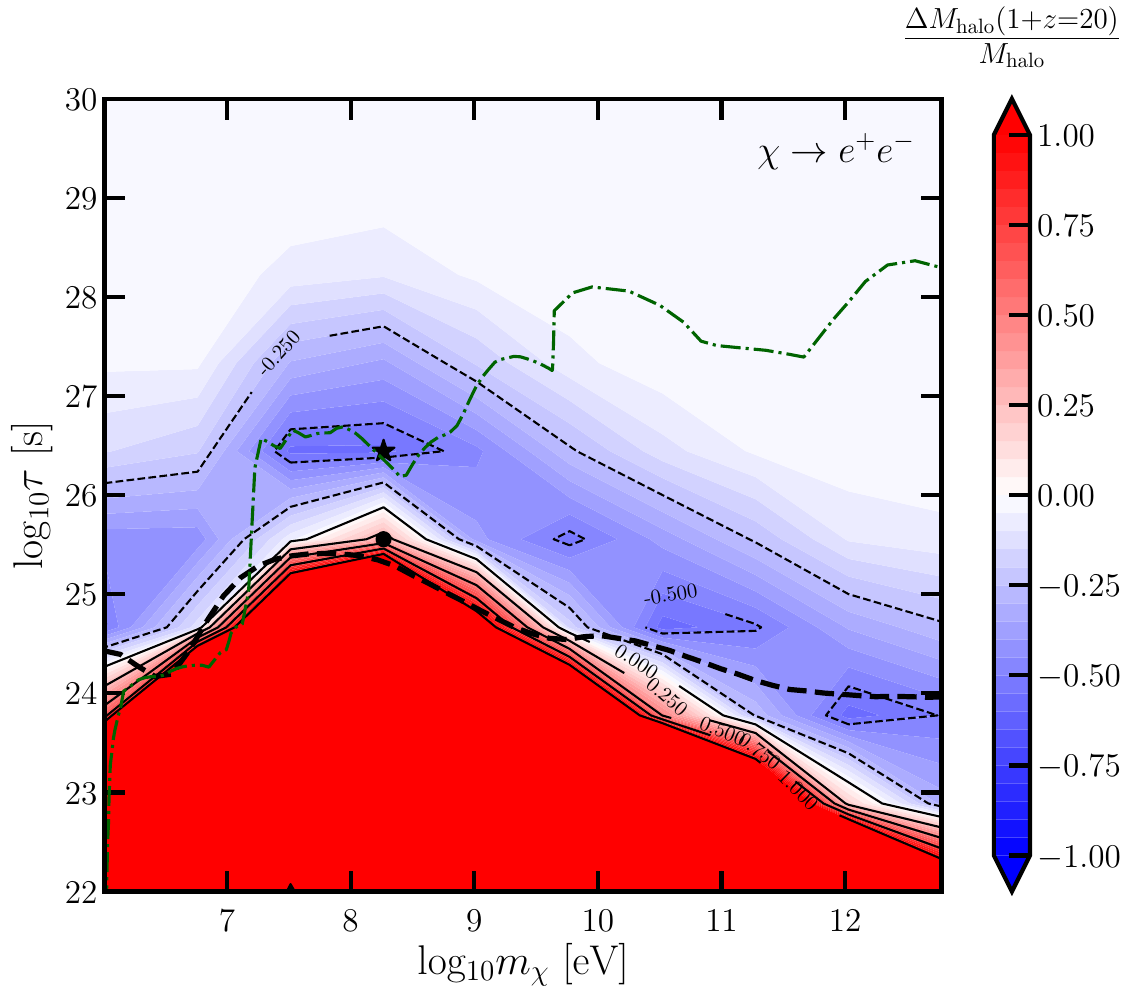} \\
    \includegraphics[scale=0.4]{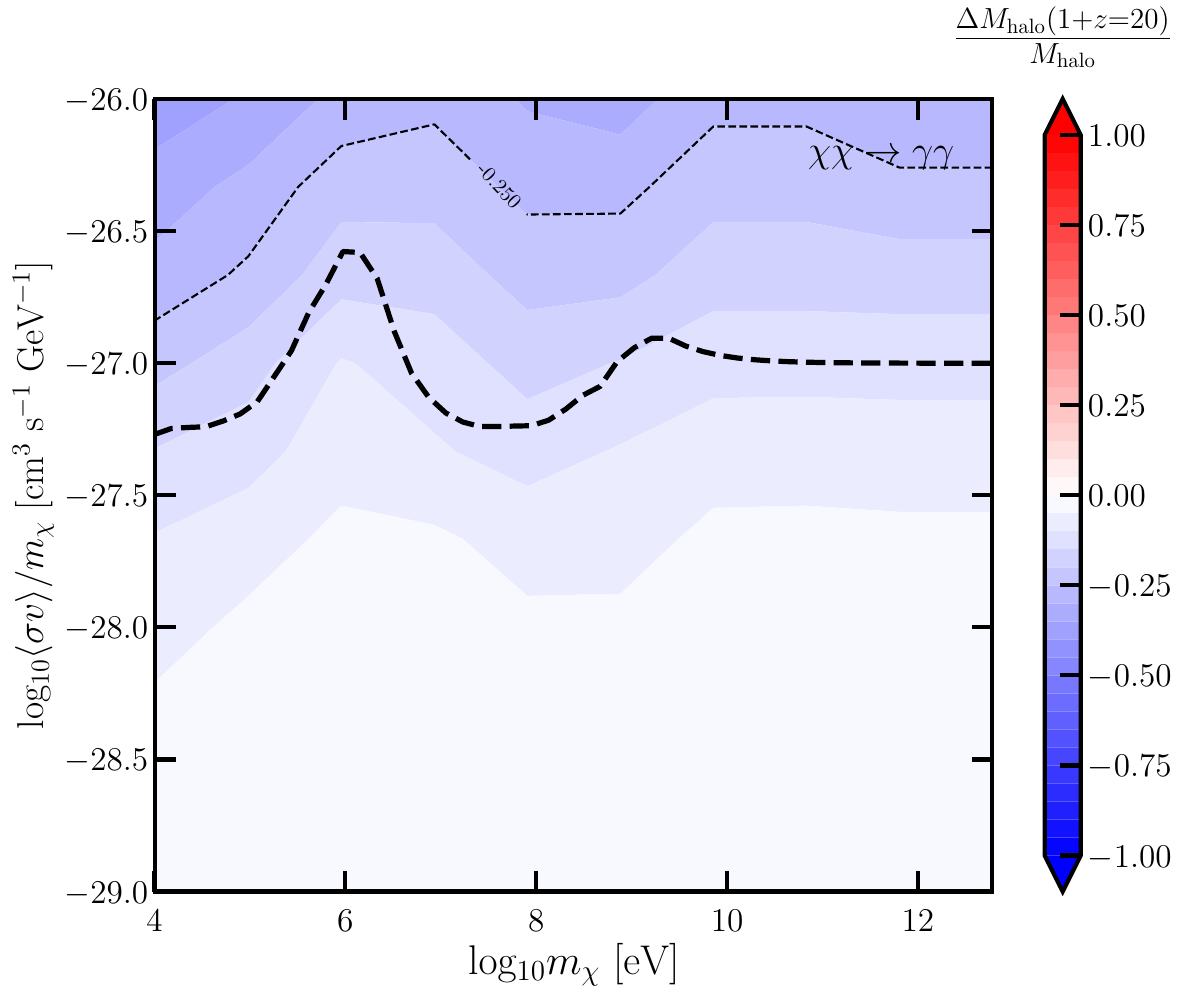}
    \includegraphics[scale=0.4]{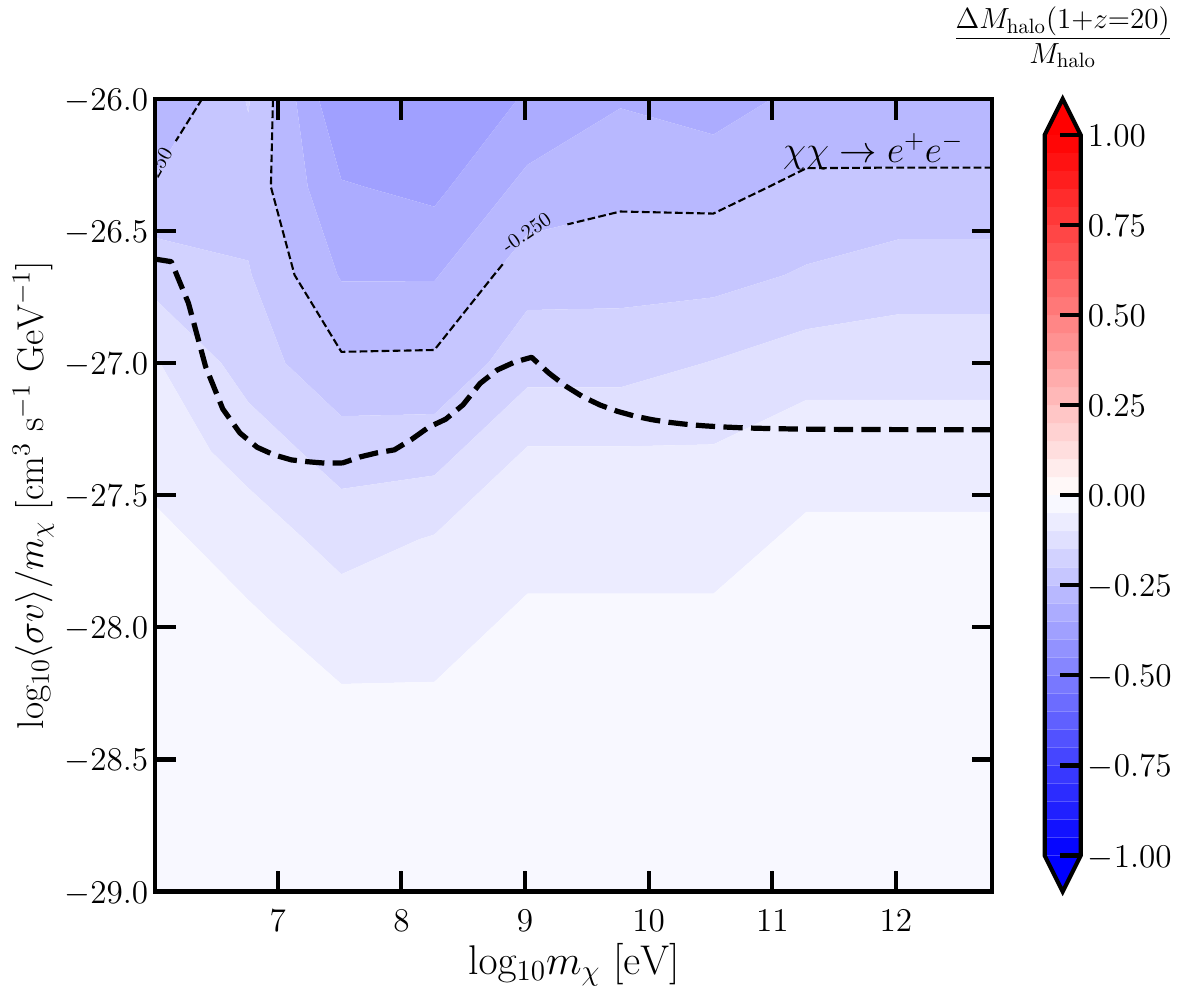} \\
    \includegraphics[scale=0.4]{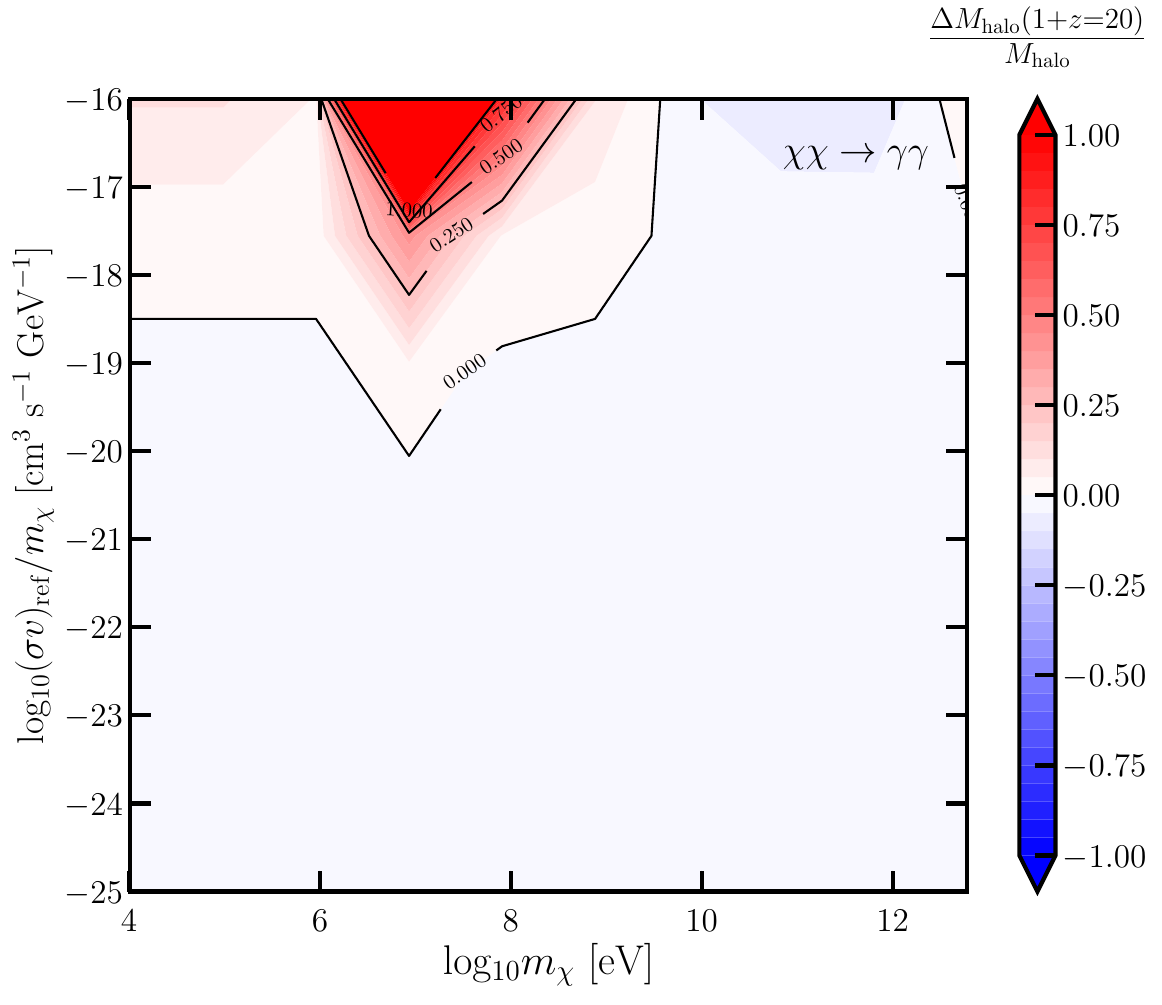}
    \includegraphics[scale=0.4]{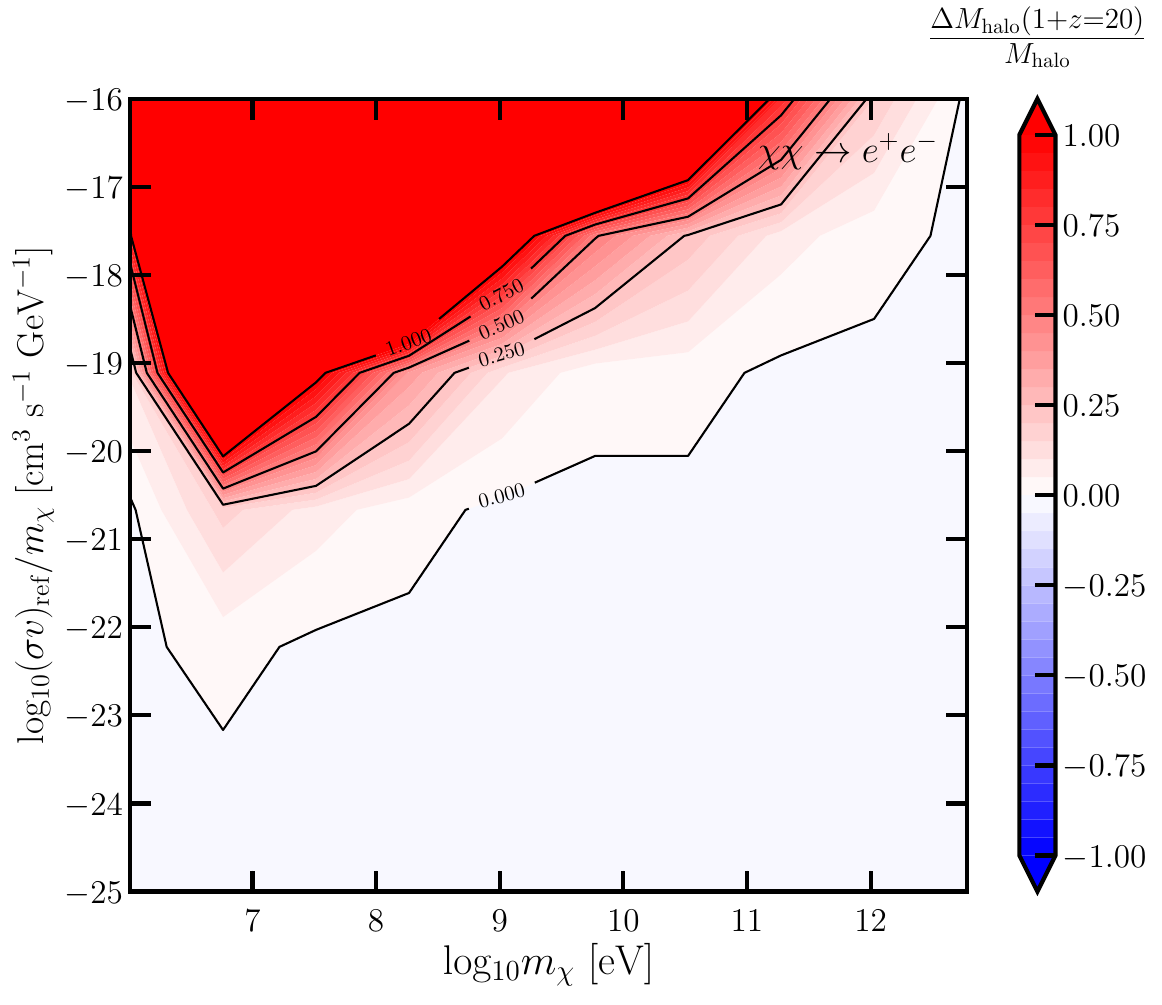}
    \caption{
    Change to the minimum halo mass $M_\mathrm{halo}$ necessary for star formation at $1+z=20$.
    From top row to bottom, the channels are decay, $s$-wave annihilation, and $p$-wave annihilation.
    For $p$-wave annihilation, the cross-section is defined using a reference velocity of 100 km/s as in Ref.~\cite{Liu:2016cnk}.
    In the left panels, the final state particles are photons; in the right, the final state particles are $e^+ e^-$ pairs.
    The thick black dashed lines show existing constraints from CMB data~\cite{Slatyer:2016qyl,Planck:2018vyg}.
    We also include constraints from X/$\gamma$-ray telescopes~\cite{Cadamuro:2011fd,Boddy:2015efa,Fermi-LAT:2013thd,Essig:2013goa, Massari:2015xea, Cohen:2016uyg,Laha:2020ivk,Cirelli:2020bpc,Cirelli:2023tnx}, where we have assumed $v = 220$ km s$^{-1}$ in the Milky Way, and Voyager I~\cite{Boudaud:2016mos, Boudaud:2018oya} (green dot-dashed). 
    For $p$-wave results, constraints lie below the bottom of the plot.
    }
    \label{fig:scans_z20}
\end{figure*}
Fig.~\ref{fig:scans_z20} shows the change in the minimum halo mass necessary for star formation at $1+z=20$ relative to the standard cosmology value over the parameter space for the energy injection channels described in Sec.~\ref{sec:scans}; here we assume self-shielding is very efficient and hence the effect of LW photons is suppressed, since this is closer to recent results from hydrodynamical simulations~\cite{Schauer:2020gvx,Kulkarni:2020ovu,2020MNRAS.492.4386S}.
We overlay existing constraints from CMB anisotropies~\cite{Slatyer:2016qyl,Planck:2018vyg} for decay and $s$-wave annihilation, as well as constraints from X/$\gamma$-ray telescopes~\cite{Cadamuro:2011fd,Boddy:2015efa,Fermi-LAT:2013thd,Essig:2013goa, Massari:2015xea, Cohen:2016uyg,Laha:2020ivk,Cirelli:2020bpc,Cirelli:2023tnx} and Voyager I~\cite{Boudaud:2016mos,Boudaud:2018oya} for decay, and mark Models $\bullet$\, and $\star$. 

For decays to photons, we show for illustration a selection of some of the strongest existing limits on photon lines from indirect detection~\cite{Cadamuro:2011fd,Boddy:2015efa,Fermi-LAT:2013thd}; we observe that these limits are generally markedly stronger than the CMB constraints. 
However, note that these bounds can only be applied directly to decays to $\gamma \gamma$ exclusively; the indirect constraints on photon-rich final states with continuum spectra are often considerably weaker. 
In contrast, we expect our parameter space to be sensitive primarily to the total injected energy (similar to the CMB limits), rather than the details of the photon spectrum; thus we expect the effects on star formation to be similar for the simple decay to $\gamma \gamma$ that we show and injection of continuum photons with similar total energy.

For $p$-wave annihilation, constraints lie below the bottom of the plot, and the $y$-axis on the bottom panels shows the value of $(\sigma v)_\mathrm{ref}$.
This velocity dependence means that the $p$-wave results are dominated by late-time structure formation.
Hence, these results should be used cautiously, since our assumption that the energy deposition fractions are equal between the IGM and the halo may be less reliable for annihilation, particularly $p$-wave annihilation.

Starting with the decay channels (top panels of Fig.~\ref{fig:scans_z20}), we see that for both photon and $e^+ e^-$ final states, most of the regions where the net effect is to raise the threshold for a halo to collapse are ruled out by current constraints; for decay to $e^+ e^-$ pairs, there still exist small regions just above the minimum allowed lifetime where the critical $M_\mathrm{halo}$ could be slightly raised; one is located around $m_\chi = 4$ MeV and the other just above $m_\chi = 1$ GeV.
There is also much unconstrained parameter space where the net effect is to \textit{lower} the threshold for collapse.
That is, DM can produce both positive and negative feedback in the first star formation, through its effect on the H$_2$ abundance of early galaxies.
The fiducial models studied earlier in this work were chosen such that one came from each of these regions.
For $s$-wave annihilation (middle panels), these models can only lower the threshold for collapse, but the regions with the largest effect are ruled out by CMB constraints.
For $p$-wave annihilation (bottom panels), these models only raise the threshold for collapse, but the regions where there is any significant effect are strongly ruled out by existing limits.

\begin{figure*}
    \includegraphics[scale=0.4]{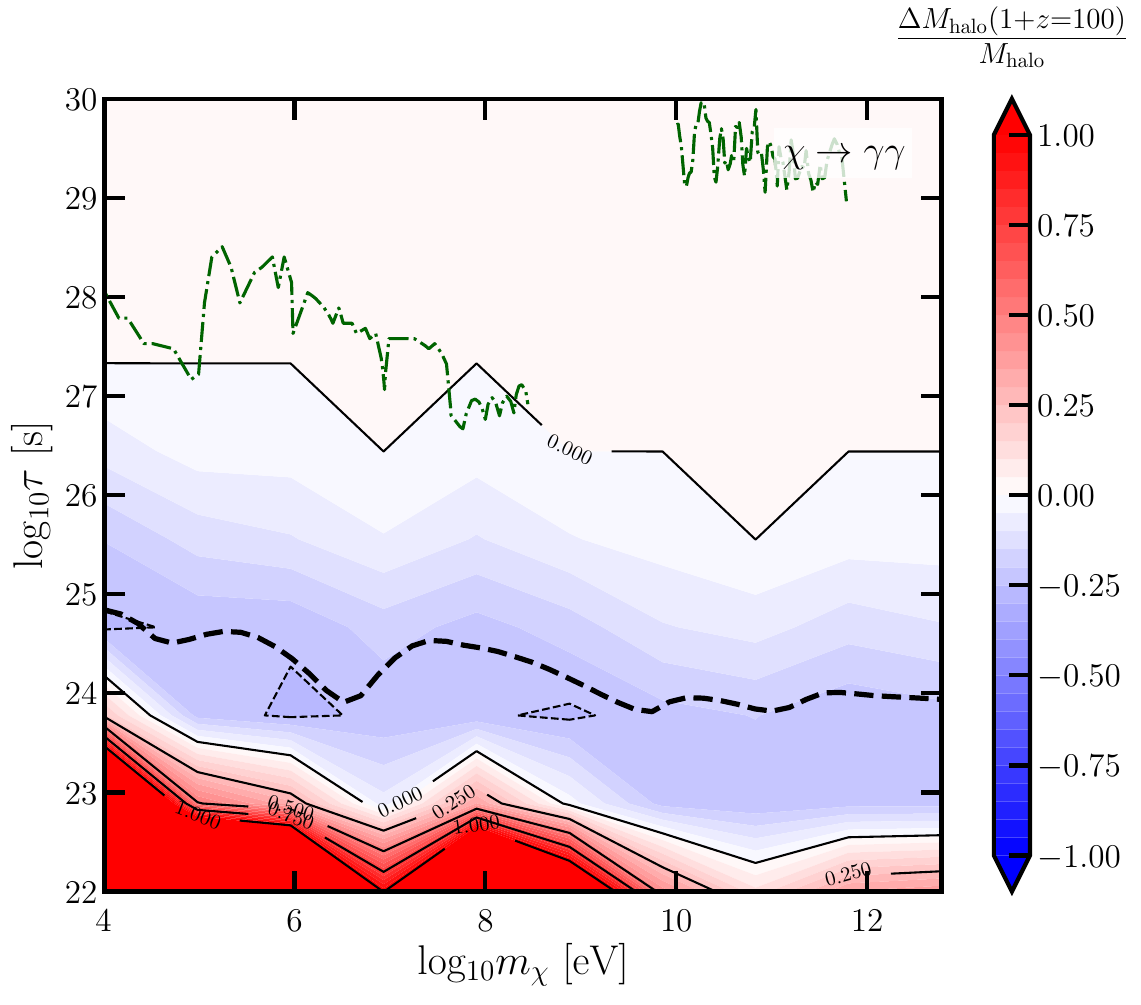}
    \includegraphics[scale=0.4]{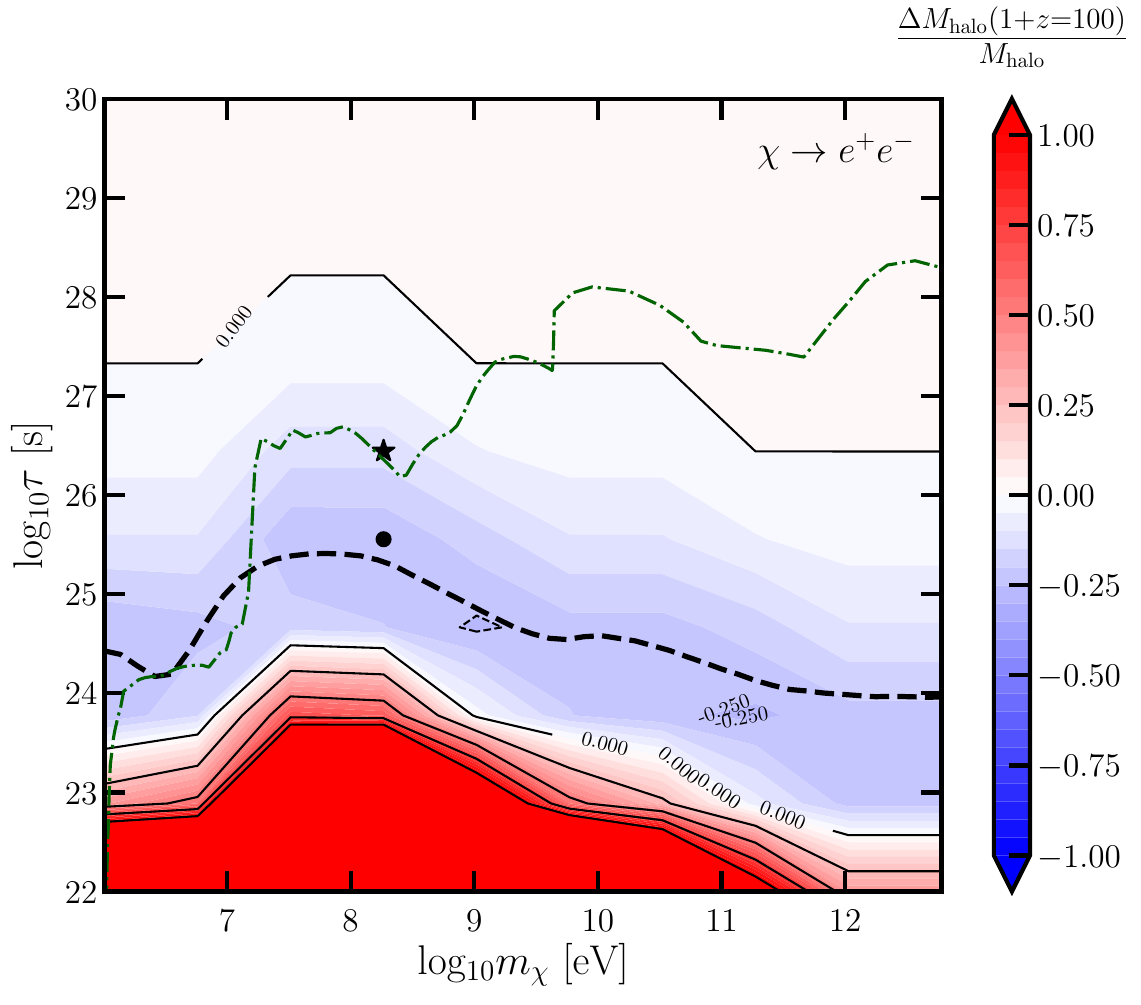} \\
    \includegraphics[scale=0.4]{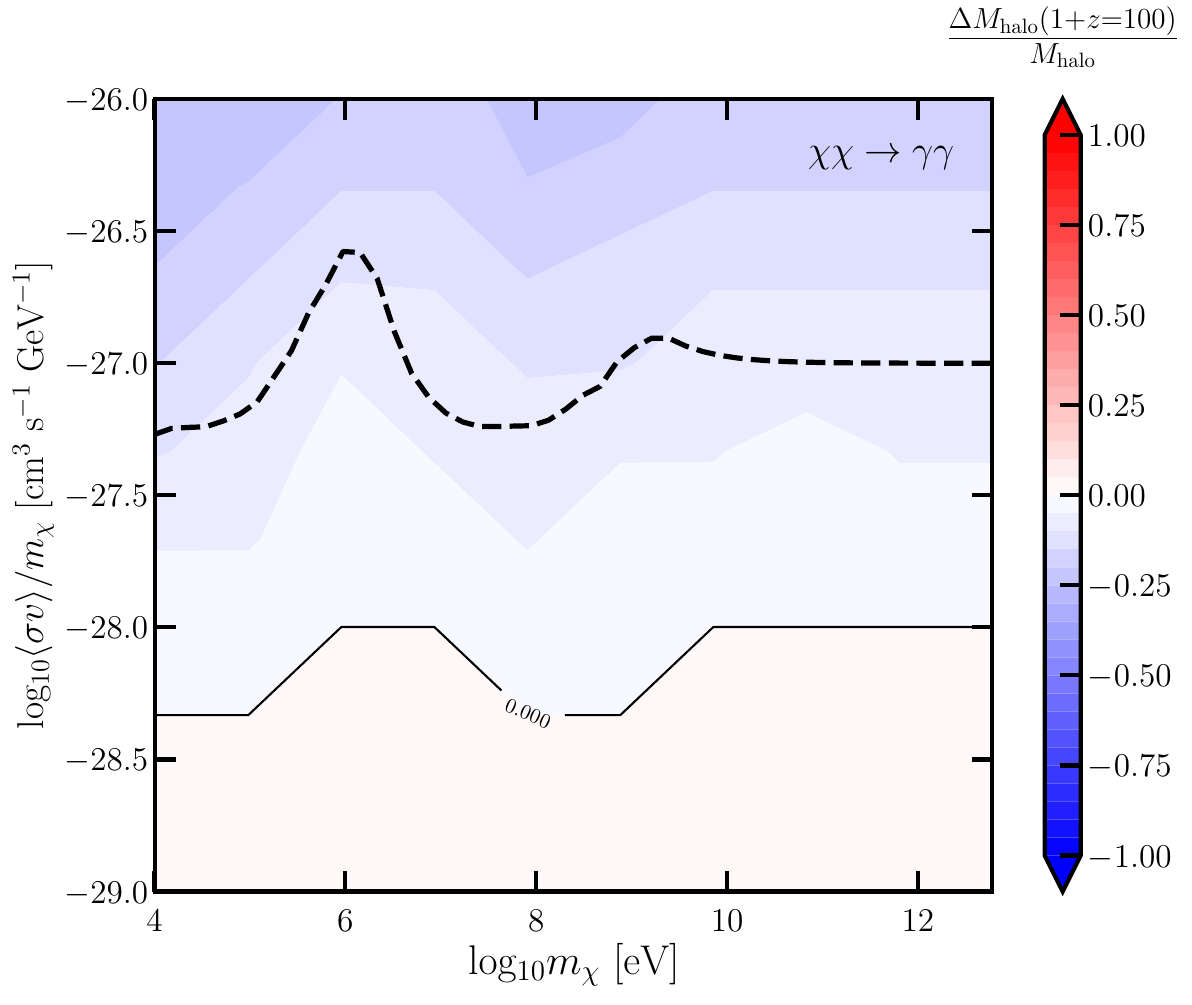}
    \includegraphics[scale=0.4]{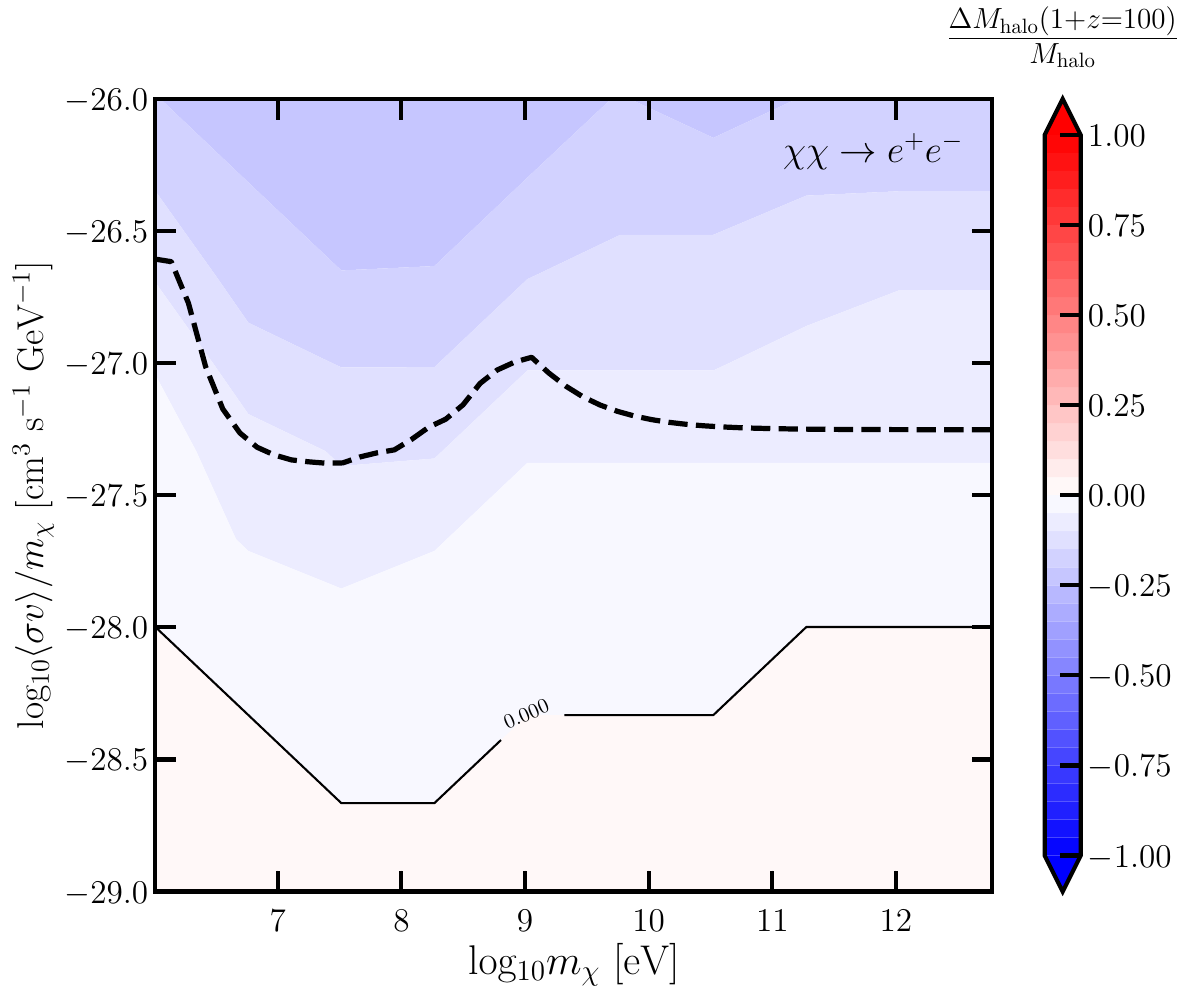} \\
    \includegraphics[scale=0.4]{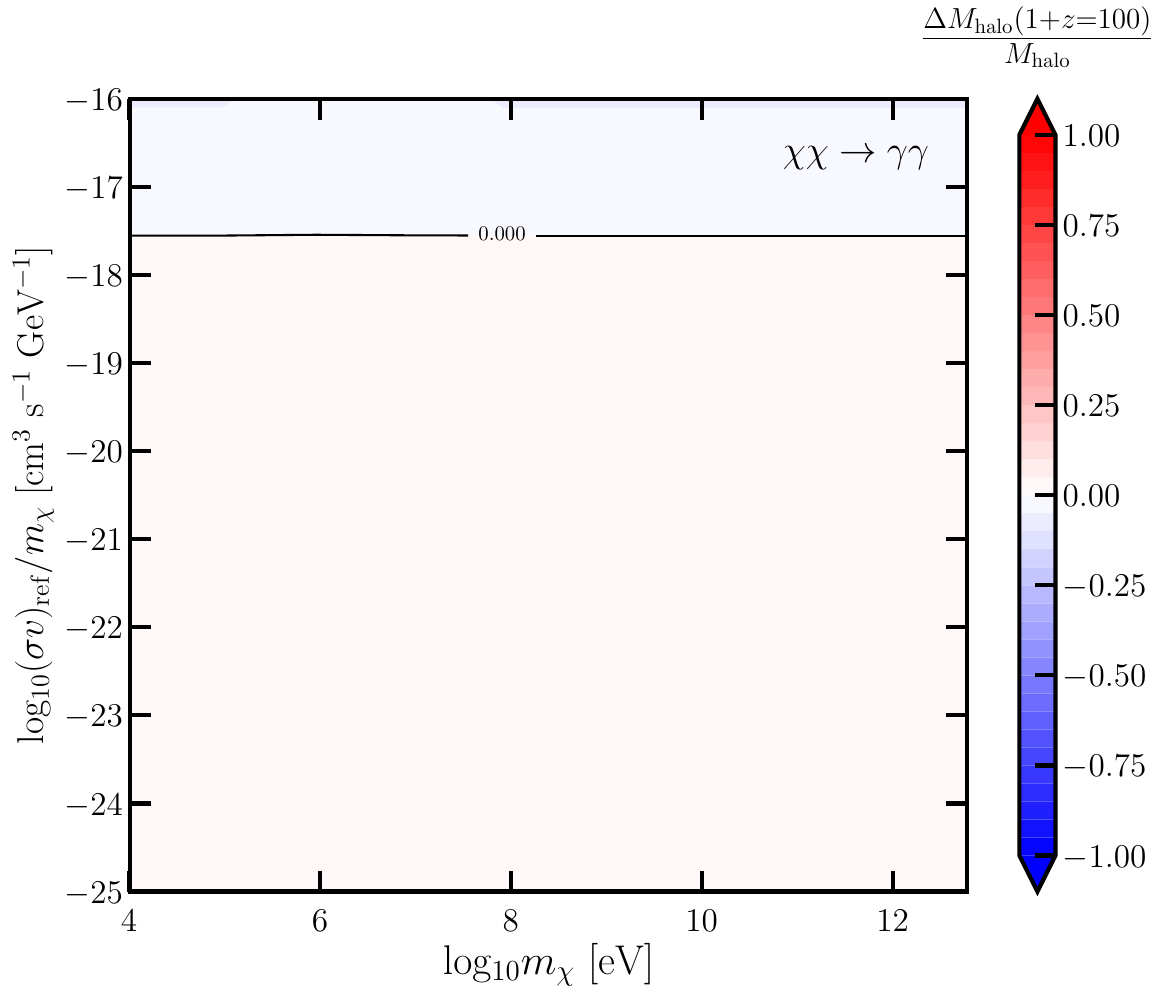}
    \includegraphics[scale=0.4]{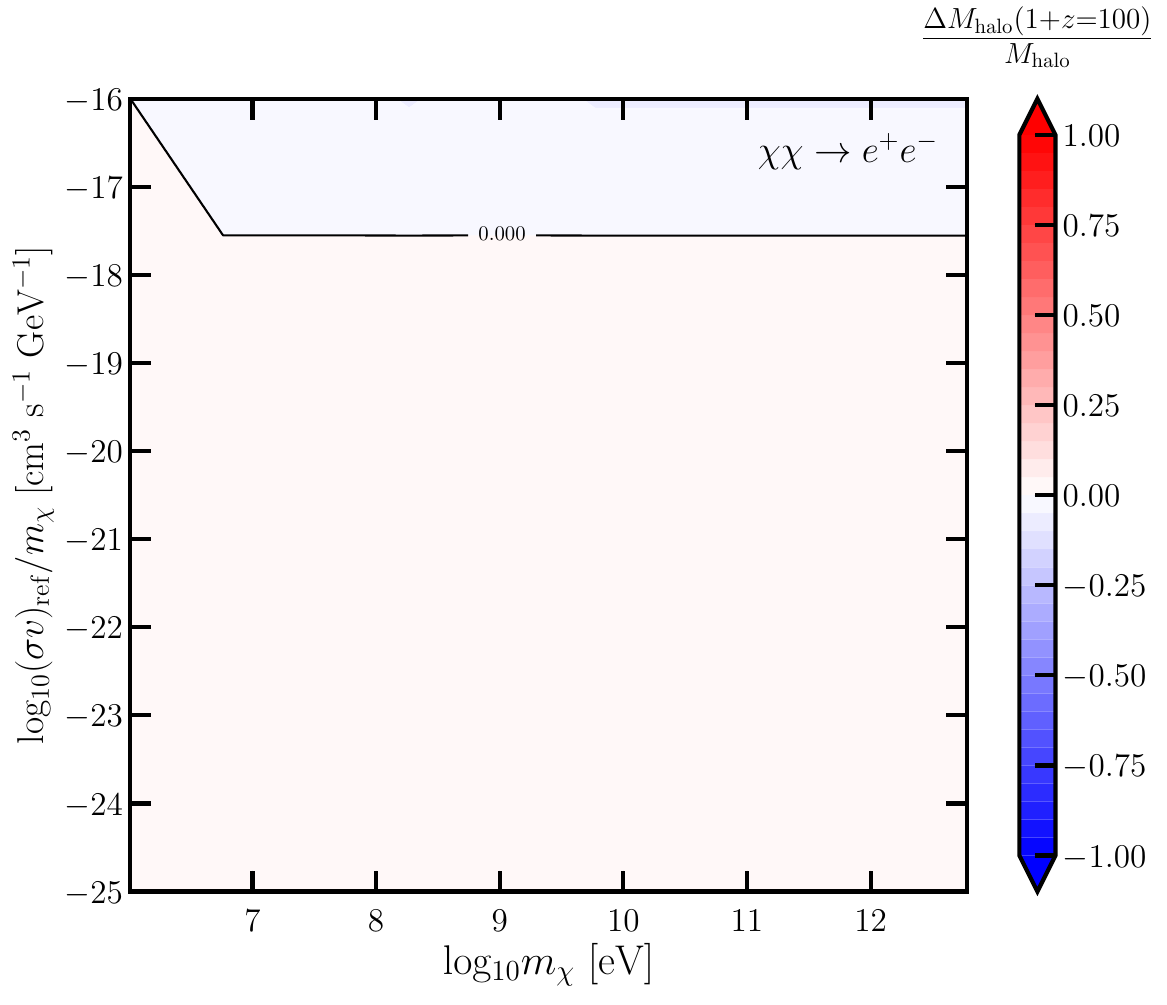}
    \caption{
    Same as Fig.~\ref{fig:scans_z20}, but showing the results at $1+z=100$.
    While the shape of the contours is broadly similar to Fig.~\ref{fig:scans_z20}, the contours are shifted to lower lifetimes/higher cross-sections, and the depth of the contours is reduced.
    }
    \label{fig:scans_z100}
\end{figure*}

The net effect of a particular energy injection model is redshift dependent.
Fig.~\ref{fig:scans_z100} shows an analogous plot to Fig.~\ref{fig:scans_z20}, but at a much earlier redshift of $1+z=100$; within the standard $\Lambda$ Cold Dark Matter ($\Lambda$CDM) cosmological model, we do not expect sufficient halos to form at this redshift, but we show these results for illustrative purposes.
The contours of the relative change to the critical $M_\mathrm{halo}$ change dramatically at this redshift.
For decaying DM, all regions that raise the critical $M_\mathrm{halo}$ at this redshift are ruled out by CMB constraints, and the region where $M_\mathrm{halo}$ can be lowered moves to smaller lifetimes.
Moreover, the shape of the contours at higher energies better matches the shape of the CMB constraints compared to the case at $1+z=20$.
This is because over time, the universe becomes increasingly transparent to photons with energy between 10 keV and 1 TeV~\cite{Chen:2003gz,Slatyer:2009yq}.
Prior to $1+z \sim 100$, most photons in this energy range will scatter and deposit their energy, so energy deposition results for these redshifts are relatively flat across the high mass range for decaying DM.
At lower redshifts, much of the particle cascade from high mass DM ends up in the transparency window and energy deposition becomes less efficient, hence the strength of the effects on star formation become weaker with increasing mass.

The panels showing the effect of $s$-wave annihilation at this redshift are similar to the case at $1+z=20$.
However, for $p$-wave annihilation, there is nearly no effect at all on the critical value for halo collapse; this is because the energy injection from $p$-wave annihilation scales even more steeply with velocity than for $s$-wave annihilation, hence the effects of $p$-wave annihilation are strongly suppressed for the redshifts just before structure formation.

\subsection{Bracketing Lyman-Werner effects}
\label{sec:shielding}

\begin{figure*}
    \includegraphics[scale=0.4]{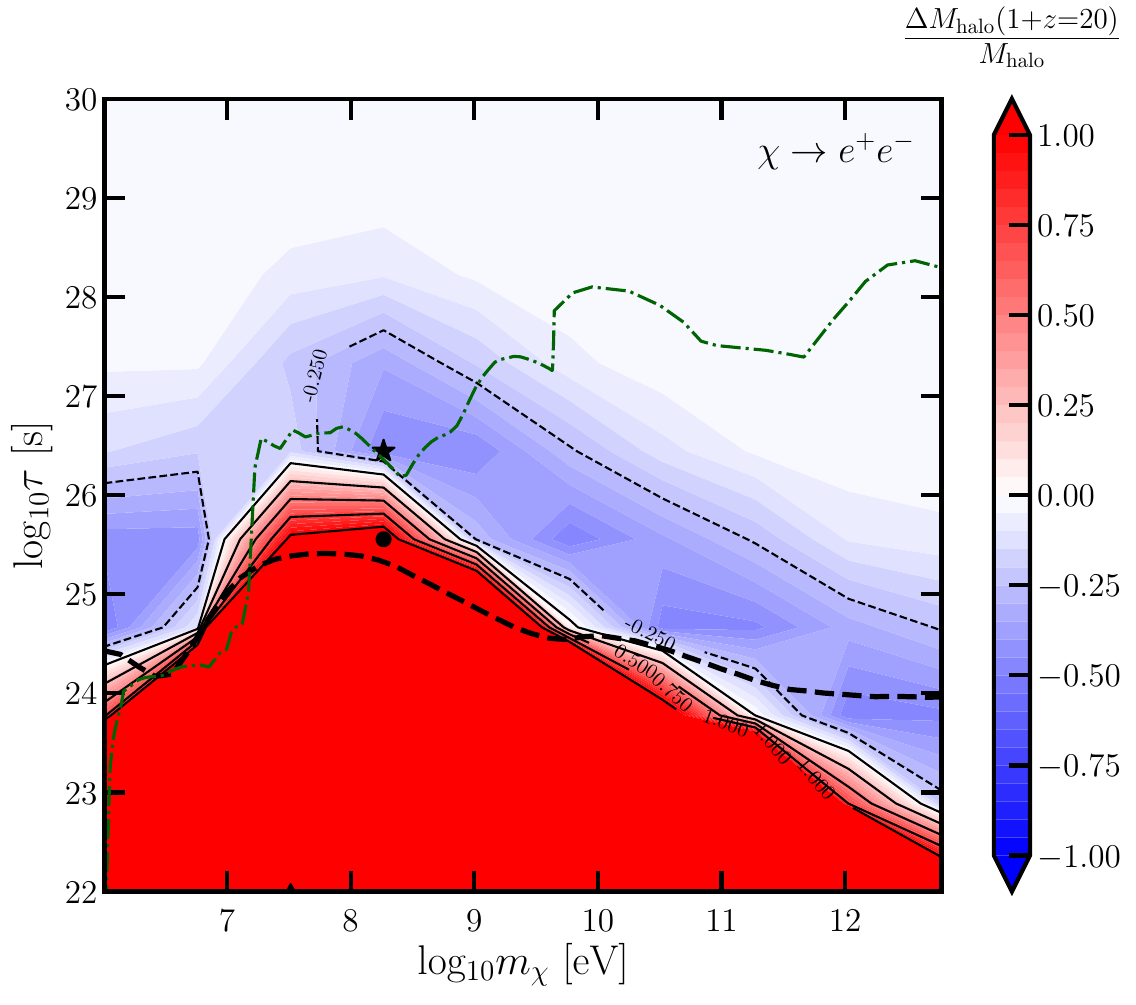}
    \includegraphics[scale=0.4]{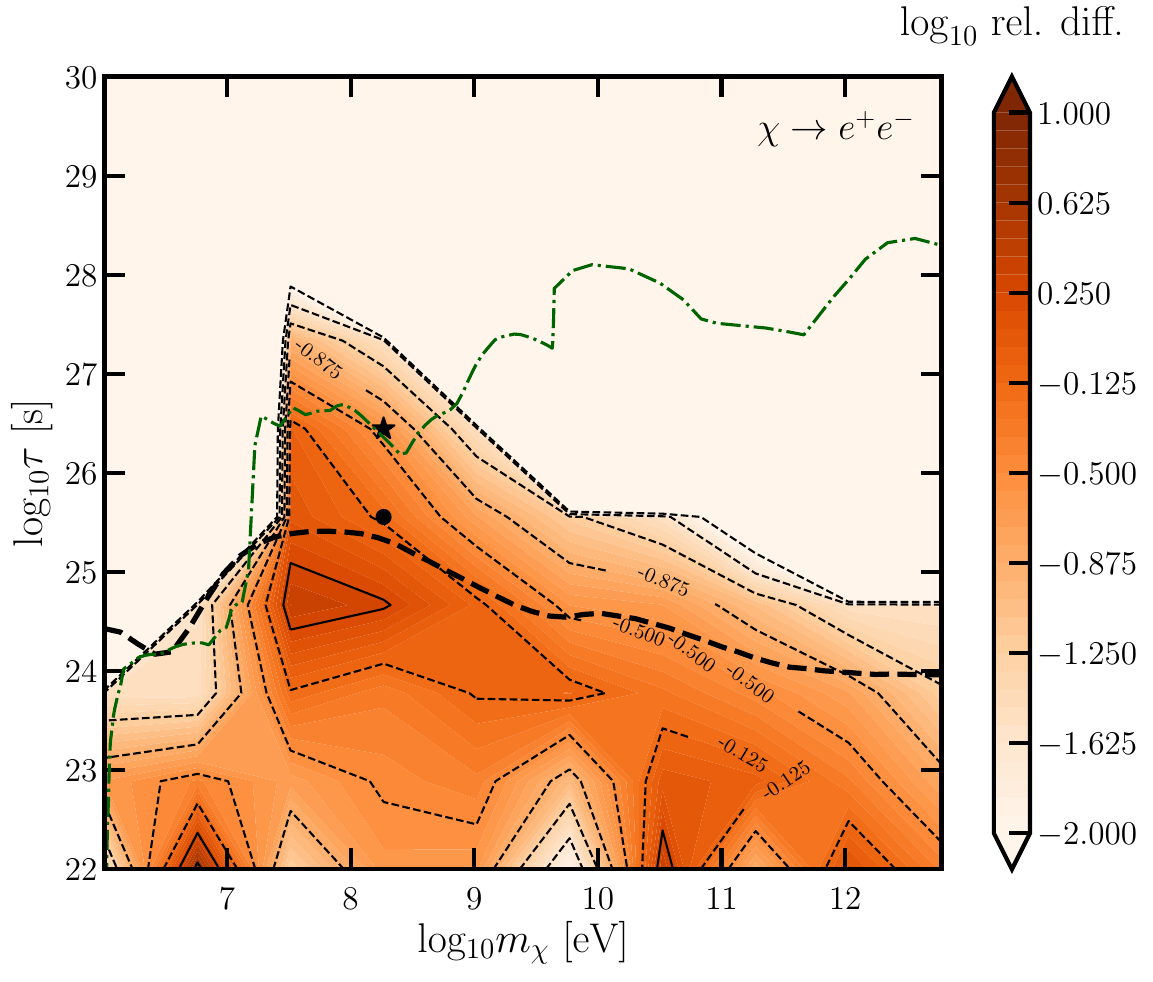} \\
    \includegraphics[scale=0.4]{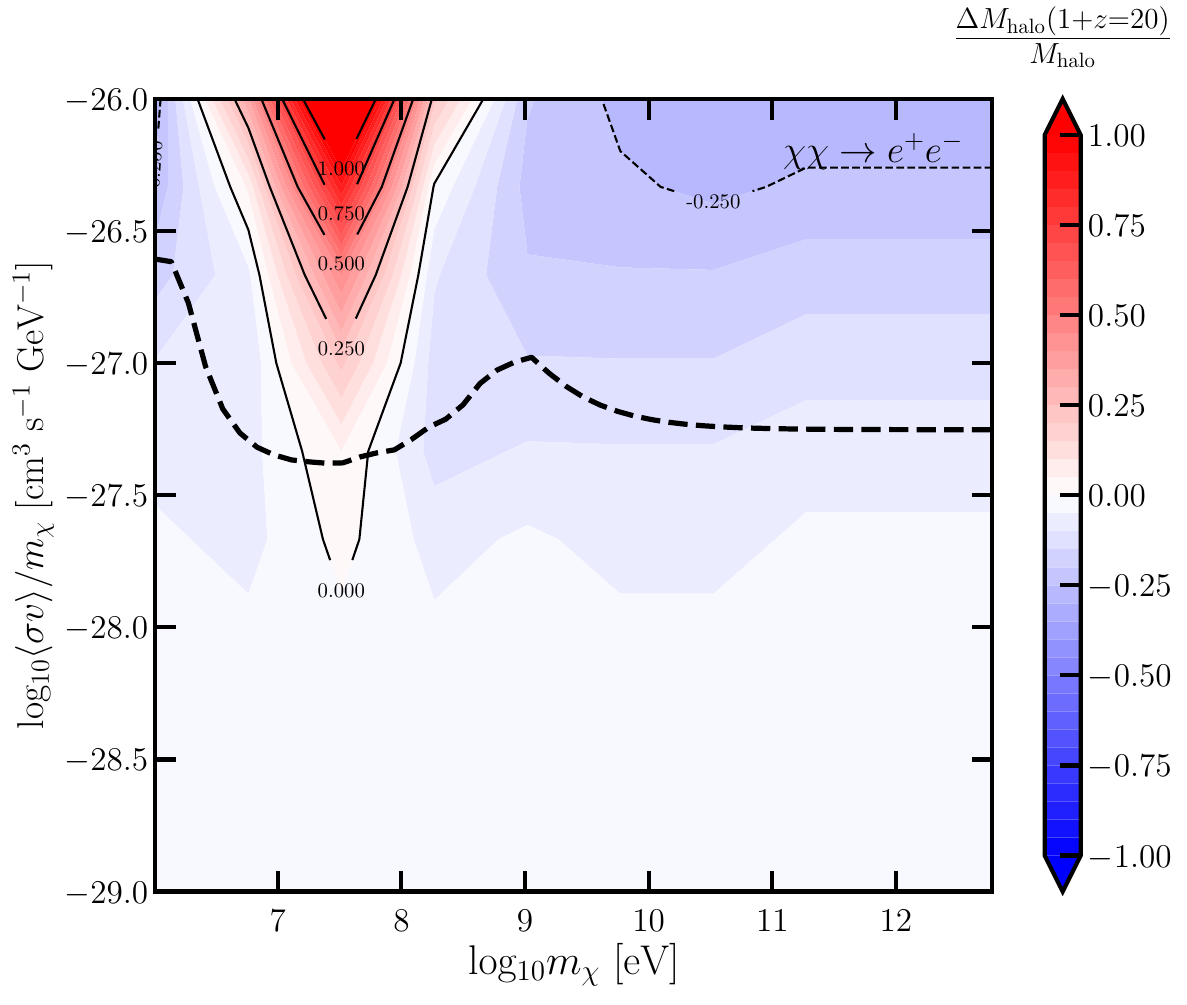}
    \includegraphics[scale=0.4]{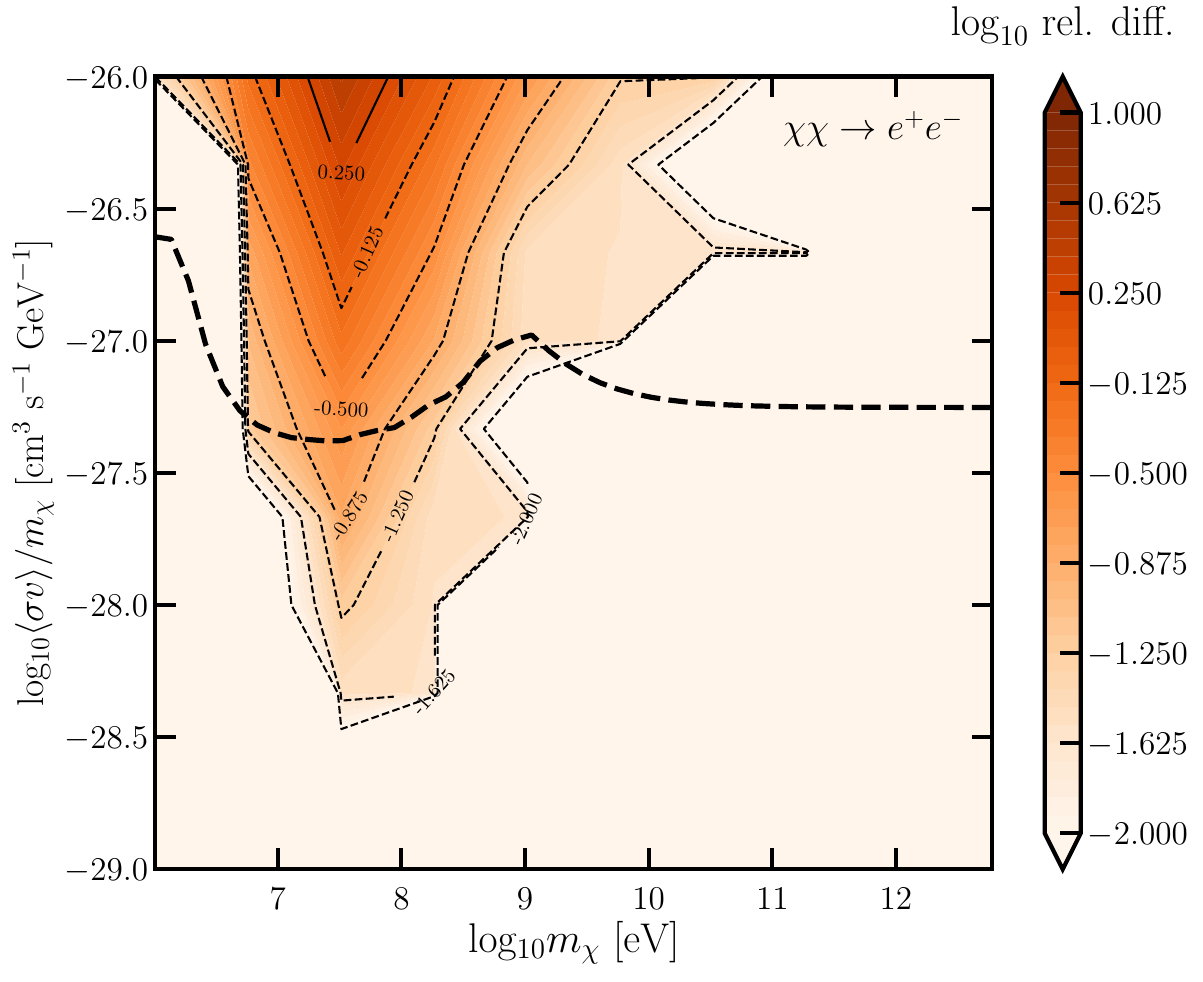}
    \caption{
    \textit{(Left)} Change to minimum halo mass $M_\mathrm{halo}$ necessary for star formation at $1+z=20$ when self-shielding is inefficient and a LW flux can have large effects.
    \textit{(Right)} $\log_{10}$ of the difference between the results for no/total self-shielding, divided by the total self-shielding results.
    In other words, the lightest contour shows where the results differ relative to each other by less than a percent.
    The top row shows the parameter space for DM decaying to $e^+ e^-$ pairs, and the bottom row is for $s$-wave annihilation to $e^+ e^-$.
    }
    \label{fig:scans_z20_LW}
\end{figure*}

In the previous subsection, we assumed that self-shielding is very efficient, such that LW photons have a negligible effect on the collapsing halo.
We now explore the opposite limit in order to bracket the effect of self-shielding.

Fig.~\ref{fig:scans_z20_LW} shows the results for decay (top row) and $s$-wave annihilation (bottom row) to $e^+ e^-$ pairs; the left panels show the same types of contours as in Fig.~\ref{fig:scans_z20} and the right shows the difference in the collapse threshold when assuming no/complete self-shielding of the halo.
See Appendix~\ref{app:LW} for discussion of other channels.
In both cases, reducing the efficacy of self-shielding raises the mass threshold for collapsing halos; the magnitude of the effect increases for larger energy injections (at even larger injections, the difference starts to decrease once the energy injection is large enough that heating matters more than the LW background).
This means that the effect on the mass threshold is increased for injection models that delay star formation, such as Model~$\bullet$, and slightly decreased for models which accelerated star formation, such as Model~$\star$.

The effect of self-shielding is most dramatic in the $s$-wave annihilation panel; whereas in the case of efficient self-shielding, all injection models shown decreased the threshold for collapse, neglecting self-shielding gives rise to a region of parameter space where it is now possible to raise the threshold.
This occurs around dark matter masses of tens of MeV; at this mass, the injected electrons are at the correct energy to upscatter CMB photons into the LW band through inverse Compton scattering (ICS).

Since inefficient self-shielding further slows the collapse of halos, this leads to the intriguing possibility that exotic heating and enhanced LW backgrounds may prevent gas fragmentation and subsequent star formation, but facilitate the collapse of the gas cloud directly into a black hole~\cite{Friedlander:2022ovf}.
We leave further study of the impact of dark matter processes on direct collapse black holes to future work.

\subsection{Signals in 21\,cm}
\label{sec:21cm}
\begin{figure}
    \includegraphics[scale=0.5]{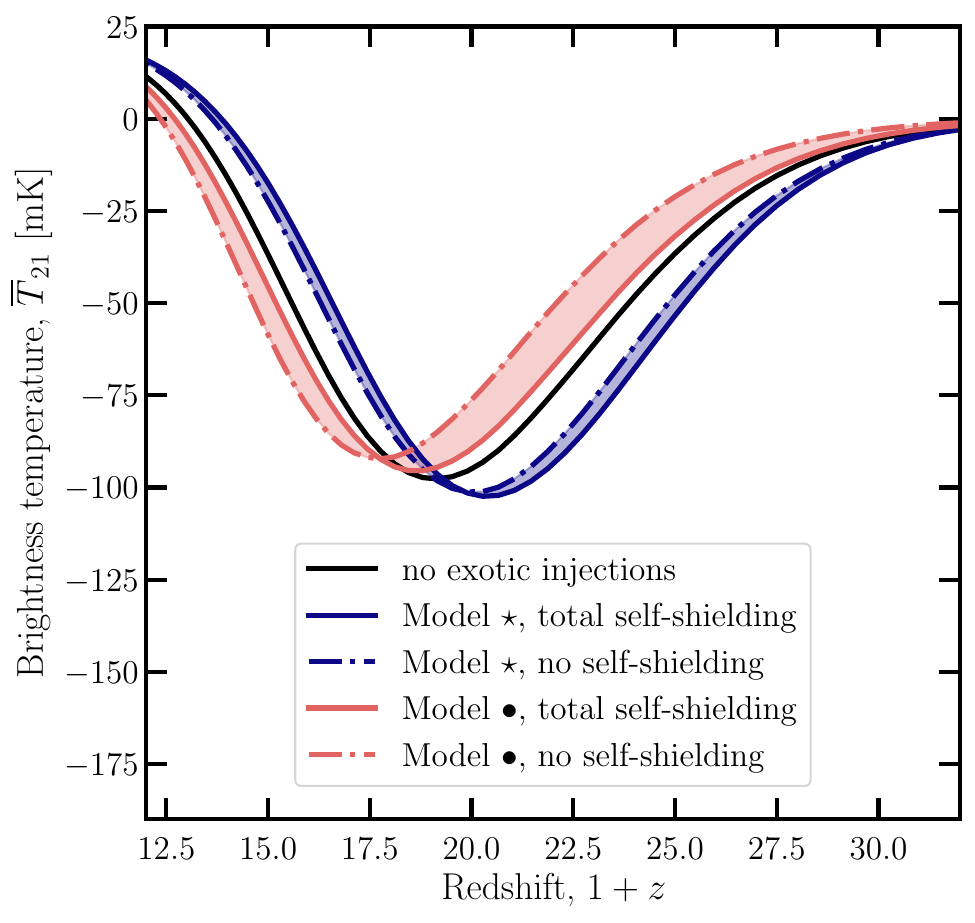}
    \includegraphics[scale=0.5]{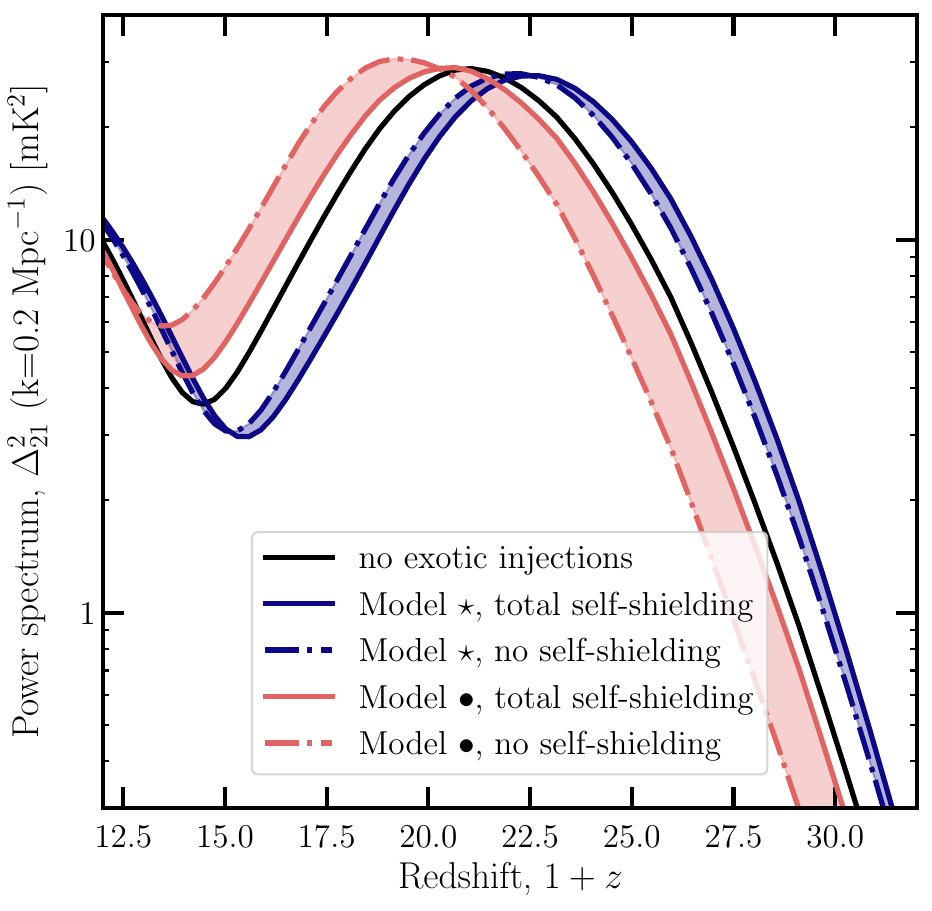}
    \caption{
    The 21\,cm global signal (top) and power spectrum at $k = 0.2$ Mpc$^{-1}$ (bottom) as a function of redshift, for our standard cosmological model and two fiducial DM models.
    The shaded contours bracket the effect of H$_2$ self-shielding.
    }
    \label{fig:21cm}
\end{figure}

One of the most promising ways to determine the timing of the first stellar formation is through the 21\,cm transition of neutral hydrogen at high redshifts (for a thorough review, see e.g.\ Ref.~\cite{Pritchard:2011xb}).
Here we briefly discuss how the DM models we have considered would affect an example 21\,cm signal.
We will focus on the timing of cosmic dawn through the enhancement or suppression of H$_2$ from exotic energy injection.
As such, we will not model other sources of feedback (e.g.\ stellar LW emission~\cite{Machacek:2000us}, DM-baryon relative velocities~\cite{Tseliakhovich:2010bj}, or their combination~\cite{Schauer:2020gvx,Kulkarni:2020ovu}, as well as heating~\cite{Valdes:2012zv,Sitwell:2013fpa,Evoli:2014pva,Liu:2018uzy}), and will focus on a few fiducial scenarios as a showcase.
We defer a detailed study of the 21\,cm signal, varying astrophysical parameters and including reionization bubbles, to future work, as this will require the establishment of hydrodynamical simulations of H$_2$ formation with DM decay or annihilation.

We use the public 21\,cm code {\tt Zeus21} \githubZeus~\cite{Munoz:2023kkg}, which we modify to include stellar formation in molecular-cooling halos.
We take a toy model where we enable star formation in halos below the atomic-cooling threshold with some constant star-formation efficiency $f_*^{\rm mol}$, and we keep the baseline {\tt Zeus21} model otherwise.
That is, we take the model for the star-formation efficiency from Ref.~\cite{Munoz:2023kkg},
\begin{equation}
    f_\star \equiv \frac{\dot M_\star}{f_b\,\dot M_\mathrm{halo}},
\end{equation}
where $\dot M_*$ is the star-formation rate, $\dot M_\mathrm{halo}$ is the mass-accretion rate, and $f_b$ is the baryon fraction, which we enhance as
\begin{equation}
    \Delta f_\star(M_\mathrm{halo}) = f_*^{\rm mol} e^{-M_{\rm mol}/M_\mathrm{halo}} \, e^{-M_\mathrm{halo}/M_{\rm atom}},
\end{equation}
between the atomic- and molecular-cooling thresholds ($M_{\rm atom}$ and $M_{\rm mol}$, respectively).
The former is set by a constant virial temperature of $T_{\rm atom}=10^4$ K~\cite{Oh:2001ex}, and we obtain the latter from our results for each DM scenario (as shown in Fig.~\ref{fig:critical_collapse_Mhalo}).
Following Refs.~\cite{Qin:2021gkn,Munoz:2021psm}, we set an amplitude of star-formation efficiency of $f_*^{\rm mol}=10^{-2.5}$, in broad agreement with current constraints to reionization.
In this simple model we assume the same stellar properties (e.g.\ Lyman-$\alpha$ and X-ray photons emitted per star-forming baryon) for all galaxies, though a more realistic model would take low-mass (H$_2$-cooling) galaxies to host older (PopIII) stars, with different spectra~\cite{Bromm:2003vv}.

With these caveats in mind, we can now estimate the degree to which we expect DM decays to affect the timing of the first stellar formation.
We show the predicted 21\,cm signals from this toy model in Fig.~\ref{fig:21cm}; the top panel shows the global signal, while the bottom panel shows the amplitude of the power spectrum at $k = 0.2$ Mpc$^{-1}$, both as a function of redshift.
Since Model $\bullet$\, raises the mass threshold for stars to form, the 21\,cm signals are slightly delayed, shifting to smaller values of $z$.
Conversely, Model $\star$\, allows smaller mass halos to host stars, so cosmic star formation begins earlier and the 21\,cm signals are accelerated.
In this latter case, the signal peaks/troughs can be shifted by as much as $\Delta z \sim 2$.

The degree to which the signals are shifted also depends on the efficiency of self-shielding; we bracket this effect using shaded contours in Fig.~\ref{fig:21cm}.
The possible variation of the signal due this effect is larger for Model $\bullet$, since this model injects more energy and can therefore contribute a larger LW background.
At lower redshifts, the astrophysical contribution to the LW background will also become important; once stars begin to form, they will emit their own LW radiation, and we have not modelled this feedback here.

While a full detectability study is beyond the scope of this work, we note in passing that a fiducial signal like our case without exotic injections in Fig.~\ref{fig:21cm} is expected to be detectable by the currently operating HERA interferometer~\cite{DeBoer:2016tnn}, boasting a signal-to-noise ratio of SNR $\approx 100$~\cite{Munoz:2021psm}. 
As such, this generation of telescopes may be sensitive to the delay/acceleration of the first galaxies due to decaying DM. 

Note that we have focused on the timing of the 21\,cm signal, and not its depth. 
In a future analysis, this and other potential effects of exotic energy injection should be studied together, which will pave the way to find the full effects of DM on the cosmic dawn.

\section{Conclusion}
\label{sec:conclusion}

We have performed an initial study of the effect of homogeneous energy deposition on early star formation.
We use \texttt{DarkHistory} to calculate how energy is deposited by decaying or annihilating DM and then track the effect of this exotic energy injection on the temperature, ionization, and H$_2$ abundance of a toy halo model.
We find that energy injection from decaying or annihilating DM can both raise and lower the critical mass/virial temperature threshold for galaxies to form, and the direction of this effect can depend on the redshift at which the halos virialize.
Hence, exotic energy injection can both accelerate and delay the onset of star formation, and this can in turn alter the timing of signals in 21\,cm cosmology.

The most interesting unconstrained parameter space comes from decaying dark matter, where the mass threshold for collapse can be lowered by as much as 60\%; while there are also some decaying DM models consistent with CMB anisotropy limits that can raise the mass threshold, most of these are ruled out by other indirect detection constraints.
Regarding annihilations, there is some unconstrained parameter space where $s$-wave annihilations can lower the mass threshold, although only by about 10\%; the regions where $p$-wave annihilation has any affect are ruled out by indirect detection limits.

We ignore a number of subdominant effects, including other H$_2$ formation pathways, the LW background, H$_2$ self-shielding, and baryon streaming velocities.
Moreover, although we include the boost factors for annihilation from structure formation, we neglect energy injection from within the halo itself for all channels; however, energy deposition by decays or annihilations from within a halo can also be significant, and even dominate the IGM signal in the case of annihilations.
A complete study of exotic energy injection on star formation would require modeling both contributions; this is left for future work.

Our results are obtained with a spherically collapsed model for the halo, so in order to make more precise statements, hydrodynamical simulations will be required.
However, it would be far too computationally expensive to scan over many DM models with such simulations in order to find those with the most significant effects.
In this work we have performed such a scan with a simpler semi-analytic model, and found the regions of DM parameter space that would be most interesting to simulate, including decays to photons between lifetimes of $\log_{10} (\tau / [\mathrm{s}]) = 24$ to 27, and decays to $e^+ e^-$ pairs between lifetimes of $\log_{10} (\tau / [\mathrm{s}]) = 24$ to 28, both for masses less than about 10 GeV (at larger masses, these lifetimes are excluded by indirect detection limits~\cite{Fermi-LAT:2013thd,Essig:2013goa, Massari:2015xea, Cohen:2016uyg,Laha:2020ivk,Cirelli:2020bpc,Cirelli:2023tnx,Boudaud:2016mos,Boudaud:2018oya}).
Within this range we find a very rich phenomenology, with DM both helping the formation of the first stars (by catalyzing H$_2$ formation), as well as hampering it (through heating and photodissociation).
We conclude that the cosmic-dawn era will not only teach us about the astrophysics of the first galaxies, but will also shed light onto the nature of dark matter.
\\

\section*{Acknowledgements}

We thank Katie Mack, Omer Katz, and Yash Aggarwal for useful discussions. 
WQ was supported by the National Science Foundation Graduate Research Fellowship under Grant Nos. 1745302 and 2141064.
TRS was supported by the Simons Foundation (Grant Number 929255, T.R.S) and by the National Science Foundation under Cooperative Agreement PHY-2019786 (The NSF AI Institute for Artificial Intelligence and Fundamental Interactions, \url{http://iaifi.org/}).
WQ and TRS were both supported by the U.S. Department of Energy, Office of Science, Office of High Energy Physics of U.S. Department of Energy under grant Contract Number DE-SC0012567. 
JBM is supported by the UT Austin Department of Astronomy Board of Visitors, and thanks the ITC at the Harvard-Smithsonian Center for Astrophysics for their hospitality during this work.
HL is supported by the DOE under Award Number DE-SC0007968, NSF grant PHY2210498, and the Simons Foundation.

This work made use of 
\texttt{NumPy}~\cite{Harris:2020xlr}, 
\texttt{SciPy}~\cite{2020NatMe..17..261V}, 
\texttt{Jupyter}~\cite{Kluyver2016JupyterN}, 
\texttt{matplotlib}~\cite{Hunter:2007ouj}, 
and \texttt{tqdm}~\cite{daCosta-Luis2019},
as well as Webplotdigitizer~\cite{Rohatgi2022}.

\appendix 
\onecolumngrid

\section{Comparison of IGM and halo contributions}
\label{app:IGM_vs_halo}

In this work, we have assumed that energy deposited per particle in the halo can be approximated by the energy deposited per particle from decays and annihilations in the IGM.
A natural question to ask is how the homogeneous contribution (i.e. from the IGM) compares to 
to the contributions of decays and annihilations within the halo itself.
For example, for Milky Way-like halos, the high density within the halo means that the annihilations within the halo typically contribute much more to indirect-detection signals than the IGM, whereas for decays the contributions are comparable~\cite{Ibarra:2009nw,Essig:2013goa}.
However, for earlier halos which are much smaller than the Milky Way halo, it is not as clear which contribution will dominate.

In the case where particles propagate with long path lengths, one can estimate the intensity of particles sourced by exotic energy injections within a system by calculating the $J$-factor for annihilations or the $D$-factor for decays.
For a spherically-symmetric system, these are given by
\begin{gather}
    J = \int d\ell \, \rho^2 (r (\ell, \theta, \psi)), \\
    D = \int d\ell \, \rho (r (\ell, \theta, \psi)) ,
\end{gather}
where $r$ denotes the distance from the center of the density profile, $\ell$ is the line-of-sight distance, and $\theta$ and $\psi$ specify the angle of the source relative to the observer.
These expressions can be derived by calculating the flux from annihilating or decaying dark matter and isolating the factors that depend on astrophysics~\cite{Lisanti:2016jxe,Slatyer:2021qgc}.

The effect of annihilations within early halos has been studied in e.g.\ Refs.~\cite{Schon:2014xoa,Schon:2017bvu}, hence we will compare to typical halos used in their work.
Consider a halo at redshift $1+z \sim 30$ with a mass of about $10^6 M_\odot$.
We will assume it has an Einasto density profile,
\begin{equation}
    \rho_\mathrm{Ein} = \rho_0 \exp \left\{ -\frac{2}{\alpha} \left[ \left( \frac{r}{r_0} \right)^\alpha - 1 \right] \right\}
\end{equation}
with $\alpha = 0.17$ and concentration parameter of approximately $c \sim 7$.
Given $M_\mathrm{halo}$, one can infer the virial radius $r_\mathrm{vir}$, which is related to the Einasto scale radius by $r_\mathrm{vir} = c r_0$; by calculating the mass within $r_\mathrm{vir}$, one can also determine the correct normalization $\rho_0$ for the density profile.
With these parameters, we find that at the center of the halo, $J \sim 10^{30}$ GeV$^2$ cm$^{-5}$ and $D \sim 10^{24}$ GeV cm$^{-2}$.

Turning to the IGM contribution, the meaning of the $D$ and $J$-factors becomes somewhat ambiguous, since when integrating out to cosmological distances, the expression for flux cannot be so easily factored into ``particle physics" and ``astrophysics" contributions due to redshifting of the emitted spectrum.
However, for the sake of an estimate, we can follow the discussion in Section 3.3 of Ref.~\cite{Slatyer:2021qgc} to effectively factor out the model-dependent terms, modifying this derivation appropriately for decays.
Then the $D$-factor from a homogeneous universe is
\begin{equation}
    D_\mathrm{IGM} = (1+z_\mathrm{obs})^3 \frac{\rho_\mathrm{DM,0}}{H_0} \int_{z_\mathrm{obs}}^\infty \frac{1}{\sqrt{\Omega_m (1+z)^3 + \Omega_\Lambda}} \, dz ,
\end{equation}
where $\rho_\mathrm{DM,0}$ is the average energy density of dark matter today and $\Omega_\Lambda$ is the density parameter for dark energy.
At $1+z_\mathrm{obs} \sim 30$, we find $D \sim 10^{26}$ GeV cm$^{-2}$; hence for early halos at this redshift, the IGM contribution can dominate by a factor up to a few orders of magnitude.

For annihilations, there is an extra factor of $\overline{\rho} (1+z)^3$; however, this integral will not converge if we integrate out to $1+z \rightarrow \infty$.
One could introduce an effective cutoff that accounts for the redshifting of the spectrum, as well as the potential absorption of emitted particles.
\begin{equation}
    J_\mathrm{IGM} = (1+z_\mathrm{obs})^6 \frac{\rho_\mathrm{DM,0}^2}{H_0} \int_{z_\mathrm{obs}}^{z_\mathrm{cut}} \frac{(1+z)^3}{\sqrt{\Omega_m (1+z)^3 + \Omega_\Lambda}} \, dz .
\end{equation}
However, this integral is likely an overestimate since the biggest contributions to $J_\mathrm{IGM}$ come from high redshifts, when densities are higher and secondary particles are more likely to be absorbed, or the emitted photons get redshifted out of observable wavelengths.
Given these uncertainties, it is less clear for the case of annihilations whether or not the contributions of the IGM and halo are comparable.

\section{Impact of the halo on dark matter energy deposition}
\label{app:fs_halo}

In reality, not all particles have long path lengths relative to the halo, and we also need to account for the enhanced gas density of the halo to understand how particles deposit their energy within it. Eqns.~\eqref{eqn:T_inj} and~\eqref{eqn:x_inj} describe the effect from DM energy deposition on the ionization and temperature within the halo. In this work, we have made the simplifying assumption that the per-baryon effect of DM is identical to that expected assuming homogeneity, allowing us to make use of results computed using \texttt{DarkHistory}~\cite{DH}. 
In this appendix, we will argue that this assumption can be justified throughout most of our parameter space of interest, including for the two fiducial models examined in detail in the main body. 
Even though the enhanced density of the halo leads to an increase in the rate of dark matter processes in the halo, and also provides additional targets for cascading particles to scatter off and cool, in most cases, we will show that the intensity of energy-depositing particles seen by targets inside the halo is essentially identical to the intensity of particles present in the homogeneous IGM. 
This means that the energy deposited \emph{per particle} is similar in both the halo and in the homogeneous IGM, justifying the use of the expressions shown in Eqns.~\eqref{eqn:T_inj} and~\eqref{eqn:x_inj}. 

\begin{figure}[t]

    \tikzset{every picture/.style={line width=0.75pt}} 
    
    \begin{tikzpicture}[x=0.75pt,y=0.75pt,yscale=-1,xscale=1]
    
    \draw  [fill={rgb, 255:red, 0; green, 0; blue, 0 }  ,fill opacity=1 ] (36,155.33) .. controls (36,145.94) and (43.61,138.33) .. (53,138.33) .. controls (62.39,138.33) and (70,145.94) .. (70,155.33) .. controls (70,164.72) and (62.39,172.33) .. (53,172.33) .. controls (43.61,172.33) and (36,164.72) .. (36,155.33) -- cycle ;
    \draw    (69.67,138.33) -- (125.57,108.61) ;
    \draw [shift={(127.33,107.67)}, rotate = 152] [color={rgb, 255:red, 0; green, 0; blue, 0 }  ][line width=0.75]    (10.93,-3.29) .. controls (6.95,-1.4) and (3.31,-0.3) .. (0,0) .. controls (3.31,0.3) and (6.95,1.4) .. (10.93,3.29)   ;
    \draw    (71,171) -- (124.22,198.09) ;
    \draw [shift={(126,199)}, rotate = 206.98] [color={rgb, 255:red, 0; green, 0; blue, 0 }  ][line width=0.75]    (10.93,-3.29) .. controls (6.95,-1.4) and (3.31,-0.3) .. (0,0) .. controls (3.31,0.3) and (6.95,1.4) .. (10.93,3.29)   ;
    \draw [color={rgb, 255:red, 245; green, 166; blue, 35 }  ,draw opacity=1 ] [dash pattern={on 4.5pt off 4.5pt}]  (109.33,50.98) -- (110,261.33) ;
    \draw  [fill={rgb, 255:red, 74; green, 144; blue, 226 }  ,fill opacity=1 ] (130,103) .. controls (130,98.58) and (133.58,95) .. (138,95) .. controls (142.42,95) and (146,98.58) .. (146,103) .. controls (146,107.42) and (142.42,111) .. (138,111) .. controls (133.58,111) and (130,107.42) .. (130,103) -- cycle ;
    \draw  [fill={rgb, 255:red, 74; green, 144; blue, 226 }  ,fill opacity=1 ] (129,202.33) .. controls (129,197.92) and (132.58,194.33) .. (137,194.33) .. controls (141.42,194.33) and (145,197.92) .. (145,202.33) .. controls (145,206.75) and (141.42,210.33) .. (137,210.33) .. controls (132.58,210.33) and (129,206.75) .. (129,202.33) -- cycle ;
    \draw [color={rgb, 255:red, 245; green, 166; blue, 35 }  ,draw opacity=1 ] [dash pattern={on 4.5pt off 4.5pt}]  (158.67,50.67) -- (158.67,260.7) ;
    \draw    (151.33,204.67) -- (178.08,212.44) ;
    \draw [shift={(180,213)}, rotate = 196.21] [color={rgb, 255:red, 0; green, 0; blue, 0 }  ][line width=0.75]    (10.93,-3.29) .. controls (6.95,-1.4) and (3.31,-0.3) .. (0,0) .. controls (3.31,0.3) and (6.95,1.4) .. (10.93,3.29)   ;
    \draw [color={rgb, 255:red, 245; green, 166; blue, 35 }  ,draw opacity=1 ] [dash pattern={on 4.5pt off 4.5pt}]  (208.67,50.63) -- (209.33,260.67) ;
    \draw    (150,106) -- (172.03,109.98) ;
    \draw [shift={(174,110.33)}, rotate = 190.23] [color={rgb, 255:red, 0; green, 0; blue, 0 }  ][line width=0.75]    (10.93,-3.29) .. controls (6.95,-1.4) and (3.31,-0.3) .. (0,0) .. controls (3.31,0.3) and (6.95,1.4) .. (10.93,3.29)   ;
    \draw  [fill={rgb, 255:red, 208; green, 2; blue, 27 }  ,fill opacity=1 ] (178,112.33) .. controls (178,107.92) and (181.58,104.33) .. (186,104.33) .. controls (190.42,104.33) and (194,107.92) .. (194,112.33) .. controls (194,116.75) and (190.42,120.33) .. (186,120.33) .. controls (181.58,120.33) and (178,116.75) .. (178,112.33) -- cycle ;
    \draw [color={rgb, 255:red, 245; green, 166; blue, 35 }  ,draw opacity=1 ] [dash pattern={on 4.5pt off 4.5pt}]  (259,50.97) -- (259,260.37) ;
    \draw    (195.33,119) -- (223.69,138.53) ;
    \draw [shift={(225.33,139.67)}, rotate = 214.56] [color={rgb, 255:red, 0; green, 0; blue, 0 }  ][line width=0.75]    (10.93,-3.29) .. controls (6.95,-1.4) and (3.31,-0.3) .. (0,0) .. controls (3.31,0.3) and (6.95,1.4) .. (10.93,3.29)   ;
    \draw  [fill={rgb, 255:red, 74; green, 144; blue, 226 }  ,fill opacity=1 ] (226,146.33) .. controls (226,141.92) and (229.58,138.33) .. (234,138.33) .. controls (238.42,138.33) and (242,141.92) .. (242,146.33) .. controls (242,150.75) and (238.42,154.33) .. (234,154.33) .. controls (229.58,154.33) and (226,150.75) .. (226,146.33) -- cycle ;
    \draw    (246,151.67) -- (273.44,161.02) ;
    \draw [shift={(275.33,161.67)}, rotate = 198.82] [color={rgb, 255:red, 0; green, 0; blue, 0 }  ][line width=0.75]    (10.93,-3.29) .. controls (6.95,-1.4) and (3.31,-0.3) .. (0,0) .. controls (3.31,0.3) and (6.95,1.4) .. (10.93,3.29)   ;
    \draw [color={rgb, 255:red, 245; green, 166; blue, 35 }  ,draw opacity=1 ] [dash pattern={on 4.5pt off 4.5pt}]  (319,51.65) -- (319,260.1) ;
    \draw  [fill={rgb, 255:red, 74; green, 144; blue, 226 }  ,fill opacity=1 ] (360.67,162.33) .. controls (360.67,157.92) and (364.25,154.33) .. (368.67,154.33) .. controls (373.08,154.33) and (376.67,157.92) .. (376.67,162.33) .. controls (376.67,166.75) and (373.08,170.33) .. (368.67,170.33) .. controls (364.25,170.33) and (360.67,166.75) .. (360.67,162.33) -- cycle ;
    \draw [color={rgb, 255:red, 245; green, 166; blue, 35 }  ,draw opacity=1 ] [dash pattern={on 4.5pt off 4.5pt}]  (418.67,51.33) -- (418.67,260.73) ;
    \draw [color={rgb, 255:red, 245; green, 166; blue, 35 }  ,draw opacity=1 ] [dash pattern={on 4.5pt off 4.5pt}]  (518,50.63) -- (518.67,260.03) ;
    \draw    (382,161.67) -- (452.06,143.5) ;
    \draw [shift={(454,143)}, rotate = 165.47] [color={rgb, 255:red, 0; green, 0; blue, 0 }  ][line width=0.75]    (10.93,-3.29) .. controls (6.95,-1.4) and (3.31,-0.3) .. (0,0) .. controls (3.31,0.3) and (6.95,1.4) .. (10.93,3.29)   ;
    \draw  [fill={rgb, 255:red, 208; green, 2; blue, 27 }  ,fill opacity=1 ] (460,141.67) .. controls (460,137.25) and (463.58,133.67) .. (468,133.67) .. controls (472.42,133.67) and (476,137.25) .. (476,141.67) .. controls (476,146.08) and (472.42,149.67) .. (468,149.67) .. controls (463.58,149.67) and (460,146.08) .. (460,141.67) -- cycle ;
    \draw [color={rgb, 255:red, 245; green, 166; blue, 35 }  ,draw opacity=1 ] [dash pattern={on 4.5pt off 4.5pt}]  (618.67,50.9) -- (618.67,259.67) ;
    \draw    (481.33,146.33) -- (554.36,197.19) ;
    \draw [shift={(556,198.33)}, rotate = 214.85] [color={rgb, 255:red, 0; green, 0; blue, 0 }  ][line width=0.75]    (10.93,-3.29) .. controls (6.95,-1.4) and (3.31,-0.3) .. (0,0) .. controls (3.31,0.3) and (6.95,1.4) .. (10.93,3.29)   ;
    \draw  [fill={rgb, 255:red, 74; green, 144; blue, 226 }  ,fill opacity=1 ] (559.33,202.33) .. controls (559.33,197.92) and (562.92,194.33) .. (567.33,194.33) .. controls (571.75,194.33) and (575.33,197.92) .. (575.33,202.33) .. controls (575.33,206.75) and (571.75,210.33) .. (567.33,210.33) .. controls (562.92,210.33) and (559.33,206.75) .. (559.33,202.33) -- cycle ;
    \draw    (580.67,198.33) -- (628.92,171.31) ;
    \draw [shift={(630.67,170.33)}, rotate = 150.75] [color={rgb, 255:red, 0; green, 0; blue, 0 }  ][line width=0.75]    (10.93,-3.29) .. controls (6.95,-1.4) and (3.31,-0.3) .. (0,0) .. controls (3.31,0.3) and (6.95,1.4) .. (10.93,3.29)   ;

    \draw (46.67,149.07) node [anchor=north west][inner sep=0.75pt]  [color={rgb, 255:red, 255; green, 255; blue, 255 }  ,opacity=1 ]  {$\chi $};
    \draw (128.67,232.07) node [anchor=north west][inner sep=0.75pt]    {$1$};
    \draw (39.33,232.33) node [anchor=north west][inner sep=0.75pt]   [align=left] {{Step}};
    \draw (182,208.73) node [anchor=north west][inner sep=0.75pt]    {$\cdots $};
    \draw (179.33,232.4) node [anchor=north west][inner sep=0.75pt]    {$2$};
    \draw (228,232.4) node [anchor=north west][inner sep=0.75pt]    {$3$};
    \draw (280,158.73) node [anchor=north west][inner sep=0.75pt]    {$\cdots $};
    \draw (10,50.33) node [anchor=north west][inner sep=0.75pt]   [align=left] {\begin{minipage}[lt]{63.16pt}\setlength\topsep{0pt}
    \begin{center}
    {Enhancement Factor}
    \end{center}
    
    \end{minipage}};
    \draw (126.33,59.73) node [anchor=north west][inner sep=0.75pt]    {$\eta _{1}$};
    \draw (349.67,60.4) node [anchor=north west][inner sep=0.75pt]    {$\eta _{p} =1$};
    \draw (176,59.73) node [anchor=north west][inner sep=0.75pt]    {$\eta _{2}$};
    \draw (224.33,59.73) node [anchor=north west][inner sep=0.75pt]    {$\eta _{3}$};
    \draw (630.67,160.68) node [anchor=north west][inner sep=0.75pt]    {$\cdots $};
    \draw (363.33,231.4) node [anchor=north west][inner sep=0.75pt]    {$n$};
    \draw (456.33,231.4) node [anchor=north west][inner sep=0.75pt]    {$n+1$};
    \draw (555,231.73) node [anchor=north west][inner sep=0.75pt]    {$n+2$};
    \draw (324.95,200.86) node [anchor=north west][inner sep=0.75pt]  [rotate=-269.85] [align=left] {\begin{minipage}[lt]{63.67pt}\setlength\topsep{0pt}
    \begin{center}
    {Homogeneous}
    \end{center}
    
    \end{minipage}};
    \draw (434,60.07) node [anchor=north west][inner sep=0.75pt]    {$\eta _{s}( I_{p} =I_{p,0})$};
    \draw (534,60.73) node [anchor=north west][inner sep=0.75pt]    {$\eta _{t}( I_{p} =I_{p,0})$};

    \end{tikzpicture}
    \caption{Cartoon illustrating a particle cascade and the two scenarios that we consider in order to calculate enhancement factors $\eta$: \emph{(Left)} $\eta_1$ for particles directly emitted by DM, followed by the secondary $\eta_2$ and tertiary $\eta_3$, as well as \emph{(Right)} uniform intensity of primaries at step $n$ ($I_p = I_{p,0}$ and $\eta_p = 1$), from which we can calculate the enhancement factor of step $n+1$ secondaries $\eta_s(I_p = I_{p,0})$ and step $n+2$ tertiaries $\eta_t(I_p = I_{p,0})$.}
    \label{fig:cascade_cartoon}
    
\end{figure}
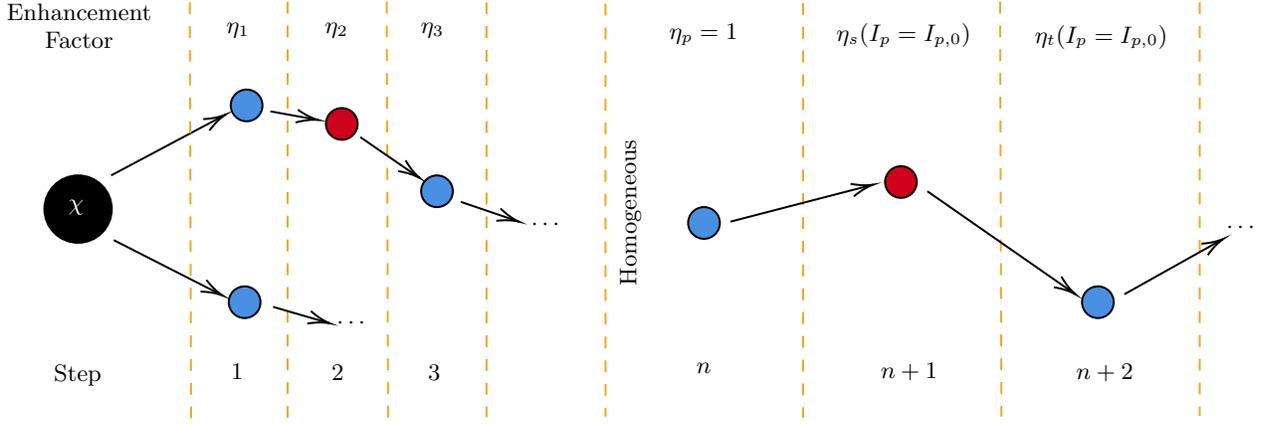

A full and precise treatment of this problem would be challenging, requiring tracking the 4D evolution of the secondary particle cascade that leads to energy deposition.  In this appendix, we instead perform a simplified analysis using the following assumptions: 
\begin{enumerate}
    \item We model the halo as a simple tophat of mass $M_\mathrm{halo}$, and radius $r_\mathrm{vir}$, with a density $\Delta$ times larger than the mean matter energy density. 
    \item The intensity of particles seen by targets in the halo is approximated by the intensity of particles seen at the center of this halo, which is spherically symmetric about this point. 
    \item All spectra produced are treated as monochromatic, with particles entirely scattering into the peak of their scattered spectra; for example, every electron with Lorentz factor $\gamma$ that inverse Compton scatters against the CMB is assumed to produce photons with energy given by the mean photon energy $2 \pi^4 \gamma^2 T_\mathrm{CMB} / [45 \zeta(3)]$ only.
    \item All scattering processes occur only in the forward direction. 
\end{enumerate}
In general, the particle cascade can undergo many steps before producing particles that are rapidly absorbed as ionization and heating, or which free-stream (in the case of sub-\SI{13.6}{\eV} photons).
Our goal is to determine the intensity of particles in the last step of the cascade, which either deposit their energy directly into ionization or heating of the gas. 
We will do this by calculating the intensity of appropriately chosen intermediate steps.
Consider a particle at step $n$ of the particle cascade that we will call the `primary' particle, cascading into a `secondary' particle at step $n+1$ and a `tertiary' particle at step $n+2$.
Under the assumptions listed above, we will find it useful to obtain the steady-state intensity of secondary and tertiary particles under the following conditions, both illustrated in Fig.~\ref{fig:cascade_cartoon}:  
\begin{enumerate}
    \item If $n = 1$, the primary particles are particles emitted directly by the DM process, and we can compute the intensity of the $n = 2$ secondary and $n = 3$ tertiary particles; 
    \item For any $n$, if the intensity of the primary particle $I_p$ is equal to its intensity expected in the homogeneous limit $I_{p,0}$,  we can obtain an expression for the intensities of $n+1$ secondary particles, denoted $I_s(I_p = I_{p,0})$, and $n+2$ tertiary particles $I_t(I_p = I_{p,0})$.
\end{enumerate}
To characterize these intensities, we define the quantity $\eta_n \equiv I_n / I_{n,0}$, i.e.\ the ratio of the intensity at step $n$ in the halo to the homogeneous, steady-state intensity that exists deep in the homogeneous IGM.
$\eta_n = 1$ means that the intensity of particles in step $n$ within the halo is given by the homogeneous, steady-state intensity, while $\eta_n \neq 1$ represents either an enhancement or a suppression inside the halo.

With a combination of the two scenarios summarized in Fig.~\ref{fig:cascade_cartoon}, we are able to determine the intensity of particles in the final step for DM decay into $e^+e^-$ and photon pairs for $m_\chi < \SI{10}{\giga\eV}$. For example, starting from particles emitted directly from the DM, we can first find that $\eta_3 \approx 1$, and then use $\eta_s(I_p = I_{p,0})$ to calculate $\eta_4(I_3 = I_{3,0})$, finding that $\eta_4 \approx 1$, allowing us to proceed down the cascade. 
$\eta_3$ and $\eta_t(I_p = I_{p,0})$ are computed so that we can skip intermediate steps that have short path length, which may result in $\eta \neq 1$. This will ultimately enable us to estimate the impact of the halo on energy deposition. 
In general, cascades will pass through a series of steps with long path lengths that have intensities that are close to the homogeneous limit; the impact of the halo is strong only when the last or last few steps have short path lengths. 
These results can easily be extended to annihilations as well.
We leave a more thorough understanding of how the intensity of particles from DM processes in a halo differs from the homogeneous intensity to future work. 

In this appendix, we will mainly focus on ionization and heating, and only note here that LW photons can also be affected by the presence of the halo. 
To model the intensity of LW photons within the halo, we would have to go significantly beyond the tophat model discussed here, since it depends sensitively on the H$_2$ abundance within the halo. 
Even within $\Lambda$CDM, the amount of self-shielding or equivalently the path length of LW photons is still somewhat unclear. 
If self-shielding of LW photons is not significant, then we expect LW photons to have path lengths a factor of a few times smaller than the Hubble radius, since the H$_2$ fraction in the IGM is negligible, and the path length is by definition much longer than the halo itself. 
We would therefore expect the LW intensity to be similar to the homogeneous, steady-state intensity in the IGM, loosely justifying our use of the LW intensity in the IGM in the limit of no self-shielding. 
For now, we set aside the question of the effect of the halo on LW photons originating from DM, leaving it to simulations to address in detail.

\subsection{Particles Directly Emitted from the Dark Matter Process}
\label{app:primaries}
We first begin by understanding the intensity of $n=1$ particles emitted directly from the DM process, and received at the center of the halo. 
We start from the radiative transport equation applied to lines pointing radially outward from the center of the halo, 
\begin{alignat*}{1}
    \frac{dI_1}{ds} = j_1(s) - \alpha_1(s) I_1(s)  \,,
\end{alignat*}
which relates the intensity of particles emitted at a distance $s$ from the halo center to the emission coefficient $j$ and the extinction coefficient $\alpha$.
We will neglect redshifting throughout this appendix for simplicity.
The emission coefficient is the energy emitted per volume, time, frequency and solid angle by the medium, which for DM processes is 
\begin{alignat*}{1}
    j_1 \equiv  \frac{dE_\omega}{dV \, dt \, d\omega \, d\Omega} = \frac{1}{4\pi} \left( \frac{dE}{dV \, dt} \right)^\mathrm{inj} \frac{d \overline{N}_1}{d \omega} \,,
\end{alignat*}
where $d \overline{N}_1 / d \omega$ is the spectrum of particles directly emitted per DM process. 
We have also made use of the isotropy of the DM process in writing down this expression. 
Outside of the halo, we denote the energy injection rate to be $(dE / dV \, dt)^\mathrm{inj}_0$, the usual injection rate under the homogeneous assumption. 
Inside the halo, however, we have
\begin{alignat*}{1}
    j_1(s < r_\mathrm{vir}) = \frac{\Delta^\beta}{4 \pi} \left( \frac{dE}{dV \, dt} \right)^\mathrm{inj}_0 \frac{d \overline{N}_1}{d\omega} \,,
\end{alignat*}
where $\beta = 1$ for decay and $\beta = 2$ for annihilation. 
The extinction coefficient is related to the optical depth of the directly emitted particle via the usual relation
\begin{alignat*}{1}
    \tau_1(s, s') = \int_s^{s'} dx \, \alpha_1(x) \,,
\end{alignat*}
where $\tau_1(s, s')$ is the optical depth between points $s$ and $s'$, with $s < s'$. 
$\alpha_1^{-1}(s)$ is the local mean free path of the directly emitted particle at $s$. 
Under the homogeneous assumption, the extinction coefficient for the particle takes on some constant value $\alpha_{1,0}$; inside the halo, however, the extinction coefficient is enhanced by a factor $\Delta_1$, the enhancement in density of the targets of the primary in the halo. 
In some cases, e.g.\ ICS of electrons off the CMB, the density of targets presented to the particle is not enhanced by the halo. 
For simplicity, we only consider the dominant process responsible for scattering of the directly emitted particle. 
The optical depth between the origin and some point $s$ away from the origin can therefore be written as
\begin{alignat}{1}
    \tau_1(0, s) = \begin{cases}
        \Delta_1 \alpha_{1,0} s \,, & s < r_\mathrm{vir} \,, \\
        \Delta_1 \alpha_{p,0} r_\mathrm{vir} + \alpha_1(s - r_\mathrm{vir})  \,, & s \geq r_\mathrm{vir}
    \end{cases}
\end{alignat}
under our simple tophat approximation. 

The radiative transfer equation can be integrated radially inward to obtain
\begin{alignat}{1}
    I_1(s) = \int_s^\infty ds'\, j_p(s') e^{-\tau_p(s, s')}  \,.
    \label{eqn:gen_pri_intensity_expr}
\end{alignat}
Under the homogeneous assumption, the intensity of directly emitted particles at all points in space is
\begin{alignat}{1}
    I_{1,0} \equiv \frac{1}{4\pi \alpha_0} \left( \frac{dE}{dV \, dt} \right)^\mathrm{inj}_0 \frac{d \overline{N}}{d\omega}  \,.
\end{alignat}

We now define the quantity $\eta_1 \equiv I_1(s = 0)/I_{1,0}$ as a measure of the impact of the halo on the intensity of directly emitted particles received inside the halo. 
If $\eta_1 \sim 1$, then the intensity of these particles is essentially equal to the intensity expected under the assumption of homogeneity. 
Otherwise, the halo plays a significant role in determining the intensity received. 
Under our tophat assumption, we can perform the integral Eqn.~\eqref{eqn:gen_pri_intensity_expr} to obtain 
\begin{alignat}{1}
    \eta_1 = \frac{\Delta^\beta}{\Delta_1} \left(1 - e^{-\Delta_1 \alpha_{1,0} r_\mathrm{vir}} \right) + e^{-\Delta_1 \alpha_{1,0}  r_\mathrm{vir}} \,.
    \label{eqn:eta_1}
\end{alignat}

Let us consider the following limits: 
\begin{itemize}
    \item $\Delta_1 \alpha_{1,0} r_\mathrm{vir} \ll 1$: since $\Delta_1 \alpha_{1,0}$ is the inverse of the local mean free path of the directly emitted particle in the halo, this corresponds to the limit where the local mean free path in the halo is much longer than the halo itself, i.e.\ the halo is optically thin. 
    We find $\eta(\Delta_1 \alpha_{1,0} r_\mathrm{vir} \ll 1) \approx 1 + \Delta^\beta \alpha_{1,0} r_\mathrm{vir} \approx 1$ for $n = 1$. 
    This demonstrates one of the key takeaways of this appendix: \emph{the center of the halo is illuminated with the same intensity of particles as in the homogeneous IGM, as long as the mean free path of the particles involved is sufficiently long.}
    
    \item $\Delta_1 \alpha_{1,0} r_\mathrm{vir} \gg 1$: the halo is optically thick to directly emitted particles, and we find $\eta(\Delta_1 \alpha_{1,0} r_\mathrm{vir} \gg 1) \approx \Delta^\beta / \Delta_1$. 
    The intensity is enhanced by the $\Delta^\beta$ enhancement in the DM process rate within the halo, but is shielded by the enhanced density of targets surrounding the center of the halo. In particular, \emph{if primaries sourced by DM decays deposit their energy promptly into ionization and heating via any process with a rate that scales as $\Delta$, e.g.\ atomic processes, the intensity within the halo remains at the homogeneous value.} An alternative way to see this result is to consider the total power injected in the halo, and divide by the number of gas particles in the halo, under the assumption that all the injected power is promptly deposited. If the both the injected power and the gas particle density are enhanced by the same factor, the effect cancels out in the ratio.
\end{itemize}
We observe that at least for the case of DM decay, these nominally opposite limits actually lead to the same behavior; this is a first hint that this behavior ($\eta\approx 1$) will be common.

\subsection{Secondaries}
\label{app:secondaries}

In most cases, particles from DM processes cool by scattering into other particles, which can further undergo subsequent interactions. 

We will first consider how to determine the intensity of secondary particles (at step $n+1$) in the limit where \emph{1)} the primary (at step $n = 1$) producing this particle is sourced directly by the DM process, with $j_1(s \leq r_\mathrm{vir}) = \Delta^\beta j_{1,0}$ and $j_1(s > r_\mathrm{vir}) = j_{1,0}$, and \emph{2)} the primary (at step $n$ for any $n$) has constant intensity $I_p = I_{p,0}$. These are two limits that we will frequently encounter in a DM process particle cascade. 
Throughout this section, we use subscript `$p$' to denote both $n = 1$ and more general primaries, and `1' only when discussing $n = 1$ directly emitted primaries.
Our goal is to compute the intensity of secondary particles at step $n+1$.

Under the simplifying assumption that all particles scatter only in the forward direction, the intensity of the primaries $I_p$ acts as a source of emission for the secondaries, i.e.\ along radial paths pointing out from the center of the halo,
\begin{alignat}{1}
    j_s(s) = \int d \omega_p \, \alpha_p(s) \frac{I_p(s)}{\omega_p} \cdot \omega_s \frac{d \overline{N}_s}{d \omega_s}(\omega_p) \,,
    \label{eqn:sec_emission_coeff}
\end{alignat}
where $\alpha_p$ is the extinction coefficient of the primaries, and $d \overline{N}_s / d \omega_s$ is the spectrum of secondaries produced per primary scattering event. Note that $\alpha_p$, $I_p$ and $d \overline{N}_s / d \omega_s$ depend on $\omega_p$, while $j_s$ depends also on $\omega_s$.
\footnote{An additional integral over $\Omega_p$ would appear without the assumption of forward scattering.}
Integrating the radiation transfer equation for secondaries gives
\begin{alignat}{1}
    I_s(s) = \int d \omega_p \, \frac{\omega_s}{\omega_p} \frac{d \overline{N}_s}{d \omega_s}(\omega_p) \int_s^\infty ds' \, \alpha_p(s') I_p(s') e^{-\tau_s(s, s')}
    \label{eqn:gen_sec_intensity_expr}
\end{alignat}

For a homogeneous medium, we have
\begin{alignat*}{1}
    I_{s,0} = \int d\omega_p \, \frac{\omega_s}{\omega_p} \frac{d \overline{N}_s}{d\omega_s} (\omega_p) I_{p,0} \frac{\alpha_{p,0}}{\alpha_{s,0}} \,,
\end{alignat*}
where $I_{p,0}$ is the homogeneous primary intensity.

First, let us consider the case where the primary intensity is given by its homogeneous value, i.e.\ $I_p = I_{p,0}$. From Eqn.~\eqref{eqn:gen_sec_intensity_expr}, the intensity of the secondaries is given by
\begin{alignat*}{2}
    I_s(I_p = I_{p,0}) &=&& \int d \omega_p \, \frac{\omega_s}{\omega_p} \frac{d \overline{N}_s}{d\omega_s} (\omega_p) I_{p,0} \int_s^\infty ds' \, \alpha_{p}(s') e^{-\tau_s(s,s')} \\
    &=&& \int d \omega_p \, \frac{\omega_s}{\omega_p}  \frac{d \overline{N}_s}{d\omega_s} (\omega_p) I_{p,0} \frac{\alpha_{p,0}}{\alpha_{s,0}} \left[ e^{-\Delta_s \alpha_{s,0} r_\mathrm{vir}} + \frac{\Delta_p}{\Delta_s} \left(1 - e^{-\Delta_s \alpha_{s,0} r_\mathrm{vir}} \right) \right] \,,
\end{alignat*}
Yet again, under the assumption of a monochromatic primary spectrum, we obtain
\begin{alignat}{1}
    \eta_s(I_p = I_{p,0}) = e^{-\Delta_s \alpha_{s,0} r_\mathrm{vir}} + \frac{\Delta_p}{\Delta_s} \left(1 - e^{-\Delta_s \alpha_{s,0} r_\mathrm{vir}} \right) \,.
\end{alignat}
Note that in general $\Delta_p \neq \Delta_s$; for example, primary electrons that undergo ICS into photoionizing photons have $\Delta_p = 1$, since ICS occurs off CMB photons which are not enhanced within a halo, but $\Delta_s = \Delta$, since photoionization occurs off neutral atoms which are enhanced in a halo. 
The limits of interest are: 
\begin{itemize}
    \item $\Delta_s \alpha_{s,0} r_\mathrm{vir} \ll 1$: this corresponds to the limit where  the halo is optically thin to secondaries. 
    We find $\eta_s(I_p = I_{p,0}) \approx 1 + \Delta_p \alpha_{s,0} r_\mathrm{vir}$, i.e. we get a potential enhancement if the primary path length is sufficiently short, due to the fact that we become dominated by primaries scattering inside the halo (and not in the homogeneous IGM), which comes with a $\Delta_p$ enhancement; 
    \item $\Delta_s \alpha_{s,0} r_\mathrm{vir} \gg 1$: the halo is optically thick to secondaries, leading to $\eta_s(I_p = I_{p,0}) \approx \Delta_p / \Delta_s$. 
    In this limit, the dominant contribution to the intensity comes from primaries scattering close to the center of the halo. 
    The scattering rate is therefore enhanced by $\Delta_p$, but suppressed by $\Delta_s$ due to the shielding that the halo overdensity provides to the secondaries. 
\end{itemize}
Another way to think about the $\Delta_p$ halo enhancement is to consider primaries with short path length. 
In this case, the assumption of a homogeneous intensity for the primaries implies that the primaries are more efficiently {\it injected} in regions of high $\Delta_p$ (where they are also more efficiently depleted). 
Since the primaries convert promptly to secondaries, the production of secondaries is similarly enhanced.

Next, we examine the case where the primaries are sourced by DM processes, i.e.\ $n = 1$ and $j_1(s \leq r_\mathrm{vir}) = \Delta^\beta j_{1,0}$, and $j_1(s > r_\mathrm{vir}) = j_{1,0}$, where $j_{1,0}$ is some constant emission coefficient. 
Substituting Eqn.~\eqref{eqn:gen_pri_intensity_expr} into Eqn.~\eqref{eqn:gen_sec_intensity_expr}, we find
\begin{alignat*}{1}
    I_2(s) = \int d \omega_1 \, \frac{\omega_2}{\omega_1} \frac{d \overline{N}_2}{d \omega_2} (\omega_1) \int_s^\infty ds' \alpha_1(s') e^{-\tau_2(s, s')} \int_{s'}^\infty ds'' j_1(s'') e^{-\tau_1(s', s'')} \,.
\end{alignat*}
The structure of this result can be understood as follows: the intensity of secondaries at $s$ is given by the sum intensity of primaries in shells of width $ds'$, multiplied by $\alpha_1$ to obtain the intensity into secondaries, and finally multiplied by the survival probability of secondaries traveling from $s'$ to $s$. 
Once again, for a monochromatic primary spectrum, we can define $\eta_2 \equiv I_2(s = 0) / I_{2,0}$, and using the fact that $I_{1,0} = j_{1,0} / \alpha_{1,0}$, we obtain
\begin{multline}
    \eta_2 = \alpha_{2,0} \left[ \int_0^{r_\mathrm{vir}} ds\, \alpha_1(s) e^{-\tau_2(0,s)} \left( \Delta^\beta \int_s^{r_\mathrm{vir}} ds' e^{-\tau_1(s, s')} + \int_{r_\mathrm{vir}}^\infty ds' \, e^{-\tau_1(s, s')} \right) \right. \\
    \left. + \int_{r_\mathrm{vir}}^\infty ds \, \alpha_1(s) e^{-\tau_2(0,s)} \int_s^\infty ds' \, e^{-\tau_1(s, s')} \right] \,.
\end{multline}
This can be evaluated with our tophat model, giving 
\begin{alignat*}{1}
    \eta_2 = e^{- \Delta_2 \alpha_{2,0} r_\mathrm{vir}} + \frac{\Delta^\beta}{\Delta_2} \left(1 - e^{-\alpha_{2,0} \Delta_2 r_\mathrm{vir}} \right) - \alpha_{2,0} (\Delta^\beta - \Delta_1) \frac{e^{- \Delta_2 \alpha_{2,0} r_\mathrm{vir}} - e^{-\Delta_1 \alpha_{1,0} r_\mathrm{vir}}}{\Delta_1 \alpha_{1,0} - \Delta_2 \alpha_{2,0}}  \,.
\end{alignat*}
Let us consider the following limits: 
\begin{itemize}
    \item $\alpha_{2,0} \to \infty$: this corresponds to a secondary with extremely short path length. 
    In this limit, the center of the halo only receives secondaries that are produced near the center.
    We find $\eta_2 = (\Delta_1 / \Delta_2) \eta_1$, i.e.\ we get an enhancement from the primary intensity itself being larger, and the fact that there can be more targets for primaries to scatter off in the halo; on the other hand, we receive a suppression due to screening of the secondaries by the dense halo; 
    \item $\alpha_{2,0} \to 0$: in this limit, secondaries have a very long path length, and the halo is optically thin. 
    One finds that also assuming $\alpha_{1,0} \to 0$, i.e.\ the halo is optically thin also to primaries, we obtain $\eta_2 = 1 + \Delta_1 \alpha_{2,0} r_\mathrm{vir}$, which is the same result as $I_s(I_p = I_{p,0})$, since $\eta_p \to 1$ as $\alpha_{p,0} \to 0$. 
    On the other hand, for $\alpha_{1,0} \to \infty$, both primaries and secondaries have a short path length. 
    We find $\eta_2 \approx 1 + \Delta^\beta \alpha_{2,0} r_\mathrm{vir}$, which is similar to the result for $\eta_1$---if the directly emitted particles have a sufficiently short path length, then $n = 2$ particles can be treated as the directly emitted particles instead. 
\end{itemize}

\subsection{Tertiaries}
\label{app:tertiaries}

Intensities of tertiaries and subsequent particles can be calculated iteratively, with increasingly more complicated integrals to perform. For our purposes, as with secondary particles, we only need to determine the intensity of tertiaries under the two assumptions of \emph{1)} the primary producing the tertiary is sourced directly by the DM process, with $j_1(s \leq r_\mathrm{vir}) = \Delta^\beta j_{1,0}$ and $j_1(s > r_\mathrm{vir}) = j_{1,0}$, and \emph{2)} the primary has constant steady-state intensity $I_p = I_{p,0}$. 

Following the same procedure as before, we can recursively obtain the intensity of tertiaries as
\begin{alignat*}{1}
    I_t(s) = \int d \omega_s \, \frac{\omega_t}{\omega_s} \frac{d \overline{N}_t}{d \omega_t} (\omega_s) \int d \omega_p \, \frac{\omega_s}{\omega_p} \frac{d \overline{N}_s}{d \omega_s} (\omega_p) \int_s^\infty ds' \, \alpha_s (s') e^{-\tau_t(s, s')}  \int_{s'}^\infty ds'' \, \alpha_p (s'') e^{-\tau_s(s', s'')} I_p(s'') \,.
\end{alignat*}
Once again, we evaluate the expected intensity in the homogeneous limit, which is
\begin{alignat*}{1}
    I_{t,0} = \int d \omega_s \, \frac{\omega_t}{\omega_s} \frac{d \overline{N}_t}{d \omega_t} (\omega_s) \int d \omega_p \, \frac{\omega_s}{\omega_p} \frac{d \overline{N}_s}{d \omega_s} (\omega_p) \frac{\alpha_{p,0}}{\alpha_{t,0}} I_{p,0} \,.
\end{alignat*}
For $n = 1$ and $j_1(s \leq r_\mathrm{vir}) = \Delta^\beta j_{1,0}$ and $j_1(s > r_\mathrm{vir}) = j_{1,0}$, assuming all cascades are monochromatic, we can again define $\eta_3 = I_3(s = 0) / I_{3,0}$ and perform the integrals over the various domains. This ultimately gives
\begin{multline}
    \eta_3 = e^{- \Delta_3 \alpha_{3,0} r_\mathrm{vir}} + \frac{\Delta^\beta}{\Delta_3} \left(1 - e^{-\Delta_3 \alpha_{3,0} r_\mathrm{vir}} \right) - \frac{\Delta_2(\Delta^\beta - \Delta_1) \alpha_{2,0} \alpha_{3,0}}{\Delta_1 \alpha_{1,0} - \Delta_1 \alpha_{1,0}} \frac{e^{-\Delta_3 \alpha_{3,0} r_\mathrm{vir}} - e^{- \Delta_1 \alpha_{1,0} r_\mathrm{vir}}}{\Delta_3 \alpha_{3,0} - \Delta_1 \alpha_{1,0}} \\ + \frac{\Delta_2 ( \Delta_1 - \Delta_2) \alpha_{2,0} \alpha_{3,0} - \Delta_1 ( \Delta^\beta - \Delta_2) \alpha_{1,0} \alpha_{3,0}}{\Delta_1 \alpha_{1,0} - \Delta_2 \alpha_{2,0}} \frac{e^{-\Delta_3 \alpha_{3,0} r_\mathrm{vir}} - e^{- \Delta_2 \alpha_{2,0} r_\mathrm{vir}}}{\Delta_2 \alpha_{2,0} - \Delta_3 \alpha_{3,0}} \,.
\end{multline}

For any $n$, in the limit where $I_p = I_{p,0}$, we can likewise define $\eta_t(I_p = I_{p,0}) \equiv I_t(s = 0) / I_{t,0}$, which is given by
\begin{multline}
    \eta_t(I_p = I_{p,0}) = e^{- \Delta_t \alpha_{t,0} r_\mathrm{vir}} + \frac{\alpha_{s,0} \Delta_p \Delta_s}{\Delta_t} \frac{1 - e^{- \Delta_t \alpha_{t,0} r_\mathrm{vir}}}{\Delta_s \alpha_{s,0} - \Delta_t \alpha_{t,0}} \\
    - \alpha_{t,0} \Delta_p \frac{1 - e^{-\Delta_s \alpha_{s,0} r_\mathrm{vir}}}{\Delta_s \alpha_{s,0} - \Delta_t \alpha_{t,0}} - \alpha_{t,0} \Delta_s \frac{e^{- \Delta_s \alpha_{s,0} r_\mathrm{vir}} - e^{- \Delta_t \alpha_{t,0} r_\mathrm{vir}}}{\Delta_s \alpha_{s,0} - \Delta_t \alpha_{t,0}} \,.
\end{multline}
This result is finite as $\Delta_s \alpha_{s,0} \to \Delta_t \alpha_{t,0}$. 
In the limit where $\alpha_{s,0} \to \infty$, i.e.\ the secondaries have an extremely short path length, we find $\eta_t \approx e^{-\Delta_t \alpha_{t,0} r_\mathrm{vir}} + (\Delta_p / \Delta_t) (1 - e^{- \Delta_t \alpha_{t,0} r_\mathrm{vir}})$. 
Comparing this with the result for the intensity of a daughter particle originating from a mother particle with homogeneous intensity, this result shows that we can simply skip the secondaries step in the cascade if the path length is sufficiently short, which matches our intuitive expectations.

\subsection{Effect of the Halo}
\label{app:effect_of_the_halo}

%
\begin{figure}
    \centering
    \includegraphics[scale=0.5]{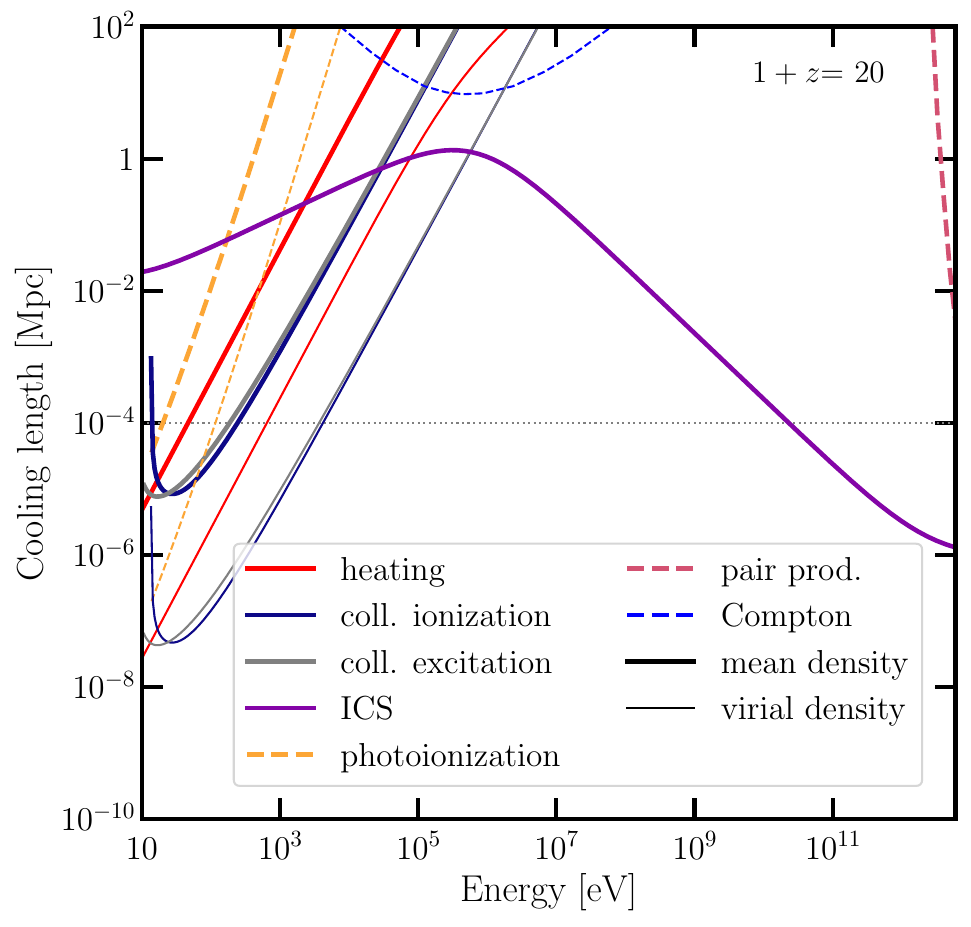}
    \caption{
    Path lengths for particles to lose a significant amount of energy by various processes.
    For cooling lengths that depend on the gas density, we show results for the mean density in solid lines and virialized halo density in dashed lines.
    The horizontal dotted line marks 0.1 kpc, which is the approximate size of $r_\mathrm{vir}$ for halos of mass $10^6 M_\odot$ virializing at $1+z=20$.
    }
    \label{fig:cooling_length}
\end{figure}
\begin{figure}
    \centering
    \includegraphics[scale=0.5]{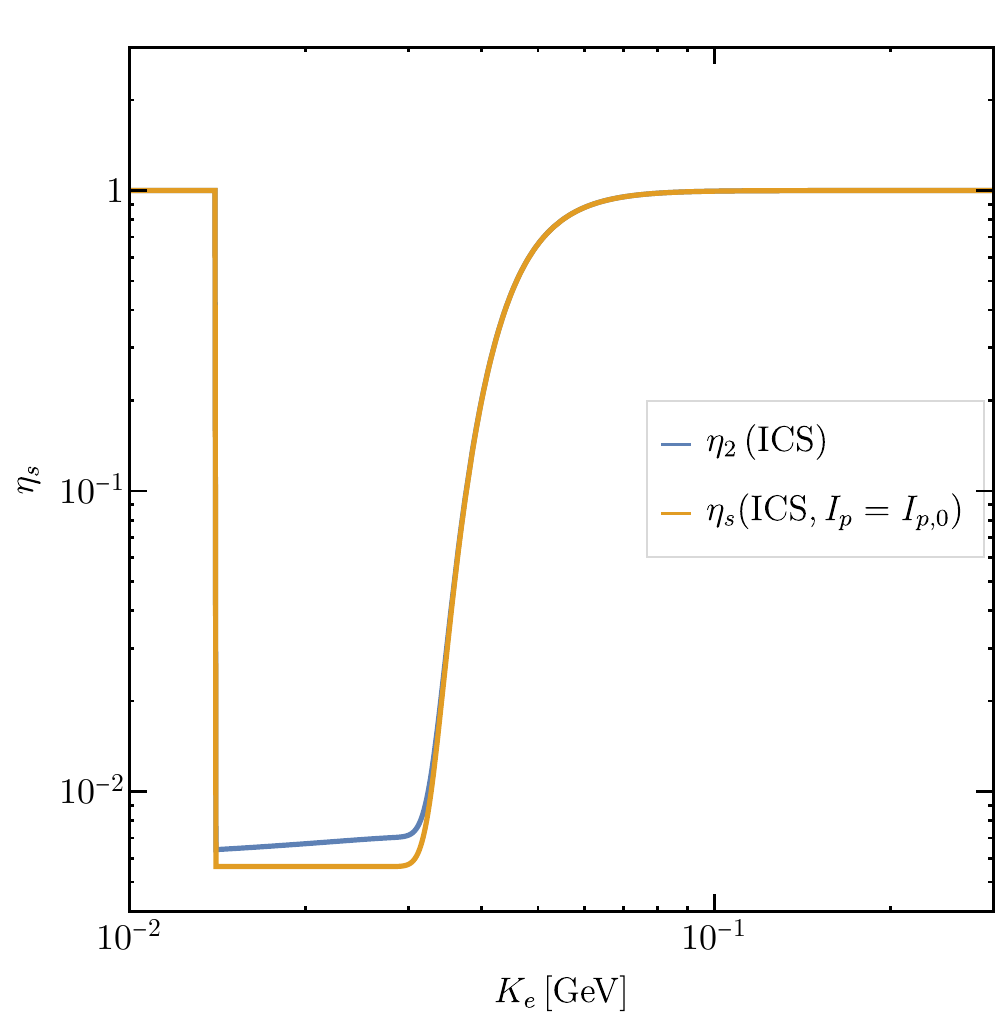}
    \hspace{5mm}
    \includegraphics[scale=0.5]{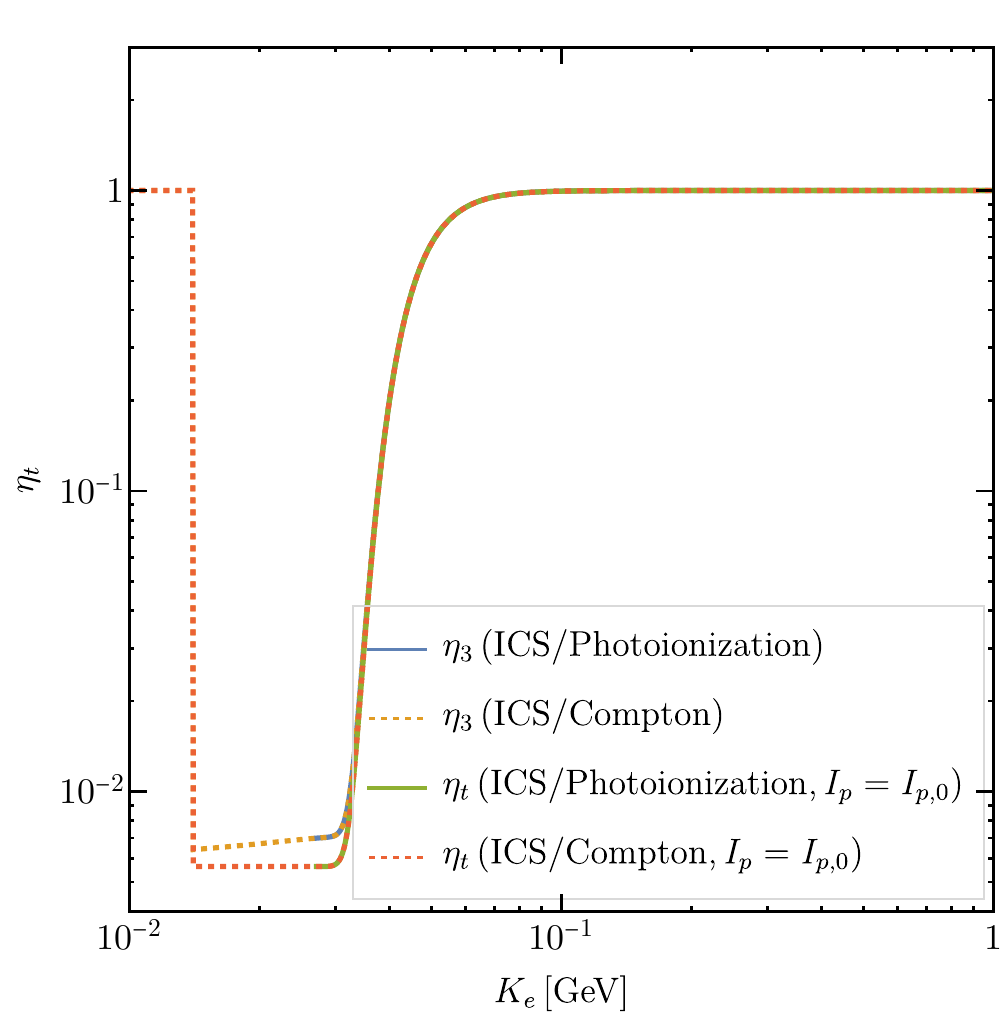} \\
    \vspace{5mm}
    \hspace{1mm}\includegraphics[scale=0.5]{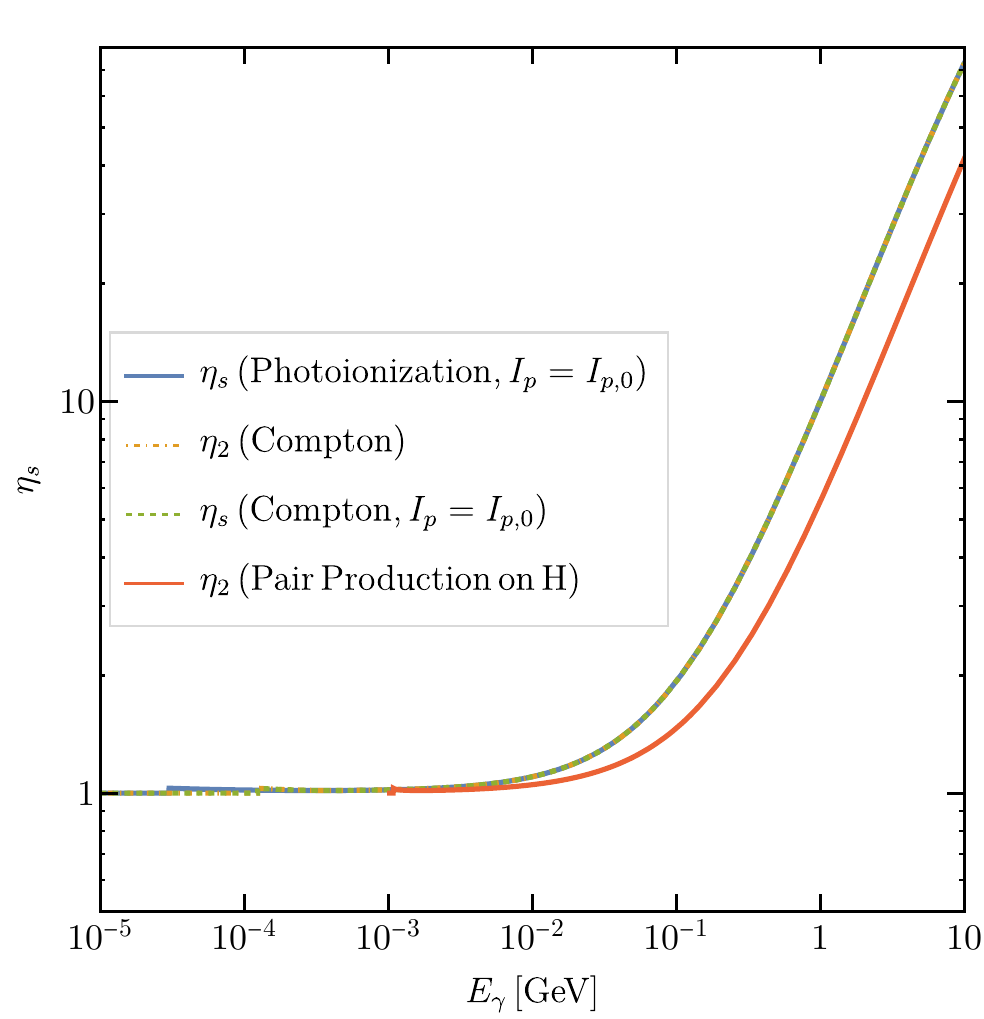}
    \hspace{4mm}
    \includegraphics[scale=0.5]{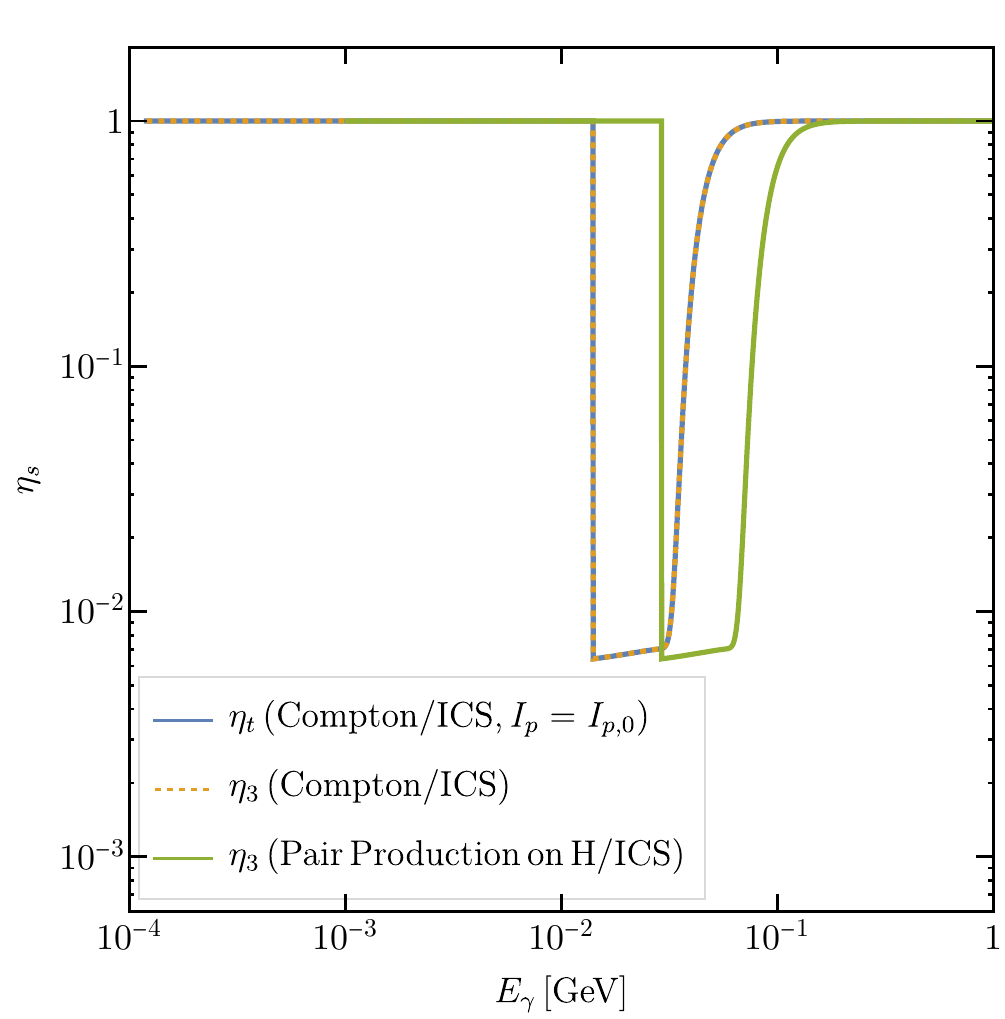} 
    \caption{
    The halo enhancement factor calculated for DM decay, under different assumptions as a function of primary particle energy, including \emph{(top left)} $\eta_2$ for secondaries produced by the ICS of electrons sourced by DM decay (blue), or $\eta_s(I_p = I_{p,0})$ for secondaries by primaries with the homogeneous, steady-state intensity (orange); \emph{(top right)} $\eta_3$ for tertiaries produced by ICS of electrons sourced by DM decay followed by photoionization of the secondary photons (blue), or Compton scattering (dashed orange), as well as $\eta_t(I_p = I_{p,0})$ for tertiaries produced by ICS of electrons with the homogeneous, steady-state intensity followed by photoionization (green), or Compton scattering (dashed red); \emph{(bottom left)} $\eta_s(I_p = I_{p,0})$ for secondaries produced by photoionization with the homogeneous, steady-state primaries (blue), $\eta_2$ for Compton scattering of photons sourced by DM decay (orange, dot-dashed), $\eta_s(I_p = I_{p,0})$ for Compton scattering of primary photons with the homogeneous, steady-state intensity (green, dashed), and $\eta_2$ for pair production of photons sourced by DM decay (red), and \emph{(bottom right)} $\eta_t(I_p = I_{p,0})$ for primary photons with the homogeneous, steady-state intensity undergoing Compton scattering, producing electrons that undergo ICS (blue), and $\eta_3$ for photons from DM decays undergoing either Compton scattering (orange, dashed) or pair production on neutral hydrogen (green), followed by ICS of the secondary electrons.
    }
    \label{fig:eta}
\end{figure}

We are now ready to examine the effect of the enhanced density in the halo on energy deposition with in the halo. 
We focus only on decaying DM in this section, since this is the process that has largest impact on star formation given existing experimental constraints.
First, we note that the energy deposited per volume per time $(dE / dV \, dt)_\mathrm{dep} \propto \int d \omega_f \alpha_f I_f$, where $f$ labels the last step in the cascade in the halo. 
Therefore, if the halo receives the same intensity as in the homogeneous limit, the energy deposited per volume per time increases by a factor of $\Delta_f$ relative to the homogeneous limit, where $\Delta_f$ is the overdensity of targets for the last step of the cascade, since $\alpha_f = \Delta_f \alpha_{f,0}$, where $\alpha_{f,0}$ is the extinction coefficient of the last step in the cascade in the homogeneous limit. 
Equivalently, the energy deposited \emph{per baryon} per time is enhanced by $\Delta_f / \Delta$, which is 1 for the final step in all cascades, since ionization and heating occurs through scattering with atoms or free electrons. In other words, receiving a homogeneous intensity at the center of the halo typically implies the same energy deposited per baryon per time as in the homogeneous limit. 

To determine what effect the halo has on the intensity of the particles in the final step, we estimate $\eta_f = I_f(s = 0) / I_{f,0}$ by obtaining $\eta$ for some of the intermediate steps, using either our ability to calculate $\eta_2$ for secondaries and $\eta_3$ for tertiaries from particles emitted directly from the DM process, or by making use of $\eta_n \approx 1$ for some intermediate step in the cascade, allowing us to calculate $\eta_{n+1}$ or $\eta_{n+2}$.
We present results for $1+z = 20$, which is the more experimentally accessible redshift, and for $M_\text{halo} = 10^6 M_\odot$, which is close to the critical halo mass at that redshift including DM effects (see Fig.~\ref{fig:critical_collapse_Mhalo}). 
Our results are relatively insensitive to halo masses within an order of magnitude of $M_\text{halo} = 10^6 M_\odot$, since the relevant parameter is $r_\text{vir} \propto M_\text{halo}^{1/3}$.

The extinction coefficients for the cooling processes in the cascade of $e^+e^-$ pairs and photons are given by the inverse of the energy loss path length, i.e. $\alpha = (1/v) (d \log E/ dt)$. 
The energy loss path lengths $\alpha^{-1}$ for relevant processes are shown in Fig.~\ref{fig:cooling_length} for $1+z = 20$ and $1+z = 100$, and are discussed and derived in detail in Refs.~\cite{Slatyer:2009yq,Liu:2023fgu}.

Tables~\ref{tab:cascade_elec} and~\ref{tab:cascade_phot} shows the approximate monochromatic cascade produced by an electron or positron, and a photon respectively, as a function of initial energy. 
Each row corresponds to a range of energies over which the cascade goes through the same processes and intermediate states, differing only in energy of the intermediate states. 
The dominant process taking particles from step $n$ to step $n+1$ in the cascade are shown under $n \to n+1$. 
The final step by which energy is deposited directly into ionization, heating or low-energy photons is shown in bold.

In order to obtain $\eta_f$ for all relevant final states, we need the following results, summarized in Fig.~\ref{fig:eta}: 
\begin{itemize}
    \item $\eta_2$ and $\eta_s(I_p = I_{p,0})$ for secondary photons produced by primary electrons undergoing ICS, assumed to either be sourced by DM decay for $\eta_2$, or to have the homogeneous, steady-state intensity, $I_p = I_{p,0}$ for $\eta_s(I_p = I_{p,0})$ (Fig.~\ref{fig:eta} top left); 
    \item $\eta_3$ and $\eta_t(I_p = I_{p,0})$ for tertiary electrons of primary electrons undergoing ICS, producing secondary photons, which subsequently produce tertiary electrons through either photoionization or Compton scattering (Fig.~\ref{fig:eta} top right); 
    \item $\eta_s(I_p = I_{p,0})$ for secondary electrons produced by primary photons undergoing photoionization or Compton scattering, and $\eta_2$ for secondary electrons produced by Compton scattering or pair production on neutral hydrogen of primary photons sourced by DM decay (Fig.~\ref{fig:eta} bottom left), and
    \item $\eta_t(I_p = I_{p,0})$ for tertiary photons produced by the ICS of secondary electrons, which is in turn produced by  primary photons undergoing Compton scattering, as well as $\eta_3$ for tertiary photons produced by the ICS of secondary electrons, coming from primary photons sourced by DM decay undergoing pair production (Fig.~\ref{fig:eta} bottom right). 
\end{itemize}

\renewcommand{\arraystretch}{1.8}

\begin{table*}[!t]
\footnotesize
\begin{center}
\begin{tabular}{C{2cm}|C{1.6cm}C{2cm}C{2.2cm}C{2.2cm}C{2cm}C{2cm}C{2cm}}
\toprule
\textbf{Electron Kinetic Energy} & $1 \to 2$  & $2$ & $2 \to 3$ & $3$ & $3 \to 4$ & $4$ & $4 \to 5$ \\
\Xhline{1\arrayrulewidth}
\SIrange{1}{14}{\mega\eV} & ICS & $< \SI{13.6}{\eV}$ $\gamma$ \tiny{\textcolor{red}{(ICS,2)}} & \bf{No Ionization /Heating} & -- & --  &-- &--\\
\SIrange{14}{60}{\mega\eV} & ICS & \SIrange{13.6}{230}{\eV} $\gamma$ & Photoionization & \SIrange{0}{215}{\eV} $e^-$ \tiny{\textcolor{red}{(ICS,3)}} & \textcolor{blue}{$e^-$ \bf{Atomic}} &-- &--\\
\SIrange{60}{350}{\mega\eV} & ICS & \SIrange{0.23}{8}{\kilo\eV} $\gamma$ & Photoionization & \SIrange{0.215}{8}{\kilo\eV} $e^-$ \tiny{\textcolor{red}{(ICS,3)}} & $e^-$ \bf{Atomic} &-- &-- \\
\SIrange{0.35}{1.37}{\giga\eV} & ICS & \SIrange{8}{120}{\kilo\eV} $\gamma$ & Compton & \SIrange{0.125}{30}{\kilo\eV} $e^-$ \tiny{\textcolor{red}{(ICS,3)}} & $e^-$ \bf{Atomic} &-- &--\\
\SIrange{1.37}{10}{\giga\eV} & ICS & \SIrange{0.12}{6.4}{\mega\eV} $\gamma$ \tiny{\textcolor{red}{(ICS,2)}} & Compton & \SIrange{0.03}{1.8}{\mega\eV} $e^-$ & ICS & $<$ \SI{13.6}{\eV} $\gamma$ \tiny{\textcolor{red}{($\gamma$,t)}}& \bf{No Ionization /Heating} \\
\botrule
\end{tabular}
\end{center}
\caption{Dominant cascade for primary electrons/positrons with energies between \SI{1}{\mega\eV} and \SI{10}{\giga\eV}. The steps in the cascade and the dominant process between steps are shown in each column. The channel through which energy is deposited in the last step of the cascade is written in bold. The relevant method to calculate $\eta$ and determine the intensity at each step in the cascade is shown in red, with `ICS' and `$\gamma$' referring to the electron ICS and photon $\eta$ results, `2' and `3' indicating the use of $\eta_2$ and $\eta_3$, while `$s$' and `$t$' indicating the use of $\eta_s(I_p = I_{p,0})$ and $\eta_t(I_p = I_{p,0})$ respectively. Final intensities which are suppressed relative to the homogeneous intensity are shown in blue.}
\label{tab:cascade_elec}
\end{table*}    

With this information, we can explain the estimate for $\eta_f$ for each type of cascade as a function of the primary particle energy.

\vspace{2mm}

\underline{$\chi \to e^+e^-$}:

\begin{itemize}
    \item \SIrange{1}{14}{\mega\eV}: $e^+e^-$ pairs mainly undergo ICS into sub-\SI{13.6}{\eV} photons with long path lengths,\footnote{\SIrange{10.2}{13.6}{\eV} photons scatter rapidly, but elastically, which we consider as having a long path length.} giving $\eta_f \approx 1$ (see Fig.~\ref{fig:eta} top left). 
    \item \SIrange{14}{60}{\mega\eV}: The primary particles undergo ICS into photons just above the ionization threshold of hydrogen, and have very short path lengths. These photons then produce low-energy electrons, which again undergo collisional ionization with short path lengths. By considering $\eta_3$ for ICS (Fig.~\ref{fig:eta} top right), we find that $\eta_f \approx 1/\Delta$;
    \item \SIrange{60}{350}{\mega\eV}: This energy range results in the same cascade as the previous range, but the photoionizing photons are of sufficiently high energy that their path lengths are much longer than the virial radius of the halo. We find $\eta_3 \approx 1$ for ICS into photons, which then photoionize neutral atoms to produce the final, tertiary final low-energy electrons (Fig.~\ref{fig:eta} top right);
    \item \SIrange{0.35}{1.37}{\giga\eV}: The photons from ICS now cool mainly by Compton scattering, and not photoionization, but $\eta_f \approx 1$ remains true in this regime, again by considering $\eta_3$ of for primary electrons undergoing ICS, producing photons that Compton cool (Fig.~\ref{fig:eta} top right);
    \item \SIrange{1.37}{10}{\giga\eV}: For this energy range, the cascade becomes longer, with the secondary photons Compton scattering into electrons that predominantly undergo ICS instead of atomic processes. We use the fact that $\eta_2 \approx 1$ for ICS into photons to show that these secondary photons have the homogeneous, steady-state intensity (Fig.~\ref{fig:eta} top left), and then use $\eta_t(I_p = I_{p,0})$ for photons that undergo Compton cooling to electrons that cool mainly via ICS to show that ultimately, $\eta_f \approx 1$ (Fig.~\ref{fig:eta} bottom right). 
\end{itemize}

\begin{table*}[!t]
\footnotesize
\begin{center}
\begin{tabular}{C{1.7cm}|C{1.5cm}C{2.2cm}C{1.2cm}C{1.9cm}C{1.6cm}C{2.2cm}C{1.2cm}C{1.6cm}C{1.6cm}}
\toprule
\textbf{Photon Energy} & $1 \to 2$  & $2$ & $2 \to 3$ & $3$ & $3 \to 4$ & $4$ & $4 \to 5$ & $5$ & $5 \to 6$ \\
\Xhline{1\arrayrulewidth}
\SIrange{10}{120}{\kilo\eV} & Compton & \SIrange{0.125}{30}{\kilo\eV} $e^-$ \tiny{\textcolor{red}{($\gamma$,2)}} & $e^-$ \bf{Atomic} & -- & --  &-- &--\\
\SIrange{0.12}{14}{\mega\eV} & Compton & \SIrange{0.03}{14}{\mega\eV} $e^-$ & ICS & $<$\SI{13.6}{\eV} $\gamma$ \tiny{\textcolor{red}{($\gamma$,3)}} & \bf{No Ionization /Heating} &-- &-- \\
\SIrange{14}{60}{\mega\eV} & Compton & \SIrange{14}{60}{\mega\eV} $e^-$ \tiny{\textcolor{red}{($\gamma$,2)}} & ICS & \SIrange{13.6}{230}{\eV} $\gamma$ & Photo- ionization & \SIrange{0}{215}{\eV} $e^-$ \tiny{\textcolor{red}{(ICS,t)}} & \textcolor{blue}{$e^-$ \bf{Atomic}} \\
\SIrange{60}{120}{\mega\eV} & H Pair Production & \SIrange{30}{60}{\mega\eV} $e^-$ \tiny{\textcolor{red}{($\gamma$,2)}} & ICS & \SIrange{58}{230}{\eV} $\gamma$ & Photo- ionization & \SIrange{43}{215}{\eV} $e^-$ \tiny{\textcolor{red}{(ICS,t)}} & \textcolor{blue}{$e^-$ \bf{Atomic}} \\
\SIrange{120}{700}{\mega\eV} & H Pair Production & \SIrange{60}{350}{\mega\eV} $e^-$ & ICS & \SIrange{0.145}{8}{\kilo\eV} $\gamma$ \tiny{\textcolor{red}{($\gamma$,3)}} & Photo- ionization & \SIrange{0.13}{8}{\kilo\eV} $e^-$ \tiny{\textcolor{red}{($\gamma$,s)}} & $e^-$ \bf{Atomic} \\
\SIrange{0.7}{2.8}{\giga\eV} & H Pair Production & \SIrange{0.35}{1.4}{\giga\eV} $e^-$ & ICS & \SIrange{8}{120}{\kilo\eV} $\gamma$ \tiny{\textcolor{red}{($\gamma$,3)}} & Compton & \SIrange{0.125}{30}{\kilo\eV} $e^-$ \tiny{\textcolor{red}{($\gamma$,s)}} & $e^-$ \bf{Atomic} \\
\SIrange{2.8}{10}{\giga\eV} & H Pair Production & \SIrange{1.4}{5}{\giga\eV} $e^-$ & ICS & \SIrange{120}{450}{\kilo\eV} $\gamma$ \tiny{\textcolor{red}{($\gamma$,3)}} & Compton & \SIrange{30}{400}{\kilo\eV} $e^-$ & ICS & $<$ \SI{10.2}{\eV} $\gamma$ \tiny{\textcolor{red}{($\gamma$,t)}} & \bf{No Ionization /Heating}\\
\botrule
\end{tabular}
\end{center}
\caption{Dominant cascade for primary photons with energies between \SI{1}{\mega\eV} and \SI{10}{\giga\eV}. The steps in the cascade and the dominant process between steps are shown in each column. The channel through which energy is deposited in the last step of the cascade is written in bold. The relevant method to calculate $\eta$ and determine the intensity at each step in the cascade is shown in red, with `ICS' and `$\gamma$' referring to electron ICS and photon $\eta$ results, `2' and `3' indicating the use of $\eta_2$ and $\eta_3$, and `$s$' and `$t$' indicating the use of $\eta_s(I_p = I_{p,0})$ and $\eta_t(I_p = I_{p,0})$ respectively. Final intensities which are suppressed relative to the homogeneous intensity are shown in blue.}
\label{tab:cascade_phot}
\end{table*}    

Note that our benchmark Models $\bullet$\, and $\star$\, directly emit \SI{92}{\mega\eV} electrons and positrons, and therefore have $\eta_f \approx 1$.

\vspace{2mm}

\underline{$\chi \to \gamma \gamma$}:

\begin{itemize}
    \item \SIrange{10}{120}{\kilo\eV}: The primary photons Compton scatter to produce low-energy electrons, which also rapidly lose all their energy through atomic processes. We find $\eta_f \approx 1$ by calculating $\eta_2$ of primary photons which Compton scatter (Fig.~\ref{fig:eta} bottom left), finding that the enhancement of $\Delta$ due to enhanced production of primaries in the halo is exactly canceled by the shielding of secondaries due to the same enhancement in density; 
    \item \SIrange{0.12}{14}{\mega\eV}: Primary photons Compton scatter once again, but produce electrons that predominantly cool via ICS into sub-\SI{13.6}{\eV} photons. $\eta_f \approx 1$ by considering $\eta_t$ for photons that Compton scatter (Fig.~\ref{fig:eta} bottom right); 
    \item \SIrange{14}{60}{\mega\eV}: Once again, we have Compton scattering into electrons ($\eta_2 \approx 1$ from Fig.~\ref{fig:eta} bottom left, so these electrons have homogeneous intensity) that ICS into photons, that are now just above the hydrogen ionization threshold; they photoionize hydrogen, producing low-energy electrons with a short path length. We find $\eta_f \approx 1 / \Delta$ by looking at $\eta_t(I_p = I_{p,0})$ for electrons undergoing ICS and subsequent photoionization (Fig.~\ref{fig:eta} bottom right); 
    \item \SIrange{60}{120}{\mega\eV}: Similar to the above, except the first step is pair production on neutral hydrogen of the primary photons. $\eta_f \approx 1 / \Delta$; 
    \item \SIrange{120}{700}{\mega\eV}: Similar to the above, except that the photoionizing photons in the 3rd step of the cascade have a long path length. Because of this, $\eta_f \approx 1$. We can deduce this by starting with $\eta_3$ for photons to show that the intensity of tertiary photons is the homogeneous intensity (Fig.~\ref{fig:eta} bottom right), and then looking at $\eta_s(I_p = I_{p,0})$ for photoionizing photons (Fig.~\ref{fig:eta} bottom left); 
    \item \SIrange{0.7}{2.8}{\giga\eV}: Similar to the above, except that the tertiary photons cool by Compton scattering, which also has a long path length, leading to $\eta_f \approx 1$ as before; 
    \item \SIrange{2.8}{10}{\giga\eV}: Finally, in this energy range, photons undergo pair production on neutral hydrogen, and follows the same cascade as \SIrange{1.4}{5}{\giga\eV} electrons do as described above. We find that $\eta_f \approx 1$ by using $\eta_3 \approx 1$ of Compton scattering photons going into electrons that ICS, and then applying $\eta_t(I_p = I_{p,0})$ on these tertiary photons (which produces electrons that again ICS into $<\SI{13.6}{\eV}$ photons) (see Fig.~\ref{fig:eta} bottom right). 
\end{itemize}

In summary, we find that $\eta_f \approx 1$ across most of the parameter space of interest for DM decays into $e^+e^-$ and $\gamma \gamma$, except for $\SI{28}{\mega\eV} \lesssim m_\chi \lesssim \SI{120}{\mega\eV}$ for $\chi \to e^+e^-$, and $\SI{28}{\mega\eV} \lesssim m_\chi \lesssim \SI{240}{\mega\eV}$ for $\chi \to \gamma \gamma$, where $\eta_f \approx 1 / \Delta$. 
This implies a reduction of $1/\Delta$ with respect to the homogeneous $f_c(z)$ calculated in \texttt{DarkHistory} for these narrow ranges of parameter space. 
One way to summarize the physical origin of this reduction is that in all these cases, near the end of the cascade, homogenized electrons with a relatively long path length undergo ICS to produce photons that promptly deposit their energy. 
The production of these photons is not enhanced by the presence of the halo, and their energy must be divided between the larger density of halo particles, leading to the observed suppression in deposited power per particle. 
\emph{In all other regimes, we have argued that $\eta_f \approx 1$ implies $f_c / n_\mathrm{H}$ in both the IGM and the halo are approximately equal, justifying our approach in the main body of the paper.}

Although the same formalism applies to annihilation, we do not perform the same in-depth analysis, since the potential effect of DM annihilation on the formation of the first stars appears to be strongly constrained by existing experimental probes such as the CMB power spectrum. 
We note briefly that in the case of annihilation, we can get both enhancement and suppression of the intensity of particles at the final step, likely violating the assumption that the same energy is deposited per baryon per time in the halo as in the homogeneous IGM over a broader swathe of parameter space. 

\begin{figure}
    \centering
    \includegraphics[scale=0.5]{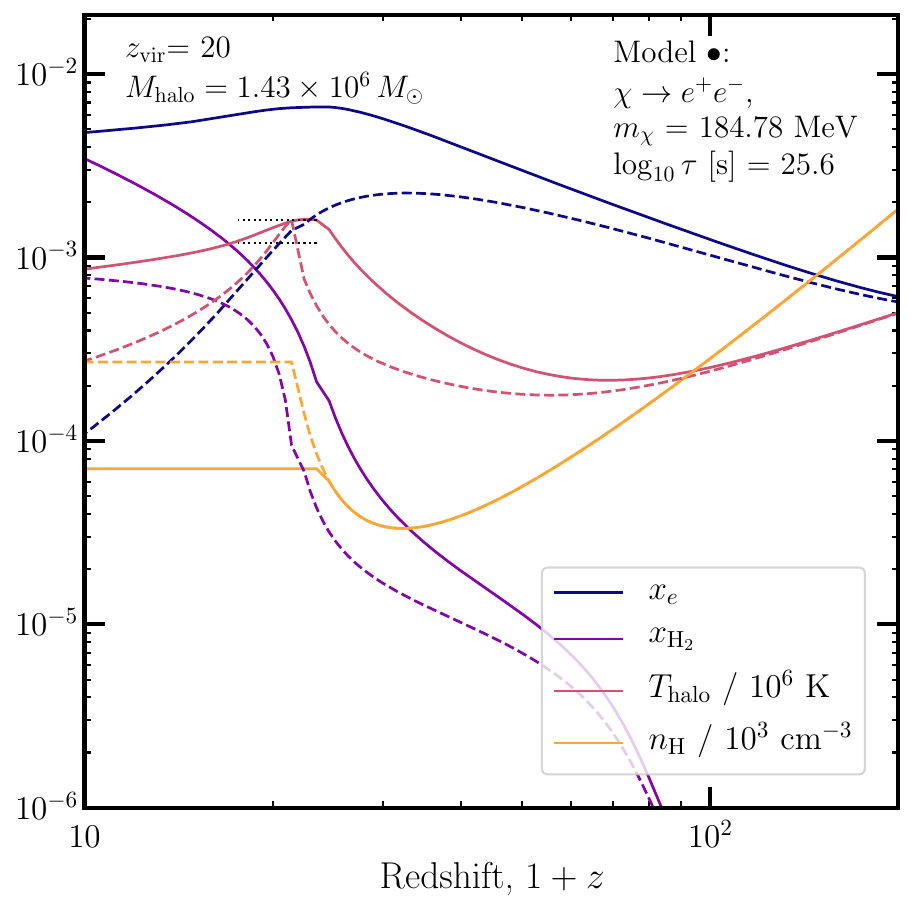}
    \hspace{5mm}
    \includegraphics[scale=0.5]{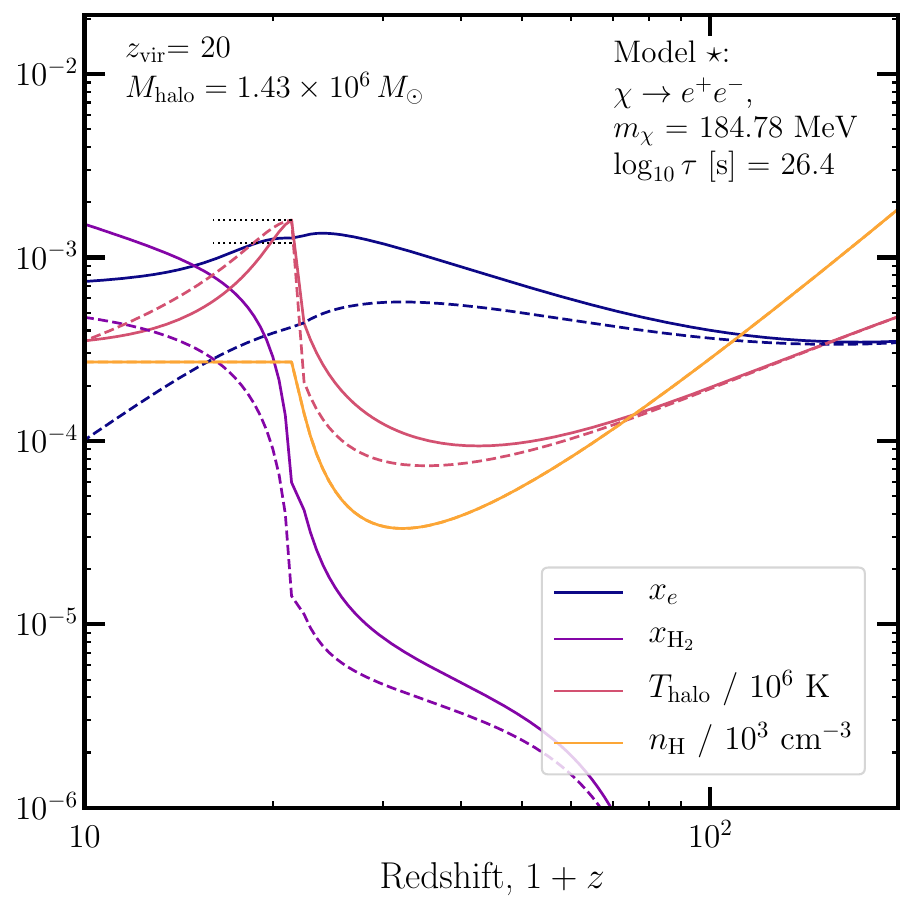}
    \caption{
    Halo evolution for Model $\bullet$\, (left) and Model $\star$\, (right), with the same halo as in Fig.~\ref{fig:halo_examples}.
    Solid lines indicate results assuming the same $f_c$'s as calculated in the IGM; dashed lines show results where the $f_c$'s are suppressed by a factor of the halo number density, i.e. the worst case scenario for energy deposition.
    The horizontal line segments on each panel indicate $T_\mathrm{vir}$ and $0.75 T_\mathrm{vir}$, and span the redshift range $(z_\mathrm{vir}, 0.75 z_\mathrm{vir})$; if the temperature curve crosses the lower line segment after virialization, the halo passes the criterion for collapsing and forming stars.
    }
    \label{fig:halo_f_suppressed}
\end{figure}

Fig.~\ref{fig:halo_f_suppressed} shows the difference in the halo evolution when we assume that the $f_c$'s are the same as in the IGM, or when we assume the $f_c$'s is suppressed by an additional factor of the overdensity $\Delta$.
For Model $\bullet$, this mild suppression of energy deposition reduces the effect of heating, such that the halo is no longer pressure supported when it virializes and is able to reach the virial density.
After virialization, the reduced heating rate means the halo cools much faster and is now able to pass the star-forming condition.
For Model $\star$, where the cooling rate of the halo is enhanced compared to the case with no exotic injection, reducing energy deposition causes the halo to cool more slowly.

From examining these two models, we can infer the ways in which Figs.~\ref{fig:scans_z20} and \ref{fig:scans_z100} would change if the $f_c$'s were suppressed for certain dark matter masses.
Overall, the contours would shift towards smaller lifetimes/larger cross-sections, since in order to have the same effect on a halo, one would have to dial up the rate of energy injection to counteract the reduced deposition rate.

\section{Bracketing Lyman-Werner effects for other channels}
\label{app:LW}

%
\begin{figure*}
    \includegraphics[scale=0.4]{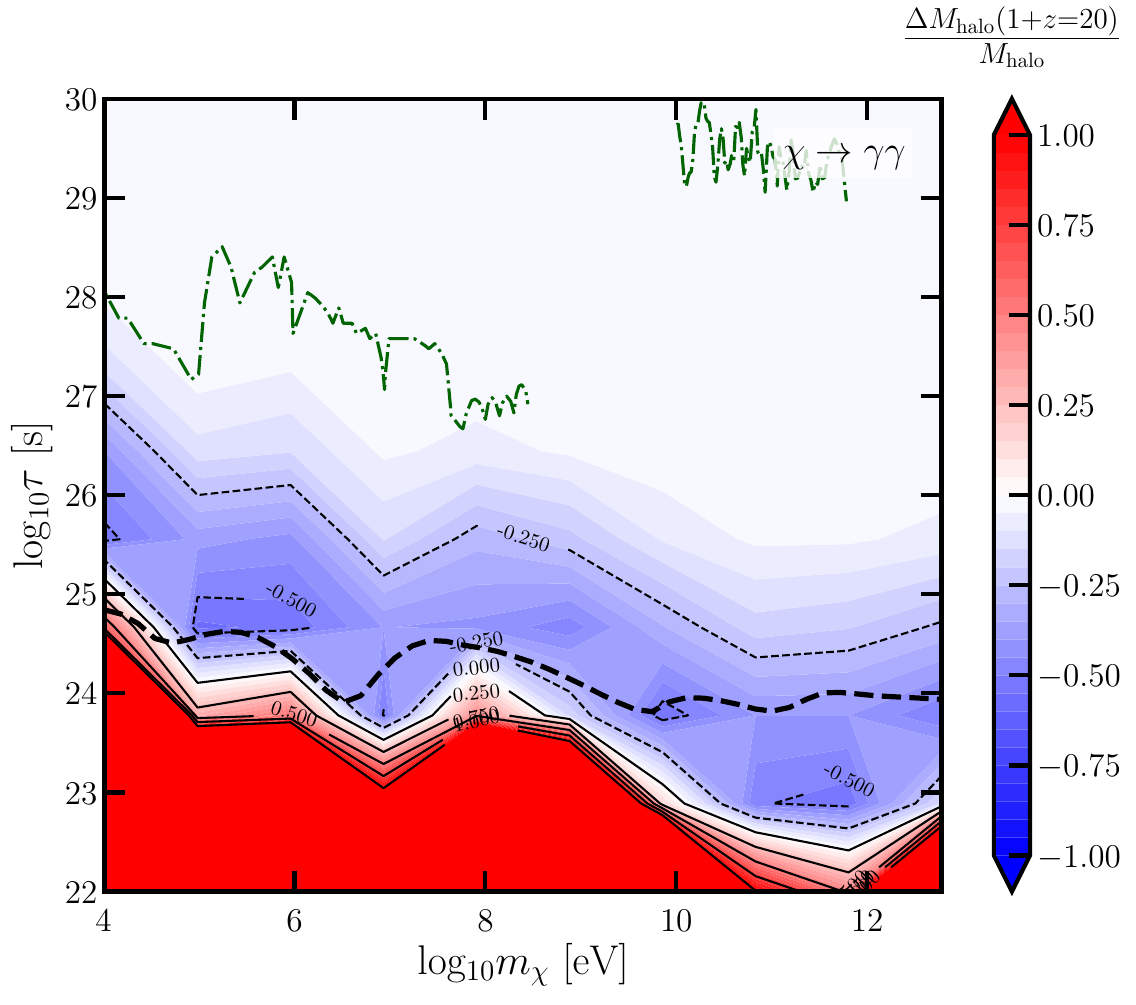}
    \includegraphics[scale=0.4]{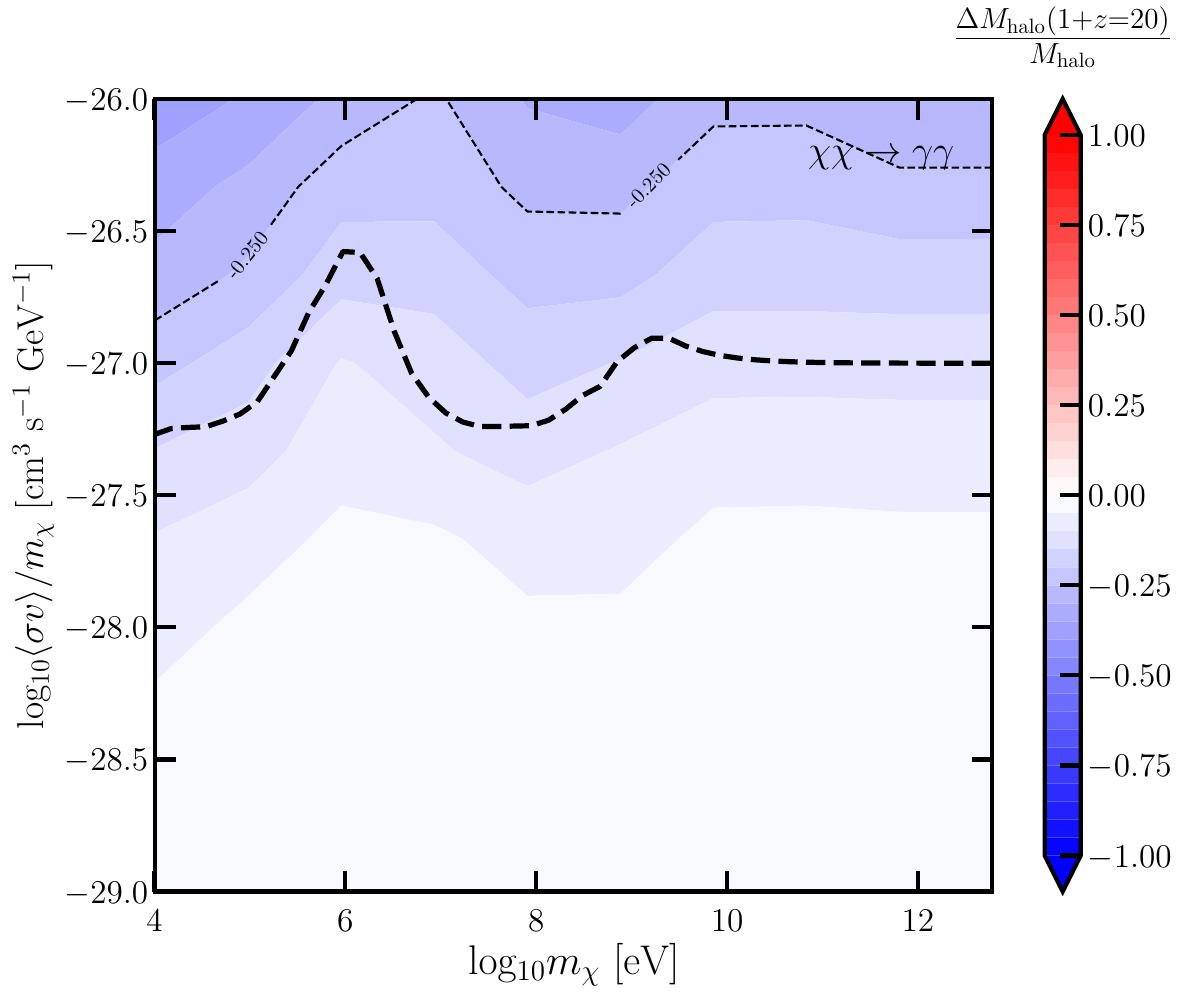}
    \includegraphics[scale=0.4]{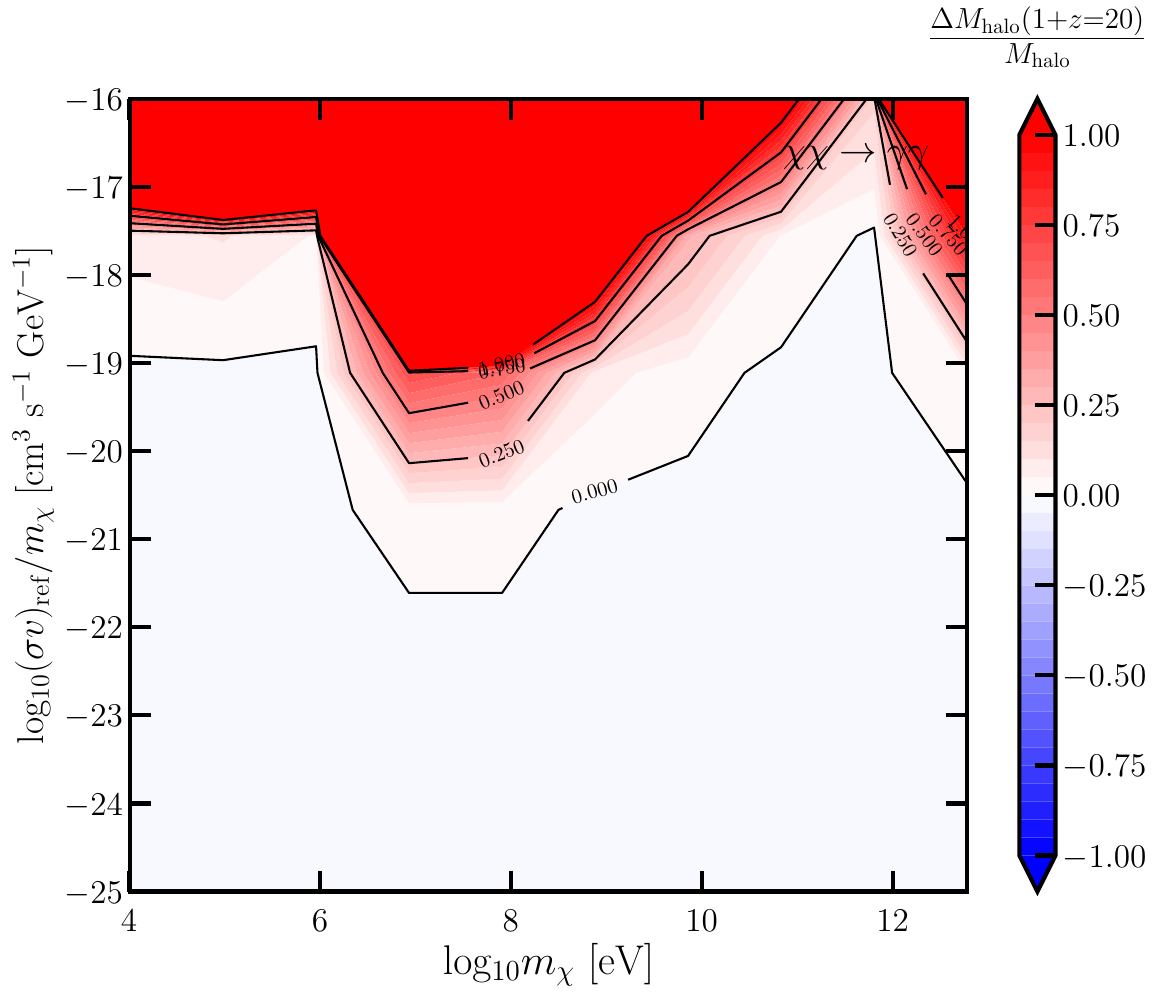}
    \includegraphics[scale=0.4]{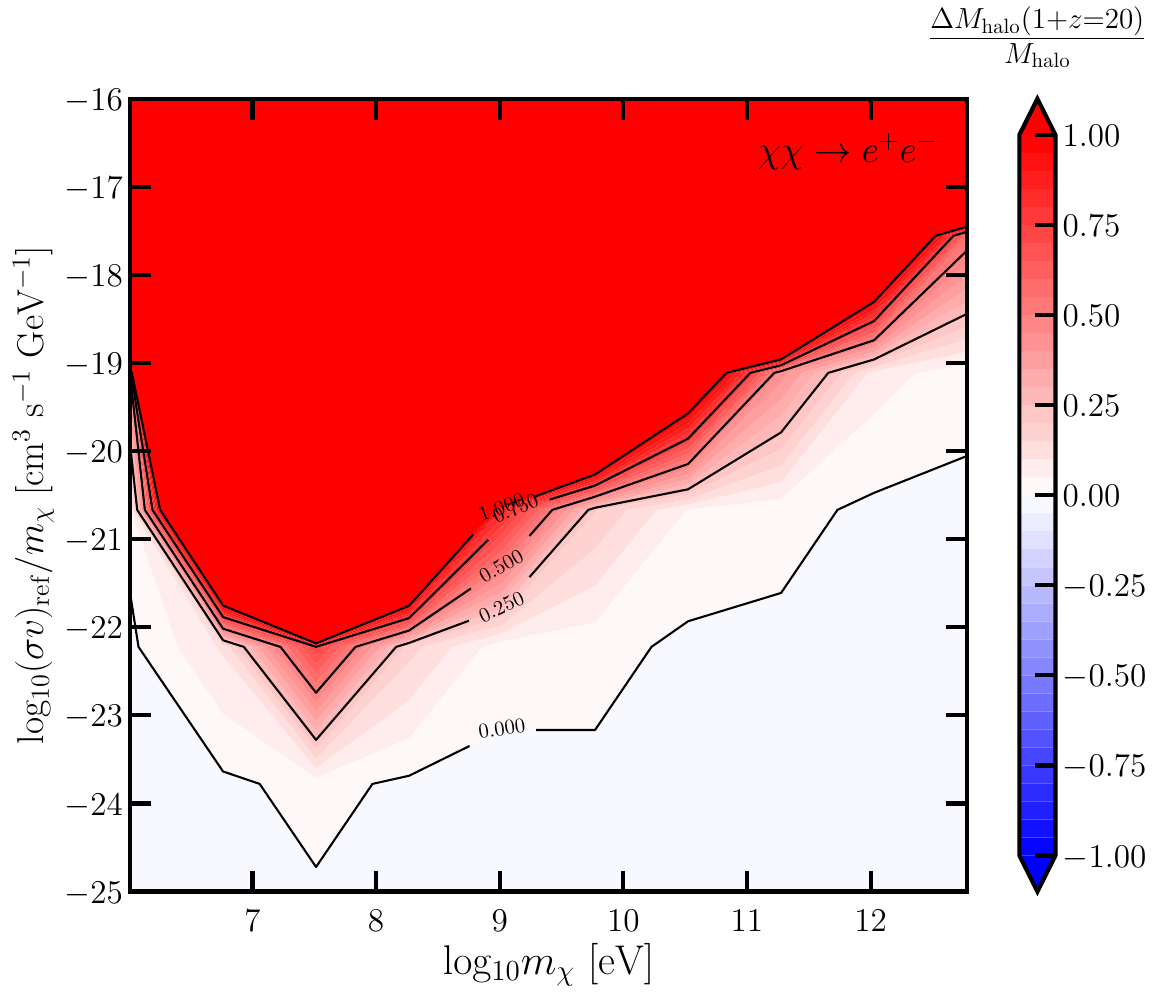}
    \caption{
    Same as Fig.~\ref{fig:scans_z20}, but assuming inefficient H$_2$ self-shielding.
    Clockwise from the top left, each panel shows the parameter space for decay to photons, $s$-wave annihilation to photons, $p$-wave annihilation to $e^+ e^-$ pairs, and $p$-wave annihilation to photons.
    }
    \label{fig:other_scans_z20_LW}
\end{figure*}

In Sec.~\ref{sec:shielding}, we showed the changes to the critical mass threshold in the parameter space for decay and $s$-wave annihilation to $e^+ e^-$ pairs, when self-shielding of the halo is inefficient.
Here, we discuss the other channels.
Fig.~\ref{fig:other_scans_z20_LW} shows the same results as in Fig.~\ref{fig:scans_z20_LW}, but for decay to photons, $s$-wave annihilation to photons, and $p$-wave annihilation to both $e^+ e^-$ pairs and photons.

For decay to photons, the depth of the blue contours is slightly reduced relative to the results assuming strong self-shielding, with the largest differences reaching to about 20\%.
The $s$-wave annihilation to photon results are only marginally impacted by self-shielding assumptions and differ by less than a percent in most of the parameter space shown.

For both $p$-wave annihilation channels, the depth of the red contours is significantly increased such that there is much more parameter space where we would likely see a significant delay to star formation.
We see that that $p$-wave to $e^+ e^-$ results are especially enhanced at masses of tens of MeV.
This is for the same reason as discussed in Sec.~\ref{sec:shielding} for the $s$-wave results; the primary electrons are injected at the right energy to upscatter CMB photons through ICS into the LW band.
However, as mentioned in Sec.~\ref{sec:results}, our assumption that energy deposition is very similar between the IGM and the halo is most likely to break for $p$-wave annihilation, where the energy injection is dominated by the largest halos with the highest velocity dispersions---we leave a more accurate calculation of the $p$-wave results to future study.

\bibliography{refs}
\end{document}